\documentclass[11pt,notitlepage]{article}

\usepackage{booktabs} 

\usepackage{fullpage}
\usepackage{theorem}
\usepackage{graphicx}
\usepackage{subfig}
\usepackage{float}
\usepackage{afterpage}
\usepackage{amsmath,amssymb}
\usepackage{color}
\usepackage{url}
\usepackage{bbm}
\usepackage{epsfig}
\usepackage{epstopdf}

\newtheorem{definition}{Definition}
\newtheorem{theorem}{Theorem}

\newcommand{\abs}[1]{\left| #1 \right|}

\renewcommand{\Pr}{\mathbf{Pr}}

\newcommand{\D}{\mathcal{D}}
\renewcommand{\O}{\mathcal{O}}

\renewcommand{\H}{\mathcal{H}}
\newcommand{\F}{\mathcal{F}}
\renewcommand{\S}{\mathcal{S}}

\newcommand{\genoa}{{\tt GEN50kS}}
\newcommand{\genoaa}{{\tt GEN50kS-aligned}}
\newcommand{\genob}{{\tt GEN20kS}}
\newcommand{\genoc}{{\tt GEN20kM}}
\newcommand{\genod}{{\tt GEN20kL}}
\newcommand{\genoe}{{\tt GEN80kS}}
\newcommand{\genof}{{\tt GEN320kS}}
\newcommand{\trec}{{\tt TREC}}
\newcommand{\uniref}{{\tt UNIREF}}

\newcommand{\ebdjoin}{{\tt EmbedJoin}}

\newcommand{\pass}{{\tt PassJoin}}
\newcommand{\edjoin}{{\tt EDJoin}}
\newcommand{\adpjoin}{{\tt AdaptJoin}}
\newcommand{\qchunk}{{\tt QChunk}}
\newcommand{\sft}{{\tt sft}}

\renewcommand{\paragraph}[1]{\medskip \noindent {\bf #1.}}

\usepackage{algpseudocode}
\usepackage{algorithmicx}
\usepackage{algorithm}

\algnewcommand\algorithmicforeach{\textbf{for each}}
\algdef{S}[FOR]{ForEach}[1]{\algorithmicforeach\ #1\ \algorithmicdo}

\usepackage{multirow}
\usepackage{array}
\usepackage[mathscr]{eucal}
\usepackage{chngpage}

\usepackage{authblk}
\usepackage{blindtext}

\begin{document}
\title{EmbedJoin: Efficient Edit Similarity Joins via Embeddings}

\author{Haoyu Zhang\thanks{Email: hz30@umail.iu.edu} }
\author{Qin Zhang\thanks{Email: qzhangcs@indiana.edu} }
\affil{Department of Computer Science, Indiana University}
\date{}
\maketitle

\begin{abstract}
We study the problem of edit similarity joins, where given a set of strings and a threshold value $K$, we want to output all pairs of strings whose edit distances are at most $K$.  Edit similarity join is a fundamental problem in data cleaning/integration, bioinformatics, collaborative filtering and natural language processing, and has been identified as a primitive operator for database systems.  This problem has been studied extensively in the literature. However, we  have observed that all the existing algorithms fall short on long strings and large distance thresholds.  

In this paper we propose an algorithm named \ebdjoin+ which scales very well with string length and distance threshold.  Our algorithm is built on the recent advance of metric embeddings for edit distance, and is very different from all of the previous approaches.  We demonstrate via an extensive set of experiments that \ebdjoin+ significantly outperforms the previous best algorithms on long strings and large distance thresholds. 
\end{abstract}

\maketitle

\section{Introduction}
\label{sec:intro}

Given a collection of strings, the task of similarity join is to find all pairs of strings whose similarities are above a predetermined threshold, where the similarity of two strings is measured by a specific distance function. Similarity join is a fundamental problem in data cleaning and integration (e.g., data deduplication), bioinformatics (e.g., find similar protein/DNA sequences), collaborative filtering (e.g., find user pairs of similar interests), natural language processing (e.g., automatic spelling corrections), etc. It has been studied extensively in the literature (see \cite{JLFL14} for a survey), and has been identified as one of the primitive operators for database systems~\cite{CGK06}.

In this paper we study similarity join under edit distance.  The edit distance between two strings $x$ and $y$, denoted by $\text{ED}(x, y)$, is defined to be the minimum number of edit operations (insertion, deletion and substitution) to transfer $x$ to $y$.  Formally, given a collection of strings $\S = \{s_1, s_2, \ldots, s_n\}$ over alphabet $\Sigma$, a similarity threshold $K$, edit similarity (self)join outputs
$$\{(s_i, s_j)\ |\ s_i, s_j \in \S; i \neq j; \text{ED}(s_i, s_j) \le K\}.$$ For example, given strings  {\tt ACCAT}, {\tt CCAAT}, {\tt GCCCT}, {\tt CACGA}, {\tt AACGG} and $K = 2$, the output pairs will be  ({\tt ACCAT, CCAAT}), ({\tt ACCAT, GCCCT}), ({\tt CACGA, AACGG}).

Compared with the Hamming distance
and token-based distances such as Cosine,
Jaccard,
Overlap
and Dice,
edit distance retains the information of the orderings of characters, and captures the best alignment of the two strings, which is critical to applications in bioinformatics, natural language processing and information retrieval.  On the other hand, edit distance is computationally more expensive than Hamming and token-based distances: computing edit distance takes at least quadratic time under the SETH conjecture~\cite{BI15}, while Hamming, Cosine, Jaccard, Overlap and Dice can be computed in linear time.  

Due to its difficulty and usefulness, a large portion of the similarity join literature has been devoted to edit distance~\cite{GJKMS01,AGK06,BMS07,BHSH07,LLL08,XWL08,WLF10,QWL11,WQX13,LDW11,WLF12}.  However, we have observed that all the existing approaches fall short on long strings and relatively large thresholds.  In the recent string similarity search/join competition, it was reported that ``an error rate of $20\% \sim 25\%$ pushes today's techniques to the limit''~\cite{WDG14}.  By  $20\%$ errors we mean that the distance threshold is set to be $20\%$ of the string length.  In fact the limit is reached much earlier on strings that are longer than those tested in the competition.  

However, long strings and large thresholds are critical to many applications. For example, documents can contain hundreds of thousands of characters; the lengths of DNA sequences range from thousands to billions of bases. If we set a threshold that is too small, then we may end up getting zero output pair which is certainly not interesting. 

\paragraph{Our Contribution}
The main contribution of this paper is a novel approach of computing edit similarity joins that scales very well with the string length and the distance threshold.  Different from all previous approaches which directly perform computations on the edit distance, we first embed the input strings from the edit space to the Hamming space, and then perform a filtering in the Hamming space using locality sensitive hashing.  

Our main algorithm, named \ebdjoin+, is randomized and may introduce a small number of errors (95\% - 99\% recall, 100\% precision in all of our experiments), but it significantly outperforms all the previous algorithms in both running time and memory usage on long strings and large thresholds.  In particular,  \ebdjoin+ scales very well up to error rate 20\% on large datasets which is beyond the reach of existing algorithms.

\paragraph{Overview of Our Approach}  
Given two strings $x, y \in \Sigma^N$, the Hamming distance between $x$ and $y$ is defined to be $\text{Ham}(x, y) = \sum_{i=1}^N \mathbf{1}(x_i \neq y_i)$.
Our approach is built on the recent advance of metric embeddings for edit distance, and is very different from all of the previous approaches.  In \cite{CGK16}, it has been shown that there exists an embedding function $f : \Sigma^N \to \Sigma^{3N}$ such that given $x, y \in \Sigma^N$,  we have with probability $1 - o(1)$ that~\footnote{The analysis in \cite{CGK16} in fact only gives ${\text{ED}(x, y)}/2 \le  \text{Ham}(f(x), f(y))$. However, as we shall describe in Algorithm~\ref{alg:CGK}, if we pad the embedded strings using a character that is not in the dictionary, then it is easy to show that $\text{ED}(x, y) \le \text{Ham}(f(x), f(y))$ with probability $1 - o(1)$.}
\begin{eqnarray*}
\label{eq:a-1}
{\text{ED}(x, y)} &\le&  \text{Ham}(f(x), f(y)),
\end{eqnarray*}
and with probability at least $0.999$ that
\begin{eqnarray*}
\label{eq:a-2}
\text{Ham}(f(x), f(y)) &\le& O\left((\text{ED}(x, y))^2\right).
\end{eqnarray*}
We call this scheme the {\em CGK-embedding}, named after the initials of the authors in \cite{CGK16}.  The details of the embedding algorithm will be illustrated in Section~\ref{sec:CGK}.  We call 
$$D(x, y) = \text{Ham}(f(x), f(y)) / \text{ED}(x, y)$$
the {\em distortion} of the CGK-embedding on input $(x, y)$.  Note that 
if $\text{ED}(x, y) \le K$, then $1 \le D(x, y) \le O(K)$ with probability at least $0.99$.  


The high level idea of our approach is fairly simple: we first embed using CGK all the strings from the edit space to the Hamming space, and then perform a filtering step on the resulting vectors in the Hamming space using locality sensitive hashing (LSH) \cite{IM98,GIM99}.  LSH has the property that it will map a pair of items of small Hamming distance to the same bucket in the hash table with good probability, and map a pair of items of large Hamming distance to different buckets with good probability.  The final step is to verify for each hash bucket $B$, and for all the strings hashed into $B$, whether their pairwise edit distances are at most $K$ or not, by an exact dynamic programming based edit distance computation.  This finishes the high level description of our basic algorithm which we name \ebdjoin. \ebdjoin\ works very well on datasets where there is a non-trivial gap between distances of similar pairs and dissimilar pairs, but does not give satisfactory accuracy on datasets where the gap is very small (e.g., random reads of DNA sequences).  We thus further improve \ebdjoin\  by adding a couple of new ideas to deal with string shifts, and obtain \ebdjoin+ which works well on all the datasets that we have tested.
  
One may observe that the worst-case distortion of the CGK-embedding can be fairly large if the threshold $K$ is large. However, we have observed that the practical performance of CGK-embedding is much better. We will give more discussions on this phenomenon in Section~\ref{sec:CGK}.   To further reduce the distortion, we choose to run the embedding multiple times, and then for each pair of strings we choose the run with the minimum Hamming distance for the filtering.  This minimization step does not have to be performed explicitly since we do not have to compute Ham$(f(x), f(y))$ for all pairs of strings which is time consuming. We instead integrate this step with LSH for a fast filtering.  

Finally, we note that since LSH is a dimension reduction step, LSH-based filtering naturally fits long strings  (e.g., DNA sequences) which are our main interest.  For short strings LSH-based filtering may not be the most effective approach and one may want to use different filtering methods.  We also note Satuluri et al.~\cite{SP12} used LSH-based filtering for computing similarity joins under the Jaccard distance and the Cosine distance.   Unfortunately there is no efficient LSH for edit distance, which is the motivation for us to first embed the strings to vectors in the Hamming space and then perform LSH.  




A preliminary version of this article appeared in \cite{ZZ17}, where only the basic version of \ebdjoin+, namely \ebdjoin, was proposed.   
Compared with \cite{ZZ17}, Section~\ref{sec:exact-ED} is newly added where have changed the algorithm for computing exact edit distance in the verification phase of \ebdjoin\ and \ebdjoin+. Part of Section~\ref{sec:speedup} has been rewritten. 
Section~\ref{sec:improved} for \ebdjoin+ is entirely new.  All the experiments in Section~\ref{sec:exp} have been redone, in particular, for the new algorithm \ebdjoin+.

\paragraph{Roadmap} The rest of this paper is organized as follows.  In Section~\ref{sec:related} we survey related work on edit similarity joins.  In Section~\ref{sec:tool} we describe a set of tools that we make use of in our algorithms.
In Section~\ref{sec:algo} we describe \ebdjoin\ which is a basic version of \ebdjoin+, and then in Section~\ref{sec:improved} we show our main algorithm \ebdjoin+.  We present experimental studies in Section~\ref{sec:exp}, and conclude the paper in Section~\ref{sec:conclude}.

\section{Related Work}
\label{sec:related}

\paragraph{Similarity Joins for Edit Distance}
The edit similarity join problem has been studied extensively in the literature.  We refer the readers to \cite{JLFL14} for a comprehensive survey.  A widely adopted approach to this problem is to first generate for each string a set of signatures/substrings. For example, in the $q$-gram signature, we generate all substrings of length $q$ (e.g., when $q=2$, the $2$-grams of {\tt ACCAT} is \{{\tt AC, CC, CA, AT}\}). We then perform a filtering step based on the frequencies, positions and/or the contents of these substrings.  The filtering step will give a set of candidate (similar) pairs, for each of which we use a dynamic programming algorithm for edit distance to verify its {\em exact} similarity.  Concrete algorithms of signature-based approach include {\tt GramCount} \cite{GJKMS01}, {\tt AllPair} \cite{BMS07}, {\tt FastSS} \cite{BHSH07}, 
{\tt ListMerger} \cite{LLL08}, {\tt EDJoin} \cite{XWL08}, {\tt QChunk} \cite{QWL11}, {\tt VChunk} \cite{WQX13}, {\tt PassJoin} \cite{LDW11}, and {\tt AdaptJoin} \cite{WLF12}.  We will briefly describe in Section~\ref{sec:setup} the best ones among these algorithms which we use as competitors to \ebdjoin+ in our experiments.

While different signature-based algorithms use different filtering methods, their common feature is to first compute some upper or lower bounds, and then prune those pairs $(x, y)$ for which $g(sig(x), sig(y))$ is above or below the predetermined upper/lower bounds, where $g$ is a predefined function, and $sig(x), sig(y)$ are signatures of $x$ and $y$ respectively.  The main drawback of signature-based approach is that the information about the sequence ordering is somewhat lost when converting strings to a set of substrings.  Another issue is that the precomputed upper/lower bounds may be too loose for effective pruning.  

There are a few other approaches for computing edit similarity joins, such as trie-based algorithm {\tt TrieJoin}~\cite{WLF10}, tree-based algorithm {\tt M-Tree}~\cite{CPZ97}, enumeration-based algorithm {\tt PartEnum}~\cite{AGK06}.  However, as reported in \cite{JLFL14}, these algorithms are not very effective on datasets of long strings.


\paragraph{Similarity Joins for Other Metrics} 
Similarity joins have been studied for a number of other metrics~\cite{GJKMS01,AGK06,BMS07,LLL08,XWLY08,WLF12,LDW11,ZLG11}, including Cosine, Jaccard, Overlap and Dice.  A survey of these works is beyond the scope of this paper, and we again refer reader to \cite{JLFL14} for an overview.

\paragraph{Other Related Work on Edit Distance}
Edit distance is also a notoriously difficult metric for sketching and embeddings, and very little is known in these frontiers.  As mentioned, embedding enables us to study the similarity join problem in an easier metric space. On the other hand, if we can efficiently obtain small sketches of the input strings, then we can solve the similarity join problem on smaller inputs.
Ostrovsky and Rabani proposed an embedding from the edit metric to the $\ell_1$ metric with an $\exp(O(\sqrt{\log N\log\log N}))$ distortion~\cite{OR07} where $N$ is the length of the string. A corresponding distortion lower bound of $\Omega(\log N)$ has been obtained by Kraughgamer and Rabani~\cite{KR09}.  Recently Chakraborty et al.\ gives a weak embedding to the Hamming space~\cite{CGK16} with an $O(K)$ distortion,\footnote{In a weak embedding, the distortion holds for {\em each} pair of strings with constant probability, say, 0.99. In contrast, in a strong embedding, with probability $0.99$ the distortion holds for {\em all} pairs of strings simultaneously.} which serves as the main tool in our algorithm.   For sketching, very recently Belazzougui and Zhang~\cite{BZ16} proposed the first almost linear time sketching algorithm that gives a sketch of sublinear size (more precisely, $O(K^8 \log^5 N)$), which, unfortunately, is still too large to be useful in practice in its current form. 

We will briefly survey algorithms for computing edit distance in the RAM and simultaneous streaming models in Section~\ref{sec:exact-ED}.


\section{Tools}
\label{sec:tool}

Before presenting our algorithm, we would like to introduce a few tools that we shall use in \ebdjoin+, including the CGK-embedding, the LSH for the Hamming distance, and an algorithm for exact edit distance computation. We list in Table~\ref{tab:notation} a set of notations that will be used in the presentation.

\begin{table}[t]
\centering
\begin{tabular}{|p{.12\textwidth}| p{.85\textwidth}| m{.01\textwidth}|} 
\hline
Notation & Definition\\ 
\hline
$[n]$ & $[n] = \{1, 2, \ldots, n\}$ \\
\hline
$K$ & Edit distance threshold\\ 
\hline
$\S$ & The set of input strings \\ 
\hline
$s_i$ & The $i$-th string in $\S$ \\ 
\hline
$\abs{x}$ & Length of string $x$ \\ 
\hline
$n$ & Number of input strings, i.e.,  $n = \abs{\S}$\\ 
\hline
$N$ & Maximum length of strings in $\S$\\ 
\hline
$\Sigma$ & Alphabet of strings in $\S$ \\ 
\hline
$r$ & Number of CGK-embeddings for each input string\\ 
\hline
$t_i^{\ell}$ & The output string generated by the $\ell$-th CGK-embedding of $s_i$ \\ 
\hline
$z$ & Number of hash functions used in LSH for each string generated by CGK-embedding\\ 
\hline
$m$ & Length of the LSH signature\\ 
\hline
$f_j^{\ell} $ &   $f_j^{\ell} : \Sigma^N \to \Sigma^m$, the $j$-th ($j\in[z]$)  LSH function for each string generated by the $\ell$-th CGK-embedding
\\ 
\hline
$\mathcal{D}_j^{\ell}$ & The hash table corresponding to the LSH function $f_j^{\ell}$ \\ 
\hline
$\Delta$ & A parameter for dealing with shifts  \\ 
\hline
$s_{i,k}$ & The $k$-th substring of $s_i$ starting at the $((k-1)\Delta+1)$-th character\\ 
\hline
$t_{i,k}^\ell$ & The output string generated by the $\ell$-th CGK-embedding of $s_{i,k}$\\ 
\hline
$T$ & The threshold of the number of matched hash signatures for a pair of substrings\\ 
\hline
\end{tabular}
\caption{Summary of Notations}
\label{tab:notation}
\end{table}

\subsection{The CGK-Embedding}
\label{sec:CGK}

We describe the CGK-embedding in Algorithm~\ref{alg:CGK}. Below we  illustrate the main idea behind the CGK-embedding, which we believe is useful and important to understand the intuition of \ebdjoin+.  We note that the original algorithm in \cite{CGK16} was only described for binary strings, and it was mentioned that we can encode an alphabet $\Sigma$ into binary codes using $\log\abs{\Sigma}$ bits for each character.  In our rewrite (Algorithm~\ref{alg:CGK}) we choose to use the alphabet $\Sigma$ directly without the encoding.  This may give some performance gain when the size of the alphabet is small.

\begin{algorithm}[t]
\begin{algorithmic}[1]
\Require A string $x \in \Sigma^\eta$ for some $\eta \le N$, and a random string $R \in \{0,1\}^{ 3 N |\Sigma|}$
\Ensure A string $x' \in \Sigma^{3N}$
\smallskip


\State Interpret $R$ as a set of functions 

$\pi_1,\dots,\pi_{3N}:  \Sigma \rightarrow \{0,1\}$; for the $k$-th char $\sigma_k$ in $\Sigma$, 

$\pi_j(\sigma_k) = R[(j-1) \cdot \abs{\Sigma} + k]$
\State $i \leftarrow 1$
\State $x' \leftarrow \emptyset$ 
\For{$j \in [3N]$} 
	\If{$i\le \abs{x}$}
		\State $x' \leftarrow x' \odot x[i]$ 
		\Comment the ``$\odot$'' denotes concatenation
		\State $i \leftarrow i+\pi_j(x[i])$
	\Else
		\State $x' \leftarrow x' \odot \perp$
		\Comment ``$\perp$'' can be an arbitrary character outside $\Sigma$
	\EndIf
\EndFor 
\end{algorithmic}
\caption{CGK-Embedding($s$, $R$) \cite{CGK16}}
\label{alg:CGK}
\end{algorithm}

Let $N$ be the maximum length of all input strings in $\S$. The CGK-embedding maps a string $x \in \S$ to an output string $x' \in \Sigma^{3N}$ using a random bit string $R \in \{0,1\}^{3N\abs{\Sigma}}$.  We maintain a counter $i \in [1 .. \abs{x}]$ pointing to the input string $x$, initialized to be $1$.  The embedding proceeds by steps $j = 1, \ldots, 3N$. At the $j$-th step, we first copy $x[i]$ to $x'[j]$. Next, with probability $1/2$, we increase $i$ by $1$, and with the rest of the probability we keep $i$ to be the same.  At the point when $i > \abs{x}$, if $j$ is still no more than $3N$, we simply pad an arbitrary character outside the dictionary $\Sigma$ (denoted by ``$\perp$'' in Algorithm~\ref{alg:CGK}) to make the length of $x'$ to be $3N$.  In practice this may introduce quite some overhead for short strings in the case that the string lengths vary significantly. We will discuss in Section~\ref{sec:speedup} how to efficiently deal with input strings of very different lengths.

Now consider two input stings $x$ and $y$.  We use $i_0$ and $i_1$ as two counters pointing to $x$ and $y$ respectively.   At the $j$-th step, we first copy $x[i_0]$ to $x'[j]$, and $y[i_1]$ to $y'[j]$, and then decide whether to increment $i_0$ and $i_1$ using the random bit string $R$.  There are four possibilities: (1) only $i_0$ increments; (2) only $i_1$ increments; (3) both $i_0$ and $i_1$ increment; and (4) neither $i_0$ nor $i_1$ increments.  Let $d = i_0 - i_1$ be the position shift of the two counters/pointers on the two strings.  Note that if $x[i_0] = y[i_1]$, then only the cases (3) and (4) can happen, so that $d$ will remain the same.  Otherwise if $x[i_0] \neq y[i_1]$, then each case can happen with probability $1/4$ -- whether $i_0$ or $i_1$ will increment depends on the two random hash values $\pi_j(x[i_0])$ and $\pi_j(y[i_1])$. Thus with probability $1/4$, $1/2$ and $1/4$, the value $d$ will increment, remain the same, or decrement, respectively.  Ignoring the case when the value $d$ remains the same, we can view $d$ as a (different) {\em simple random walk} on the integer line with $0$ as the origin.  

We now try to illustrate the high level idea of why CGK-embedding gives an $O(K)$ distortion.  Let $u = \abs{x}$ and $v = \abs{y}$.  Suppose that at some step $j$, letting $p = i_0(j)$ (the value of $i_0$ at step $j$) and $q = i_1(j)$, we have two tails $x[p .. u] = \alpha \circ \tau$ and $y = y[q .. v] = \tau$ where $\alpha, \tau$ are two substrings and $\abs{\alpha} = k \le K$. That is, we have $k$ consecutive deletions in the optimal alignment of the two tails.  Now if after a few random walk steps, at step $j' > j$, we have $p' = i_0(j') \ge p + k$, $q' = i_1(j') \ge q$ and $p' - q' = (p - q) + k$, then the two tails $x[p'..u]$ and $y[q'..v]$ can be perfectly aligned, and consequently the pairs of characters in the output strings $x', y'$ will always be the same; in other words, they will {\em not} contribute to the Hamming distance from step $j'$. 

Now observe that since the value of $d$ changes according to a simple random walk, by the theory of random walk, with probability $0.999$ it takes at most $O(k^2)$ steps for $d$ to go from $(p - q)$ to $(p' - q')$ where $\abs{(p - q) - (p' - q')} = k$.  Therefore the number of steps $j$ where $x'[j] \neq y'[j]$ is bounded by $O(k^2)$.  
This is roughly why $\text{Ham}(x', y')$ can be bounded by $O(K^2)$ if $\text{ED}(x, y) \le K$, and consequently the distortion can be bounded by $O(K)$.


\paragraph{Small Distortion is Good for Edit Similarity Join}  We now explain why the distortion of the embedding matters. If we have an embedding $f$ such that for any pair of input strings $(x, y)$, the distortion of the embedding is upper bounded by $D$, then the set $\{(x, y)\ |\ \text{Ham}(f(x), f(y)) \le D \cdot K\}$ will include all pairs $(x, y)$ such that $\text{ED}(x, y) \le K$.  Therefore a small $D$ can help to reduce the number of false positives, and consequently reduce the verification time which typically dominates the total running time.

\medskip

\noindent{\bf Why CGK-embedding Does Better in Practice?}
Although the worst-case distortion of CGK-embedding can be large when $\text{ED}(x, y)$ is large, we have observed that its practical performance on the datasets that we have tested is much better.  While it is difficult to fully understand this phenomenon without a thorough investigation of the actual properties of the datasets, we can think of the following reasons.

First, if a set of $z$ edits fall into an interval of length $O(z)$, {\em and} the difference between the numbers of insertions and deletions among the $z$ edits is at most $O(\sqrt{z})$ (substitutions do not matter), then with probability $0.999$ after $O(z)$ walk steps the random walk will re-synchronize.  In other words, the distortion of the embedding is $O(1)$ with probability $0.999$ on this cluster of edits.  We have observed that in our protein/genome datasets (Section~\ref{sec:setup}) the edits are often clustered into small intervals; in each cluster most edits are substitutions, and consequently the difference between the numbers of insertions and deletions is small. 

Second, in the task of differentiating similar pairs of strings and dissimilar pairs of strings, as long as the distance gap between strings is preserved after the embedding, the distortion of CGK-embedding will not affect the performance by much. In particular, when the distortion of CGK-embedding is $\Theta(k)$ (which is very likely when edits are well separated), the embedding actually {\em amplifies} the distance gap between similar and dissimilar pairs, which makes the next LSH step easier.

\begin{figure}[t]
\centering
\includegraphics[height = 1.6in]{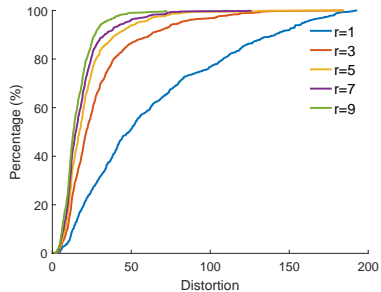}
\caption{The CDF of the best distortions of 1000 random pairs strings from the \genoa\ dataset, under different numbers of CGK-embeddings (value $r$)}
\label{fig:CGK-distortion}
\end{figure}

 
To further improve the effectiveness of the CGK-embedding, we run the embedding multiple times and then take the one with the minimum Hamming distance. That is, we choose the run with the best distortion.  This is just a heuristic, and cannot improve the distortion by much in theory, but we have observed that for the real-world datasets that we have tested, repeating and then taking the minimum does help to reduce the distortion. In Figure~\ref{fig:CGK-distortion} we depicted the best distortions under different numbers of runs of the CGK-embedding on a real-world genome dataset.

\subsection{LSH for the Hamming Distance}
\label{sec:LSH}

Our second tool is the LSH for the Hamming distance, introduced in~\cite{IM98,GIM99} for solving nearest neighbor problems.  We first give the definition of LSH.  By $h \in_r \H$ we mean sampling a hash function $h$ randomly from a hash family $\H$. 

\begin{definition}(Locality Sensitive Hashing \cite{GIM99})
Let $U$ be the item universe, and $d(\cdot,\cdot)$ be a distance function. We say a hash family $\mathcal{H}$ is $(l, u, p_1,p_2)$-sensitive if for any $x, y \in U$
\begin{itemize}
\item if $d(x, y)\le l$, then $\Pr_{h\in_r \H}[h(x) = h(y)] \ge p_1$,
\item if $d(x, y)\ge u$, then $\Pr_{h\in_r \H}[h(x) = h(y)] \le p_2$.
\end{itemize}
\end{definition}

We will make use of the following vanilla version of LSH for the Hamming distance.
\begin{theorem}(Bit-sampling LSH for Hamming  \cite{GIM99})
For the Hamming distance over vectors in $\Sigma^N$, for any $d > 0, c > 1$, the family $$\mathcal{H}_N = \{v_i: v_i(b_1, \dots, b_N) = b_i\ |\ i \in [N]\}$$ is  $(d, cd, 1-{d}/{N}, 1-{cd}/{N})$-sensitive.
\end{theorem}

We can use the standard AND-OR amplification method\footnote{See, for example, \url{https://en.wikipedia.org/wiki/Locality-sensitive_hashing}.} to amplify the gap between $p_1$ and $p_2$.
We first concatenate $m$ ($m$ is a parameter) hash functions, and define 
\begin{eqnarray*}
f = h_1 \circ h_2 \circ \ldots \circ h_m \text{ where } \forall i \in [m], h_i \in_r \H,
\end{eqnarray*}
such that for $x \in U$, $f(x) = (h_1(x), h_2(x), \ldots, h_m(x))$ is a vector of $m$ bits.  Let $\F(m)$ be the set of all such hash functions $f$.
We then define (for a parameter $z$)
\begin{eqnarray*}
g = f_1 \vee f_2 \vee \ldots \vee f_z, \text{ where } \forall j \in [z], f_j \in_r \F(m),
\end{eqnarray*}
such that for $x, y \in U$ $g(x) = g(y)$ if and only if there is at least one $j \in [z]$ for which $f_j(x) = f_j(y)$.  Easy calculation shows that $g$ is
\begin{eqnarray*}
\left(d, cd, 1 - \left(1 - \left({d}/{N}\right)^m\right)^z, 1 - \left(1 - \left( {cd}/{N} \right)^m \right)^z \right) \text{-sensitive.}
\end{eqnarray*}
 By appropriately choosing the parameters $m$ and $z$,  we can amplify the gap between $p_1$ and $p_2$ so as to reduce the numbers of false positives/negatives.  


\subsection{Exact Edit Distance Computation for Verification}
\label{sec:exact-ED}

We will use the classic algorithm by Ukkonen~\cite{Ukkonen85} for computing threshold edit distance as our verification algorithm.  In the high level, defining the diagonal $d$ of a matrix $D$ to be the set of all entries $D_{i, i+d}$,
the algorithm tries to fill {\em a subset of} the entries in the $2K+1$ diagonals $\{-K, \ldots, K\}$ in the $N \times N$ dynamic programming matrix, which are sufficient to give the final output.  The worst-case running time of this algorithm is $O(N K)$. But if one of the strings is a random string, then the algorithm only uses $O(N + K^2)$ time in expectation \cite{Myers86}.   In \cite{Myers86}, Myers also proposed another algorithm using suffix-tree whose worse-case running time is $O(N+K^2)$. However, we found that suffix-tree is computational expensive in practice and has no advantage over a ``brute force'' table filing \cite{Ukkonen85}. 

We also note that Belazzougui and Zhang~\cite{BZ16} (and independently, Chakraborty et al.~\cite{CGK16b}) showed that the $O(N + K^2)$ running time is also achievable in the simultaneous streaming model where we can only scan each string once in the coordinated fashion.  However, the algorithms in \cite{BZ16,CGK16b} still needs to use suffix-tree.  Chakraborty et al.~\cite{CGK16b} also proposed an algorithm with $N + O(K^3)$ running time in the simultaneous streaming model {\em without} using suffix-tree, but this bound would be large when the distance threshold $K$ is large, say, $20\%$ of the string length $N$.

In an earlier version of this paper~\cite{ZZ17} we used the algorithm in~\cite{LDW11} for computing edit distance in the verification step.  We later found that it is more efficient to use Ukkonen's algorithm.

\section{The \ebdjoin\ Algorithm}
\label{sec:algo}

\label{sec:embed}

Now we are ready to describe our basic algorithm \ebdjoin, which is presented in Algorithm~\ref{alg:embed-join} using Algorithm~\ref{alg:preprocessing} as a subroutine. We explain them in words below.

\begin{algorithm}[t]
\caption{Preprocessing ($\mathcal{S}$, $r$, $z$, $m$)}
\label{alg:preprocessing}
\begin{algorithmic}[1]
\Require Set of input strings $\mathcal{S} = \{s_1, \ldots, s_n\}$, and parameters $r$, $z$ and $m$ described in Table~\ref{tab:notation}
\Ensure  Strings in $\mathcal{S}$ in the sorted order, strings after CGK-embedding $\{t_i^\ell\ |\ \ell \in [r], i \in [n]\}$, and hash tables $\{\D_j^{\ell} \ |\ \ell \in [r], j \in [z]\}$.
\State Sort $\mathcal{S}$ first by string length increasingly, and second by the alphabetical order. 

\ForEach{$\ell \in [r]$}
	\ForEach{$j \in [z]$}
	\State Initialize hash table $\D_j^{\ell} $ by generating a random hash function $f_j^{\ell} \in \F(m)$  
	\EndFor
\EndFor

\ForEach{$\ell \in [r]$}
\State Generate a random string $R^{\ell} \in \{0,1\}^{3 N \abs{\Sigma}}$
\ForEach{$s_i\in \mathcal{S}$}
\State $t_i^{\ell}  \leftarrow $ CGK-Embedding($s_i$, $R^{\ell}$) 
\EndFor
\EndFor
\end{algorithmic}
\end{algorithm}

\begin{algorithm}[t]
\caption{EmbedJoin ($\mathcal{S}$, $K$, $r$, $z$, $m$)}
\label{alg:embed-join}
\begin{algorithmic}[1]
\Require Set of input strings $\mathcal{S} = \{s_1, \ldots, s_n\}$, distance threshold $K$, and parameters $r$, $z$ and $m$ described in Table~\ref{tab:notation}
\Ensure  $\O \gets \{(s_i, s_j)\ |\ s_i, s_j \in \S; i \neq j; \text{ED}(s_i, s_j) \le K\}$ 
\State Preprocessing$(\S, r, z, m)$  
\Comment Using Algorithm~\ref{alg:preprocessing}
\State  $\mathcal{C} \leftarrow \emptyset $  
\Comment Collection of candidate pairs

\ForEach{$s_i \in \mathcal{S} \text{ (in the sorted order)}$}
	\ForEach{$\ell \in [r]$}
		\ForEach{$j \in [z]$}
			\ForEach{string $s$ stored in the $f_j^\ell (t_i^\ell)\text{-th bucket of table} \ \D_j^\ell $}
				\If{$\abs{s_i} - \abs{s} \le K$} \label{ln:a-1}
				\State $\mathcal{C}  \leftarrow \mathcal{C}  \cup (s, s_i)$ 
\Else
				\State Remove $s$ from $\D_j^\ell$ \label{ln:a-2}
				\EndIf
			\EndFor
			\State Store $s_i$ in the $f_j^\ell (t_i^\ell)$-th bucket of $\D_j^\ell$
		\EndFor
	\EndFor
	\State Remove duplicate pairs in $\mathcal{C}$ \label{ln:a-3}
\EndFor

\ForEach {$(x, y) \in \mathcal{C}$}
	\If{$\text{ED}(x, y) \le K$}
	\Comment Using the algorithm in \cite{Ukkonen85}
	\State $\mathcal{O} \leftarrow  \mathcal{O} \cup (x, y)$
	\EndIf
\EndFor
\end{algorithmic}
\end{algorithm}

In the preprocessing we generate $r \times z$ hash tables $\D_j^\ell\ (\ell \in [r], j \in [z])$ implicitly by sampling  $r \times z$ random hash functions $f_j^\ell\ (\ell \in [r], j \in [z])$ from $\F(m)$ (defined in Section~\ref{sec:LSH}).  We then CGK-embed each string $s_i \in \S$ for $r$ times, getting $t_i^\ell\ (\ell \in [r])$.

Similar to previous algorithms, \ebdjoin\ has two stages: it first finds a small set of candidate pairs, and then verifies each of them using exact edit-distance computation via dynamic programming.  We use the algorithm for computing edit distance in~\cite{Ukkonen85} for the second step.  In the rest of this section we explain the first filtering step.

The main idea of the filtering step is fairly straightforward. We use LSH to find all pairs $(s_i, s_j)$ for which there exists an $\ell \in [r]$ such that $t_i^\ell$ and $t_j^\ell$ are hashed into the same bucket by at least one of the hash functions $f_j^\ell \in \F(m)\ (j \in [z])$.  In other words, for at least one of the $r$ CGK-embeddings, the output pairs corresponding to $s_i$ and $s_j$ are identified to be similar by at least one of the $z$ LSH functions.  Recall that we do $r$ repetitions of CGK-embedding to achieve a good distortion ratio (see the discussion in Section~\ref{sec:CGK}), and we use $z$ LSH functions from $\F(m)$ to amplify the gap between $p_1$ and $p_2$ in the definition of LSH to reduce false positives/negatives (see the discussion in Section~\ref{sec:LSH}).

In the actual implementation, we use a sliding window to speed-up the filtering: We first sort the input strings in $\S$ according to their lengths increasingly (breaking ties by the alphabetical orders of the strings).  We then process them one by one. If $s_i \in S$ is hashed into some bucket $B$ in the hash table, when fetching each string $s$ in $B$ we first test whether $\abs{s_i} - \abs{s} \le K$ (Line~\ref{ln:a-1}).  If not, we can immediately conclude $\text{ED}(s, s_i) > K$, and consequently $\text{ED}(s, s_{i'}) > K \ (i' > i)$ for all the future strings $s_{i'} \in \S$, since we know for sure that $\abs{s_{i'}} - \abs{s} > K$ due to the sorted order. We thus can safely delete $s$ from bucket $B$ (Line~\ref{ln:a-2}).  Otherwise we add $(s, s_i)$ to our candidate set $\mathcal{C}$.  After these we store $s_i$ in bucket $B$ for future comparisons.  Note that each pair $(s_i, s_j)$ can potentially be added into $\mathcal{C}$ multiple times by different LSH collisions, we thus do a deduplication at Line~\ref{ln:a-3}.

There are two implementation details that we shall mention.  First, in the preprocessing we do not need to generate the whole $t_i^\ell$, but just those $m$ bits that will be used by each of the $z$ LSH functions. This reduces the space usage from $3N \cdot r \cdot n$ to $z \cdot m \cdot r \cdot n$.  Second, It is time/space prohibited to generate the hash table $\D_j^\ell$ whose size is $\abs{\Sigma}^m$.  We adopt the standard two-level hashing implementation of LSH:  For a signature in $\Sigma^m$, we first convert it into a vector $u \in \{1, \ldots, \abs{\Sigma}\}^m$ in the natural way.  We then generate a random vector $v \in \{0, \ldots, P - 1\}^m$ where $P > 1,000,000$ is a prime we choose that fits our datasets in experiments.  Finally, the second level hash function returns $\langle u, v\rangle \bmod P$, where $\langle \cdot, \cdot \rangle$ denotes the inner product.

\begin{table}[t]
\centering
\begin{minipage}[]{0.3\textwidth}
\centering
\begin{tabular}{ |l|l| }
  \hline
  \multicolumn{2}{|c|}{Strings} \\
  \hline
  $s_1$ & ACGTGACGTG \\
  $s_2$ & ACGTCGCGTG\\
  $s_3$ & ACTTACCTG \\
  $s_4$ & ATCGATCGGT\\
  \hline
\end{tabular}
\centerline{(a)}
\end{minipage}
\begin{minipage}[]{0.3\textwidth}
\centering
\begin{tabular}{ |l|l| }
  \hline
  \multicolumn{2}{|c|}{LSH functions } \\
  \hline
  $f_1^1$ & ($h_2$,$h_9$) \\   \hline
  $f_2^1$ &  ($h_1$,$h_4$) \\   \hline
  $f_1^2$ &  ($h_2$,$h_5$)  \\   \hline
  $f_2^2$ &  ($h_7$,$h_3$) \\
  \hline
\end{tabular}
\centerline{(b)}
\end{minipage}
\medskip
\caption{A collection of strings and LSH functions}  
\label{tab:strs}
\end{table}

\begin{table}[t]
\centering
\begin{minipage}[]{0.65\textwidth}
\centering
\begin{tabular}{l*{10}{c}r}
\ $j$              & 1 & 2 & 3 & 4 & 5  & 6 & 7 & 8  & 9 & 10 & $\dots$  \\
\hline
$\pi_j(A)$    & 0 & 1 & 0 & 0 & 1 & 0 & 1 & 1 & 0 & 1  & $\dots$\\
$\pi_j(C)$     & 1 & 1 & 0 & 1 & 1 & 1 & 1 & 0 & 0 & 0 & $\dots$ \\
$\pi_j(G)$    & 0 & 1 & 1 & 1 & 0& 0 & 0 & 1 & 1 & 1  & $\dots$\\
$\pi_j(T)$     & 1 & 0 & 0 & 0 & 1 & 0 & 1 & 1 & 0 & 1  & $\dots$ \\
\end{tabular}
\centerline{(a) random string $R^1$}
\smallskip
\end{minipage}

\begin{minipage}[]{0.65\textwidth}
\centering
\begin{tabular}{l*{10}{c}r}
\ $j$              & 1 & 2 & 3 & 4 & 5  & 6 & 7 & 8  & 9 & 10 & $\dots$  \\
\hline
$\pi_j(A)$    & 1 & 0 & 0 & 0 & 1 & 0 & 0 & 0 & 1 & 0  & $\dots$\\
$\pi_j(C)$     & 1 & 1 & 0 & 0 & 1 & 1 & 1 & 1 & 0 & 0 & $\dots$ \\
$\pi_j(G)$    & 1 & 0 & 1 & 1 & 0& 0 & 1 & 1 & 0 & 1  & $\dots$\\
$\pi_j(T)$     & 1 & 0 & 0 & 1 & 0 & 1 & 1 & 1 & 0 & 0  & $\dots$ \\
\end{tabular}
\centerline{(b) random string $R^2$}
\end{minipage}
\medskip
 \caption{Random strings for two CGK embeddings; represented as the equivalent $\pi_j(\cdot)$'s} 
\label{tab:CGK}
\end{table}

\begin{table}[t]
\centering
\begin{minipage}[]{0.4\textwidth}
\centering
\begin{tabular}{ |l|l| } 
  \hline
  \multicolumn{2}{|c|}{Strings after embedding} \\
  \hline
  $t_1^1$ & AACCGGGGTT $\dots$ \\
  $t_1^2$ & ACGTGGGAAC $\dots$\\
  $t_2^1$ & AACCGGGGTT $\dots$\\
  $t_2^2$ & ACGTCGGCGG $\dots$\\
  $t_3^1$ & AACCTTTACC $\dots$\\
  $t_3^2$ & ACTTTTAAAC $\dots$\\
  $t_4^1$ & AATTTCGGAA $\dots$\\
  $t_4^2$ & ATTTCGGAAT $\dots$\\
  \hline
\end{tabular}
\centerline{(a)}
\end{minipage}
\begin{minipage}[]{0.6\textwidth}
\centering
\begin{tabular}{l*{3}{c}r}
\ $i$              &$f_1^1(t_i^1)$  &$f_2^1(t_i^1)$ & $f_1^2(t_i^2)$ & $f_2^2(t_i^2)$   \\
\hline
$1$    & (A,T) & (A,C) & (C,G) & (G,G) \\
$2$     & (A,T) & (A,C) & (C,C) & (G,G)  \\
$3$    & (A,C) & (A,C) & (C,T) & (A,T) \\
$4$     & (A,A) & (A,T) & (T,C) & (G,T)  \\
\end{tabular}
\centerline{(b)}
\end{minipage}
\medskip
\caption{(a) Strings after embedding; (b) Signatures of LSH functions}  
\label{tab:strs2}
\end{table}



\paragraph{A Running Example}  
Table~\ref{tab:strs} shows the collection of input strings, and the set of LSH functions we use. Set distance threshold $K=3$. We choose parameters $r=2, z=2, m=2$ for \ebdjoin.  
Table~\ref{tab:CGK} shows two random strings $R^1, R^2$ (represented as the equivalent $\pi_j(\cdot)$'s; see Algorithm~\ref{alg:CGK}) that we use for the two rounds of CGK-embeddings. Table~\ref{tab:strs2}(a) shows the strings after CGK-embedding, and Table~\ref{tab:strs2}(b) shows the signatures of LSH functions. From Table~\ref{tab:strs2}(b) we find that $f_1^1(t_1^1)=f_1^1(t_2^1)=(A,T)$, $f_2^1(t_1^1)=f_2^1(t_2^1)=f_2^1(t_3^1)=(A,C), f_2^2(t_1^2)=f_2^2(t_2^2)=(G,G)$, and thus $ (s_1,s_2),(s_1,s_3),(s_2,s_3)$ are candidate pairs.  Finally after the verification step, we output $(s_1,s_2)$, $(s_1,s_3)$ as the results of similarity joins. 

\paragraph{Choices of parameters}  There are three parameters $m, z, r$ in \ebdjoin\ that we need to specify.  Recall that $m$ is the length of the LSH signature, or, the number of primitive hash functions $h \in \H$ we use in each $f \in \F(m)$; and $z$ is the number of LSHs we use for each string generated by CGK-embedding.  The larger $z$ and $m$ are, the better LSH performs in terms of accuracy and filtering effectiveness.  The product $m \cdot z$ will contribute to the total running time of the algorithm.  On the other hand, $r$ is number of CGK-embeddings we perform for each input string.  The larger $r$ we use, the smaller distortion we will get (see Figure~\ref{fig:CGK-distortion}).  

The concrete choices of $m, z$ and $r$ depend on the data size, distance thresholds, computation time/space budget and accuracy requirements.  For our datasets we have tested a number of parameter combinations.  We refer readers to Section~\ref{sec:embed-exp} for some statistics.
We have observed that $r = z = 7$, and $m = \log_2 N - \lfloor \log_2 x \rfloor$ where $x \% = K/N$ is the {\em relative} edit distance threshold, are good choices to balance the resource usage and the accuracy.

\paragraph{Running time}  The preprocessing step takes time $O(r \cdot z \cdot P + r \cdot n \cdot 3N \abs{\Sigma})$.  The time cost of LSH-based filtering depends on the effectiveness of the sliding window pruning; in the worst case it is $O(n r z m)$ where $m$ counts the cost of evaluating a hash function $f \in \F(m)$. Finally, the verification step costs $O(N K \cdot Z)$ where $Z$ is the number of candidate pairs after LSH-based filtering.

\subsection{Further Speed-up}
\label{sec:speedup}

Note that in the CGK-Embedding (Algorithm~\ref{alg:CGK}), we always pad the output strings $x'$ up to length $3N$, where $N = \max_{i \in [n]} \{\abs{s_i}\}$.  This approach is not very efficient for datasets containing strings with very different lengths (for example, our datasets \uniref\ and \trec; see Section~\ref{sec:setup}), since we need to pad a large number of `$\perp$' to the output strings which can be a waste of time. For example, for two strings $s_1$ and $s_2$ where $\abs{s_1}, \abs{s_2} \ll N$, if we map them to bit vectors $s'_1$ and $s'_2$ of size $3N$, then most of the aligned pairs in $s'_1$ and $s'_2$ are $(\perp, \perp)$s which carry almost no information.  Then if we use bit-sampling LSH for the Hamming distance we need a lot of samples in order to hit the interesting region, that is, the coordinates of strings in $s'_1$ and $s'_2$ where at least one of the two characters is {\em not} `$\perp$'.  
This is time and space expensive. 
We propose two ways to handle this issue. 

\vspace{-0.5mm} 

\paragraph{Grouping}
We first partition the set of strings of $\S$ to $(N'/K - 1)$ groups where $N' = \lceil N/K \rceil \cdot K$.  The $i$-th group contains all the strings of lengths $((i-1) K, (i+1) K]$. Note that each string will be included in two groups (i.e., the redundancy), and every pair of strings with distance at most $k$ will both be included in at least one of the groups. We then apply \ebdjoin\ on each group, and union the outputs at the end.  Due to the redundancy this approach may end up evaluating at most twice of the total number of candidates.

\vspace{-0.5mm} 

\paragraph{Truncation}
The second method is to use truncation, that is, we truncate each embedded string to predefined threshold $L$.  We then apply \ebdjoin\ on all the truncated strings.  Note that after truncation we essentially assume that all the bits after the $L$-th position in the embedded strings are the same, and thus truncation will {\em not} increase the Hamming distance of any pair of strings, and consequently will not introduce any false negative.  It can introduce some false positives but this is not a problem since we have a verification step at the end to remove all the false positives.

From the theory of the CGK-embedding we know that the number of the embedding steps is tightly concentrated around $2N$ where $N$ is the length of the original string (and then possibly many `$\perp$' will be appended afterwards). This indicates that for a datasets of strings of different lengths,  setting $L = 2 avg(\S)$ where $avg(\S)$ is the average length of the strings in $\S$ may be a good choice.  From our experimental results (see Section~\ref{sec:embed-exp}) we noticed that we can also be a little bit more aggressive to set $L = avg(\S)$.
\smallskip

Our experimental results (see Section~\ref{sec:embed-exp}) show that truncation always has the better performance than grouping on our tested datasets.  Therefore in the rest of the paper we always use truncation.

\begin{algorithm}[t]
\caption{Preprocessing+ ($\mathcal{S}$, $K$, $r$, $z$, $m$, $\Delta$)}
\label{alg:preprocessing2}
\begin{algorithmic}[1]
\Require Set of input strings $\mathcal{S} = \{s_1, \ldots, s_n\}$, distance threshold $K$, and parameters $r$, $z$, $m$ and $\Delta$ described in Table~\ref{tab:notation}
\Ensure  Strings in $\mathcal{S}$ in the sorted order, strings after CGK-embedding $\{t_{i,k}^\ell\ |\ \ell \in [r], i \in [n], k \in [\lceil K/\Delta \rceil]\}$, and hash tables $\{\D_j^{\ell} \ |\ \ell \in [r], j \in [z]\}$.
\State Sort $\mathcal{S}$ first by string length increasingly, and second by the alphabetical order. 

\ForEach{$\ell \in [r]$}
	\ForEach{$j \in [z]$}
	\State Initialize hash table $\D_j^{\ell} $ by generating a random hash function $f_j^{\ell} \in \F(m)$  
	\EndFor
\EndFor

\ForEach{$\ell \in [r]$}
\State Generate a random string $R^{\ell} \in \{0,1\}^{3 N \abs{\Sigma}}$
\ForEach{$s_i\in \mathcal{S}$}
\ForEach{$k\in [\lceil K/\Delta \rceil]$}
\State $s_{i,k} \gets s_i[(k-1)\Delta+1, \abs{s_i}]$ \label{line:c-1}
\State $t_{i,k}^{\ell}  \leftarrow $ CGK-Embedding($s_{i,k}$, $R^{\ell}$) \label{line:c-2}
\EndFor
\EndFor
\EndFor
\end{algorithmic}
\end{algorithm}

\begin{algorithm}[]
\caption{EmbedJoin+ ($\mathcal{S}$, $K$, $r$, $z$, $m$, $\Delta$, $T$)}
\label{alg:embed-joins+}
\begin{algorithmic}[1]
\Require Set of input strings $\mathcal{S} = \{s_1, \ldots, s_n\}$, distance threshold $K$, and parameters $r$, $z$, $m$, $\Delta$ and $T$ described in Table~\ref{tab:notation}
\Ensure  $\O \gets \{(s_i, s_j)\ |\ s_i, s_j \in \S; i \neq j; \text{ED}(s_i, s_j) \le K\}$ 
\State Preprocessing+ ($\mathcal{S}$, $K$, $r$, $z$, $m$, $\Delta$) 
\Comment Using Algorithm~\ref{alg:preprocessing2}
\State  $\mathcal{C} \leftarrow \emptyset $  
\Comment Collection of candidate pairs

\ForEach{$s_i \in \mathcal{S} \text{ (in the sorted order)}$}
	\ForEach{$\ell \in [r]$}
		\ForEach{$j \in [z]$}
			\ForEach{$k\in [\lceil K/\Delta \rceil]$}
			\ForEach{tuple $(s,k') $ ($s \neq s_i$)  stored in the $f_j^\ell (t_{i,k}^\ell)\text{-th bucket of table} \ \D_j^\ell $}
				\If{$\abs{s_i} - \abs{s} \le K$} \label{ln:a-1}
				\State $\mathcal{C}  \leftarrow \mathcal{C}  \cup (s, s_i, k', k)$ \label{line:d-1}
\Else
				\State Remove $(s,k')$ from $\D_j^\ell$ \label{ln:a-2}
				\EndIf
			\EndFor
			\State Store $(s_{i},k)$ in the $f_j^\ell (t_{i,k}^\ell)$-th bucket of $\D_j^\ell$ \label{line:d-2}
		\EndFor
		\EndFor
	\EndFor
\EndFor
\State Count the frequency of each tuple $(x, y,k_x,k_y)$ in  $\mathcal{C}$
\ForEach {$(x, y,k_x,k_y) \in \mathcal{C}$ with count $\ge T$ \label{line:d-3}}
	\If{$\text{ED}(x, y) \le K$}
	\State $\mathcal{O} \leftarrow  \mathcal{O} \cup (x, y)$
	\EndIf
\State Remove all tuples $(x, y, \cdot, \cdot)$ in $\mathcal{C}$ \label{line:d-4}
\Comment We only need one pair of substrings of $(x, y)$ with count at least $T$
\EndFor
\end{algorithmic}
\end{algorithm}
\section{The \ebdjoin+ Algorithm}
\label{sec:improved}

An important application of similar joins is to find similar pairs of strings in a biological datasets that consists of {\em random} reads of the human genomes or protein sequences. A sufficient number of similar pairs of reads can be used to reconstruct the original genome or protein sequence~\cite{GYB10}.
In those datasets, for two strings $x$ and $y$ who are overall similar, there could be a long prefix of insertions at the beginning of one of the strings in the optimal alignment of $x$ and $y$, which we call the {\em shift}.  More precisely, given two strings $x[1..N]$ and $y[1..M]$, we define the shift between $x$ and $y$ to be $\sft(x, y) = \max\{\sft_1, \sft_2\}$ where 
$$\sft_1 = \max_{t \in [N]}\{\text{ED}(x[1..N], y[1..M]) = t + \text{ED}(x[t+1..N], y[1..M])\},\quad \text{and}$$ 
$$\sft_2 = \max_{t \in [M]}\{\text{ED}(x[1..N], y[1..M]) = t + \text{ED}(x[1..N], y[t+1..M])\}$$
When applying \ebdjoin\ directly to find similar pairs of strings on such datasets under large thresholds, the shift may contribute most of the edits which will be further ``amplified'' by the CGK-embedding, since consecutive errors is one of the worst cases for the distortion of the CGK-embedding.  This phenomenon may introduce a large number of false negative, and consequently reduce the accuracy of the join results.

In this section we propose an improved version of \ebdjoin\ called \ebdjoin+ to handle string shifts.  \ebdjoin+ contains several new ideas which we will illustrate below.

A natural way to handle shifts is to start the CGK-embedding from multiple positions of the strings.  Given a parameter $\Delta$ which we will set later, for each string $s_i$, we consider $\lceil K / \Delta \rceil$ substrings which are suffixes of $s_i$ with starting positions $1, \Delta+1, \ldots, (\lceil K/\Delta \rceil - 1)\Delta + 1$;  we denote these substrings by $s_{i,1}, \ldots, s_{i,{\lceil K/\Delta \rceil}}$.  By embedding all the substrings, we can guarantee that for any pair of strings $(s_i, s_j)$ such that $\text{ED}(s_i, s_j) \le K$, there is a pair of substrings $(s_{i,p}, s_{j,q})$ such that  $\sft(s_{i,p}, s_{j,q}) \le \Delta/2$.  

However, the direct implementation of this idea will cause the number of false positives in the set of candidate pairs (after the CGK-embedding and LSH) to increase significantly, and consequently make the verification the bottleneck.  In order to reduce the number of false positives, we require a candidate $(s_i, s_j)$ to have a pair of substrings $(s_{i,p}, s_{i,q})$ with at least $T \in [z]$ matched hash signatures in the process of LSH (recall that $z$ is number of hash functions we use in LSH).  Intuitively, when $T > 1$, this requirement will make it harder for a pair to be selected as a candidate.  More precisely, let $p$ be the collision probability of a pair of substrings under a single hash function, then the probability that the two substrings have at least $T$ common hash signatures is 
$$p_T = 1- \sum_{i=0}^{T-1} \binom zi p^{i} (1-p)^{z-i}.$$

\begin{figure}[t]
\centering
\includegraphics[height = 1.6in]{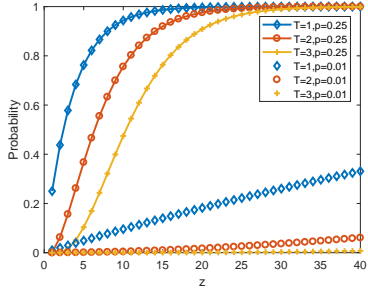}
\caption{The probability of having $T=1,2,3$ matched hash signatures for similar and dissimilar pairs under different numbers of hash functions $z$.  We assume that the collision probability for a similar pair is $0.9^{13} = 0.25$, and that for a dissimilar pair is $0.7^{13} = 0.01$, where $m = 13$ is the number of bits we use in the bit-sampling LSH for the Hamming distance.}
\label{fig:numsig}
\end{figure}

We plot $p_T$ for two different $p$ values in Figure~\ref{fig:numsig}.  It can be seen that when $T$ becomes larger, the gap of probabilities between similar and dissimilar pairs becomes bigger.  However, for larger $T$ we will need more hash functions to guarantee that the number of false negatives is small, which will increase the time of performing LSH.  In practice, we observed that when $K/\Delta > 1$ (i.e., we will produce at least $2$ substrings for each string), then setting $T = 2$ is a good choice.  Otherwise if $K/\Delta \le 1$,  then we set $T = 1$, and \ebdjoin+ degenerates to \ebdjoin.

The pseudocode of \ebdjoin+ is very similar to that of \ebdjoin; see Algorithm~\ref{alg:preprocessing2} and Algorithm~\ref{alg:embed-joins+}.  In the preprocessing (Algorithm~\ref{alg:preprocessing2}), the only difference is that we need to embed for each string $s_i$ the substrings $s_{i,1}, \ldots, s_{i,{\lceil K/\Delta \rceil}}$ (Line~\ref{line:c-1}-\ref{line:c-2}). In the main algorithm (Algorithm~\ref{alg:embed-joins+}), for each substring $s_{i,k}$ generated from string $s_i$, we record both its original string and its substring index in the hash table, that is, $(s_i, k)$ (Line~\ref{line:d-2}).  For each pair of substrings in the same hash table, we record the match using their original strings and their indices, in the form of $(s_i, s_j, k_{s_i}, k_{s_j})$ (Line~\ref{line:d-1}).  At the end we need to count and verify for each pair $(x, y)$ whether at least one of their substring pairs have at least $T$ matches (Line~\ref{line:d-3}-\ref{line:d-4}).

\paragraph{Choices of parameters}
Compared with \ebdjoin, we have one more parameter to choose in the algorithm \ebdjoin+, that is, the ``step length'' $\Delta$ for creating substrings.  From the theory of CGK-embedding, with a good probability a consecutive set of insertions of length $\Delta$ will introduce $c \Delta^2$ (for some constant $c$) Hamming errors after the embedding.  Since we truncate each string at the position $avg(\S)$, it is meaningful to ensure that $c \Delta^2 \le avg(\S)$. On the other hand, we would like to set $\Delta$ as large as possible since $\lceil K/\Delta \rceil$ substrings generated for each string $s_i$ will contribute to both time and space of the algorithm. We thus choose $\Delta \approx \sqrt{avg(\S)}$ or a bit smaller.   

As already mentioned, the variable $T$ is determined by $K$ and $\Delta$: When $\lceil K/\Delta \rceil > 1$ we set $T = 2$; otherwise \ebdjoin+ degenerates to \ebdjoin.

Similar to \ebdjoin, in \ebdjoin+ we set $r = 7$ and $m = \log_2 N - \lfloor \log_2 x \rfloor$ where $x \% = K/N$ is the relative edit distance threshold.  For the value of $z$, we set $z = 16$ when $T = 2$, and $z = 7$ when $T = 1$.
This is according to the fact that when $T$ increases, we have to increase the number of hash functions in LSH to achieve a good accuracy. 

\paragraph{Running time}
The preprocessing step takes time $O(r \cdot z \cdot P + r \cdot \lceil K/\Delta \rceil \cdot n \cdot 3N \abs{\Sigma})$.  The time cost of LSH-based filtering again depends on the effectiveness of the sliding window pruning; in the worst case it is $O(n r z m \cdot \lceil K/\Delta \rceil)$ where $m$ counts the cost of evaluating a hash function $f \in \F(m)$. Finally, the verification step costs $O(N K \cdot Z)$ where $Z$ is the number of candidate pairs after LSH-based filtering.

\section{Experiments}
\label{sec:exp}

In this section we present our experimental studies.  After listing the datasets and tested algorithms, we first give an overview of the performance of \ebdjoin+. We then compare it with the existing best algorithms.  Finally, we show the scalability of \ebdjoin+ in the ranges that the existing best algorithms cannot reach.

\subsection{The Setup}
\label{sec:setup}

\paragraph{Datasets} 
We tested the algorithms in three publicly available real world datasets.

\medskip
\noindent \uniref: a dataset of UniRef90 protein sequence data from UniProt project.\footnote{Available in \url{http://www.uniprot.org/}
} Each sequence is an array of amino acids coded in uppercase letters. We first remove sequences whose lengths are smaller than 200, and then extract the first 400,000 protein sequences. 

\medskip
\noindent \trec:  a dataset of references from Medline (an online medical information database) consisting of titles and abstracts from 270 medical journals.\footnote{Available in \url{http://trec.nist.gov/data/t9_filtering.html}
} We first extract and concatenate title, author, and abstract fields, and then convert punctuations into white spaces and letters into their upper cases. 

\medskip
\noindent \genoaa, \genoa, \genob, \genoc, \genod, \genoe, \genof:  datasets of human genomes of 50 individuals obtained from the {\em personal genomes project},\footnote{Available in \url{http://personalgenomes.org/}} and the reference sequence is obtained from GRCh37 assembly.   We choose to use Chromosome 20.  For \genoaa\ we partition the long DNA sequences into shorter substrings according to the indices of the reference sequence, so that the shift is small in similar pairs. For all other genome datasets we select substrings with {\em random} starting positions.  The names of datasets can be read as `GEN $\circ$ number of strings ($20k$ to $320k$) $\circ$ string length (S $\approx$ 5k, M $\approx$ 10k, L $\approx$ 20k)'. 

\begin{figure}[t]
\centering
\includegraphics[height = 1.6in]{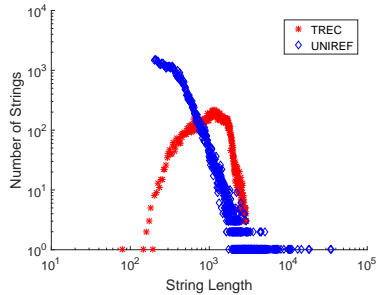}
\caption{String length distributions of \uniref\ and \trec\ datasets}
\label{fig:unidis}
\end{figure}

We summarize the statistics of our datasets in Table~\ref{tab:stat}. 
The distributions of the string lengths of the \uniref\ and \trec\ datasets are plotted in Figure~\ref{fig:unidis}.

\begin{table}[t]
\centering 
\begin{tabular}{lccccc} 
\hline
Datasets &$n$ &Avg Len &Min Len & Max Len & $|\Sigma|$\\  \hline 
\uniref &400000  &445 &200  &35213 &25\\  \hline
\trec &233435  &1217 &80   &3947 &37    \\  \hline
\genoaa &50000   &5000  &4844 & 5109 &4   \\  \hline
\genoa &50000   &5000  &4829 & 5152 &4   \\  \hline
\genob &20000   &5000  &4829 &5109 &4   \\  \hline
\genoc &20000   &10000  &9843 &10154 &4   \\  \hline
\genod &20000   &20000  &19821 &20109 &4   \\  \hline
\genoe &80000   &5000  &4814  &5109 &4   \\  \hline
\genof &320000   &5000  &4811  &5154 &4   \\  \hline
\end{tabular}
\caption{Statistics of tested datasets}
\label{tab:stat}
\end{table}

\paragraph{Tested Algorithms} 
We now list all the algorithms that we have used in our experiments.
We choose these competing algorithms based on the recommendations of the experimental study \cite{JLFL14} and the similarity search/join competition \cite{WDG14}. We believe that these are the best existing algorithms for edit similarity joins.  

\medskip
\noindent \ebdjoin, \ebdjoin+: our purposed algorithms.  Note again that when $\lceil K/\Delta \rceil = 1$ \ebdjoin+ degenerates to \ebdjoin.  We implemented our algorithms in C++ and complied using GCC 5.4.0 with O3 flag.

\medskip
\noindent \pass \cite{LDW11}: an exact algorithm for similarity joins use a partition-based framework.  The basic idea of \pass\ is to use the pigeon-hole principle: given an edit distance threshold $K$, \pass\ partitions each string into $K+1$ segments. Two similar strings must share at least one segment.  The \pass\ has the best time performance for similarly joins on long strings according to the report \cite{JLFL14} and competition \cite{WDG14}.   We obtained the implementation of \pass\ from the authors.

\medskip
\noindent \edjoin \cite{XWL08}: an exact algorithm for similarity joins based on prefix filtering.  The idea of prefix filtering is that given an edit distance threshold $K$, we generate $q$-grams for each string, sort them based on a global ordering, and then choose the first $qK+1$ grams as the string's signatures. Two similar strings must have at least one common signature. The \edjoin\ further improves the prefix filtering by {\em Position Filtering} and {\em Content Filtering}. We download the binary codes from the authors' project website.\footnote{\url{http://www.cse.unsw.edu.au/~weiw/project/simjoin.html\#\_download}.} To make the comparison fair, for each dataset and each threshold value $K$ we always report the best time performance among different parameters $q$.

\medskip
\noindent \adpjoin \cite{WLF12}:  an exact algorithm for similarity joins based on prefix filtering.  It improves the original prefix filtering by learning the tradeoff between number of signatures and the  filtering power, instead of using a fixed number of $q$-grams. We download the binary codes from the authors' project website.\footnote{\url{https://www2.cs.sfu.ca/~jnwang/projects/adapt/}.} There are three filtering methods used in \cite{WLF12}, named {\em Gram}, {\em IndexGram} and {\em IndexChunk}. We found that {\em Gram} always has the best time performance. We thus report the best time performance among different parameters $q$ using the {\em Gram} filter.

\medskip
\noindent \qchunk \cite{QWL11}: an exact algorithm for similarity joins based on prefix filtering.  It improves the prefix filter by introducing {\em $q$-chunk} which is $q$-gram with starting positions at $i \cdot q +1$ for $i\in \{0, 1, \ldots, \frac{l-1}{q}\}$, where $l$ is the string length. \qchunk\ then employs effective filters based on $q$-chunk. We download the binary codes from the authors' project website.\footnote{\url{http://www.cse.unsw.edu.au/~weiw/project/simjoin.html\#\_download} and \url{http://www.cse.unsw.edu.au/~jqin/}.} There are two filtering methods used in \cite{QWL11}, named {\em IndexGram} and {\em IndexChunk}. We found that {\em IndexChunk} always has the better time performance. We thus report the best time performance among different parameters $q$ using the {\em IndexChunk} filter.

\paragraph{Measurements}
We report three types of measurements in our experiments: {\em accuracy}, {\em memory usage} and {\em running time}.  Recall that \ebdjoin\ and \ebdjoin+ only have false negatives; the {\em accuracy} we report is number of output pairs returned by \ebdjoin\ and \ebdjoin+ divided by the ground truth returned by other exact competing algorithms.  The memory usage we report is the maximum memory usage of a program during its execution. 

As mentioned, the competing algorithms may use different filtering methods or different parameters. We always choose the {\em best} combinations for comparisons.  To make the comparison fair we have counted the time used for all the preprocessing steps.

\paragraph{Computing Environment}
All experiments were conducted on a Dell PowerEdge T630 server with 2 Intel Xeon E5-2667 v4 3.2GHz CPU with 8 cores each, and 256GB memory.

\begin{figure}[t]
\centering
\includegraphics[height = 2in]{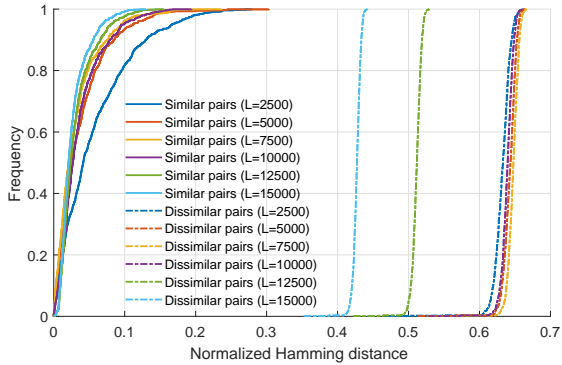}
\caption{The influence of length of Truncation $L$ on the {\em minimum} normalized Hamming distance $Ham(x,y)/L$, for $1000$ random selected similar pairs (with $ED(x,y) \le 150$) and dissimilar pairs (with $ED(x,y) > 150$) on \genoa\ dataset. The parameters are $r=z=5$.}
\label{fig:trun}
\end{figure}

\subsection{Performance Overview of \ebdjoin+}
\label{sec:embed-exp}

In this section we present an overview of the performance of \ebdjoin+.   All the results for \ebdjoin+\ are the average of five independent runs. 

\paragraph{Length of truncation}
As discussed in Section~\ref{sec:speedup}, we use {\em truncation} to speed up the CGK-embedding.  This is useful since when the distance threshold $K$ or number of candidates are small, the embedding will dominate the total running time.  In the following we show how different choices of the truncation lengths affect distance gaps between similar and dissimilar pairs after the CGK-embedding.

Figure~\ref{fig:trun} presents how the length of truncation $L$ influences the {\em minimum normalized Hamming distance} of similar and dissimilar pairs on the \genoa\ dataset. The minimum normalized Hamming distance of a pair is the minimum value of normalized Hamming distance ($Ham(x,y)/L$) over all pairs of substrings and all embeddings.  The total string length after CGK-embedding is $15000$; we thus truncate strings from $2500$ characters to $15000$ characters. 
From the plot we notice that the normalized Hamming distances of similar pairs are almost the same under different $L$ values, and increase a little when $L = 2500$. On the other hand, the normalized Hamming distances of dissimilar pairs are almost the same when $L \le 10000$, and decrease a lot when $L > 10000$; this is because most characters after the $10000$-th digit are ``$\perp$'', which do not contribute to the Hamming distances. The plot recommends us to choose $L$ between $5000$ and $10000$, or, between $avg(\S)$ and $2 avg(\S)$.  In our experiments we will use truncation length $L=avg(\S)$ on genome datasets in which the string lengths are very close, and $L=2 avg(\S)$ on other datasets where the string lengths vary.

\begin{table*}[t]
\centering
\scalebox{0.8}{
  \begin{tabular}{|c|c|c|c|c|c|c|c|c|c|}
    \hline
    \multirow{3}{*}{Accuracy} &
      \multicolumn{3}{c|}{$r=5$} &
      \multicolumn{3}{c|}{$r=7$} &
      \multicolumn{3}{c|}{$r=9$} \\

& $z=3$  &$z=5$ &  $z=7$ & $z=3$  &$z=5$ &  $z=7$ & $z=3$  &$z=5$ & $z=7$ \\
    \hline
    $m=5$ & 94.5\% & 97.4\% &98.6\% & 96.9\% &99.0\% &99.5\% & 98.5\% & 99.4\% & 99.7\% \\
    \hline
     $m=7$ & 91.6\% & 94.0\% & 95.6\% & 95.2\% & 97.2\% & 98.4\% & 96.4\% & 98.4\% & 99.1\% \\
    \hline
     $m=9$ & 90.1\% & 90.8\% & 92.9\% & 90.7\% & 94.7\% & 96.1\% & 92.9\% & 96.2\% & 97.6\% \\
    \hline
  \end{tabular}
  }
  \caption{Accuracy of \ebdjoin+, \uniref\ dataset, $K=20, \Delta = 50$}
\label{tab:uniacc}
\end{table*}

\begin{table*}[t]
\centering
\scalebox{0.8}{
  \begin{tabular}{|c|c|c|c|c|c|c|c|c|c|}
    \hline
    \multirow{3}{*}{Accuracy} &
      \multicolumn{3}{c|}{$r=5$} &
      \multicolumn{3}{c|}{$r=7$} &
      \multicolumn{3}{c|}{$r=9$} \\

& $z=3$  &$z=5$ &  $z=7$ & $z=3$  &$z=5$ &  $z=7$ & $z=3$  &$z=5$ & $z=7$ \\
    \hline
    $m=8$ & 91.3\% & 94.2\% &95.6\% & 91.3\% &94.2\% &95.6\% & 95.6\% & 95.6\% & 98.6\% \\
    \hline
     $m=10$ & 90.0\% & 92.8\% & 92.8\% & 91.3\% & 94.2\% & 94.2\% & 92.8\% & 94.2\% & 95.6\% \\
    \hline
     $m=12$ & 90.0\% & 90.0\% & 91.3\% & 90.0\% & 90.0\% & 91.3\% & 91.3\% & 92.8\% & 94.2\% \\
    \hline
  \end{tabular}
  }
  \caption{Accuracy of \ebdjoin+, \trec\ dataset, $K=40, \Delta = 50$}
\label{tab:trecacc}
\end{table*}

\begin{table*}[!ht]
\centering
\scalebox{0.8}{
  \begin{tabular}{|c|c|c|c|c|c|c|c|c|c|}
    \hline
    \multirow{3}{*}{Accuracy} &
      \multicolumn{3}{c|}{$r=5$} &
      \multicolumn{3}{c|}{$r=7$} &
      \multicolumn{3}{c|}{$r=9$} \\

& $z=8$  &$z=12$ &  $z=16$ & $z=8$  &$z=12$ &  $z=16$ & $z=8$  &$z=12$ &  $z=16$  \\
    \hline
    $m=11$ & 99.0\% & 99.2\% &99.4\% & 99.2\% &99.8\% &99.9\% & 99.7\% & 99.9\% & 100.0\% \\
    \hline
     $m=13$ & 97.4\% & 97.7\% & 98.0\% & 98.9\% & 99.6\% & 99.7\% & 99.6\% & 99.9\% & 99.9\% \\
    \hline
     $m=15$ & 96.1\% & 96.7\% &98.3\% & 98.0\% & 98.2\% & 99.3\% & 98.8\% & 99.4\% & 99.6\% \\
    \hline
  \end{tabular}
  }
  \caption{Accuracy of \ebdjoin+, \genoa\ dataset, $K=100, \Delta = 50$}
\label{tab:genoacc}
\end{table*}

\begin{table*}[!ht]
\centering
\scalebox{0.8}{
  \begin{tabular}{|c|c|c|c|c|c|c|c|c|c|}
    \hline
    \multirow{3}{*}{Accuracy} &
      \multicolumn{3}{c|}{$r=5$} &
      \multicolumn{3}{c|}{$r=7$} &
      \multicolumn{3}{c|}{$r=9$} \\

& $z=8$  &$z=12$ &  $z=16$ & $z=8$  &$z=12$ &  $z=16$ & $z=8$  &$z=12$ &  $z=16$  \\
    \hline
    $\Delta=25$ & 99.8\% & 99.9\% &100.0\% & 100.0\% &100.0\% &100.0\% & 100.0\% & 100.0\% & 100.0\% \\
    \hline
     $\Delta=34$ & 99.8\% & 99.9\% & 100.0\% & 100.0\% & 100.0\% & 100.0\%& 100.0\% & 100.0\% & 100.0\% \\
    \hline
     $\Delta=50$ & 97.4\% & 97.7\% & 98.0\% & 98.9\% & 99.6\% & 99.7\% & 99.6\% & 99.9\% & 99.9\% \\
    \hline
  \end{tabular}
  }
  \caption{Accuracy of \ebdjoin+, \genoa\ dataset, $K=100, m = 13$}
\label{tab:genoaccdel}
\end{table*}

\paragraph{Accuracy}
In Table~\ref{tab:uniacc}, \ref{tab:trecacc} and \ref{tab:genoacc} we study how different parameters $(r, z, m)$ influence the accuracy of \ebdjoin+. 
We vary $r$ in $\{5,7,9\}$, $z$ in $\{3,5,7\}$ for \trec\ and \uniref, and $z$ in $\{8,12,16\}$ for \genoa. We choose slightly different values for $m$ on different datasets (the choices of $m$ largely depend on the string length and the distance threshold $K$).  

We observe that the accuracy of \ebdjoin+\ is $90.1 \sim 99.7\%$ on \uniref, $90.0 \sim 98.6\%$ on \trec, 
and $96.1 \sim 100.0\%$ in \genoa.  

We note that the accuracy of \ebdjoin+\ increases with $r$ and $z$, and decreases with $m$. This is consistent with the theory.  When $r$ and $z$ increase, we use more hash functions (recall that the total number of hash functions used is $r \cdot z$), and thus each pair of strings have more chance to be hashed into the same bucket in at least one of the hash tables. Similarly, when $m$ decreases, each LSH function has larger collision probability.  Of course, the increase of the collision probability will always introduce more false positives, and consequently increase the verification time.  Using more hash functions/tables will also increase the space usage.

In Table~\ref{tab:genoaccdel} we study how the parameter $\Delta$ influences the accuracy of \ebdjoin+. We vary $\Delta$ in $\{25, 34, 50\}$ so that the number of substrings for each string are  $\{4, 3, 2\}$. We observe that the accuracy of \ebdjoin+\ decreases when $\Delta$ increases. This is because when $\Delta$ increases, the length of shifts between similar pairs may increase, which makes the chance of hashing them into the same bucket to be smaller.

\paragraph{Time and Space}
In Table~\ref{tab:genotime-1} we study how different parameters $(r, z, m)$ influence the running time of \ebdjoin+\ in the \genoa\ dataset.  We note that the running time increases when $r$ and $z$ increase, decreases when $m$ increases.  This is just the opposite to what we have observed for accuracy, and is consistent to the theory that increasing the collision probability will introduce more false positives/candidates and thus increase the verification time.

In Table~\ref{tab:genomem} we study how different parameters $(r, z, m)$ influence the memory of \ebdjoin+\ in the \genoa\ dataset. 
We observe that the memory usage increases when $r$ and $z$ increase. This is because when $r$ and $z$ increase we need to store more hash tables {\em and} we will have more candidate pairs to verify.  When $m$ increases, the memory usage stays the same or slightly increases.  There are two kinds of mutually exclusive forces that affect this.  On the one hand, when $m$ increases the size of each hash signature increases.  On the other hand, when $m$ increases the number of candidate pairs decreases. From what we have observed, the first force generally dominates the second.

\begin{figure*}[!t]
\centering
\begin{minipage}[d]{0.3\linewidth}
\centering
\includegraphics[width=1\textwidth]{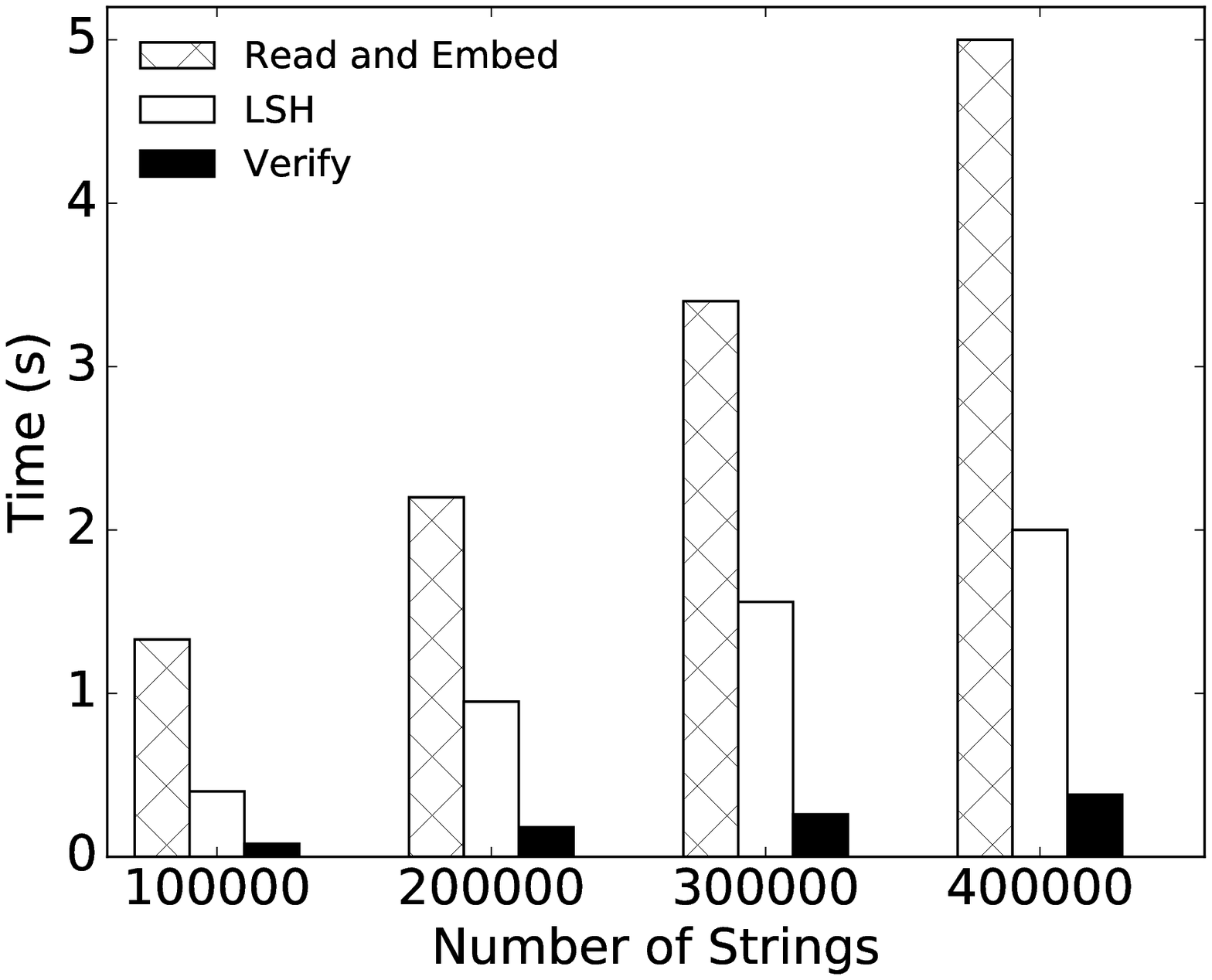}
\centerline{\uniref\ ($K=20$)}
\end{minipage}
\begin{minipage}[d]{0.3\linewidth}
\centering
\includegraphics[width=1\textwidth]{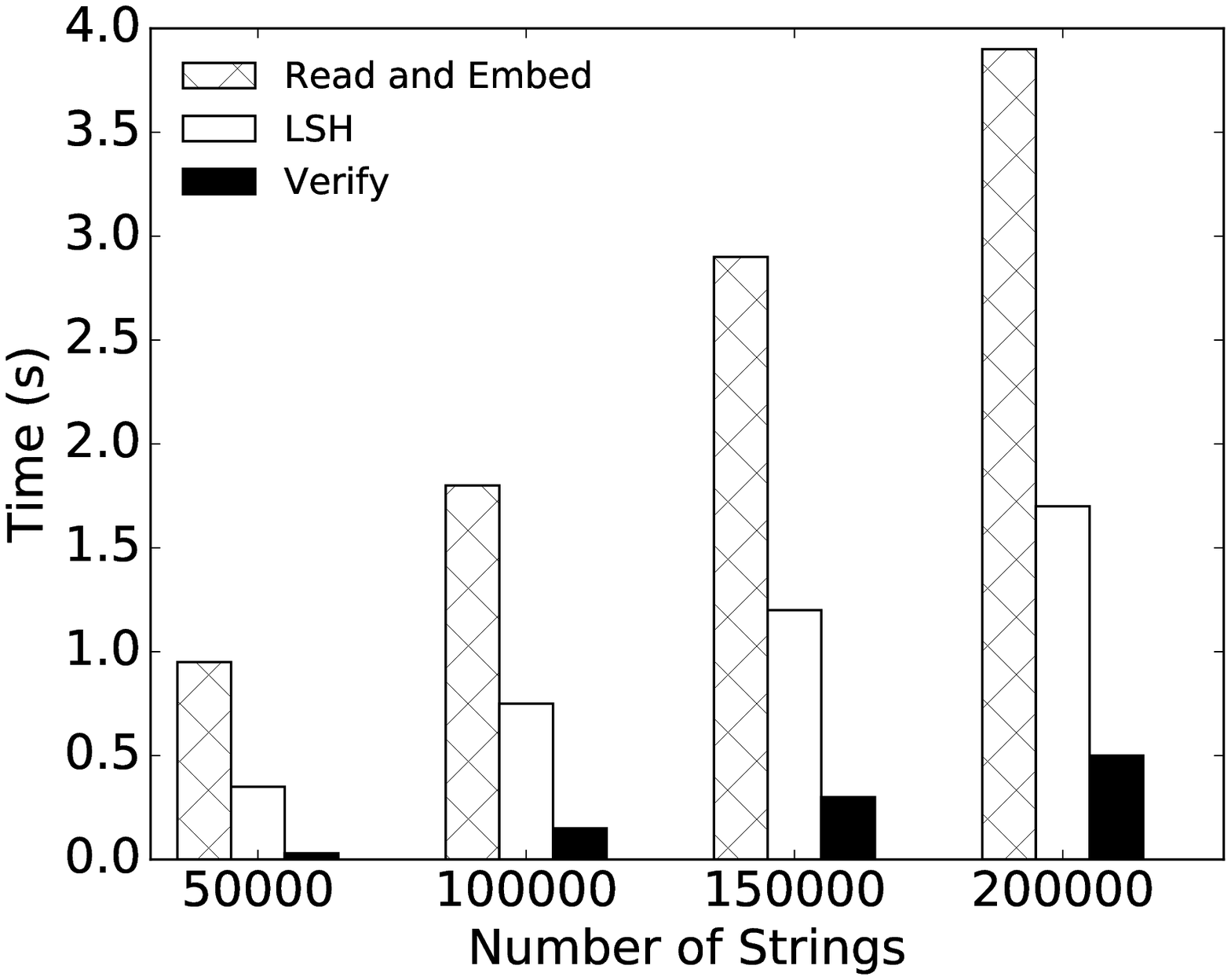}
\centerline{\trec\ ($K=40$)}
\end{minipage}
\begin{minipage}[d]{0.3\linewidth}
\centering
\includegraphics[width=1\textwidth]{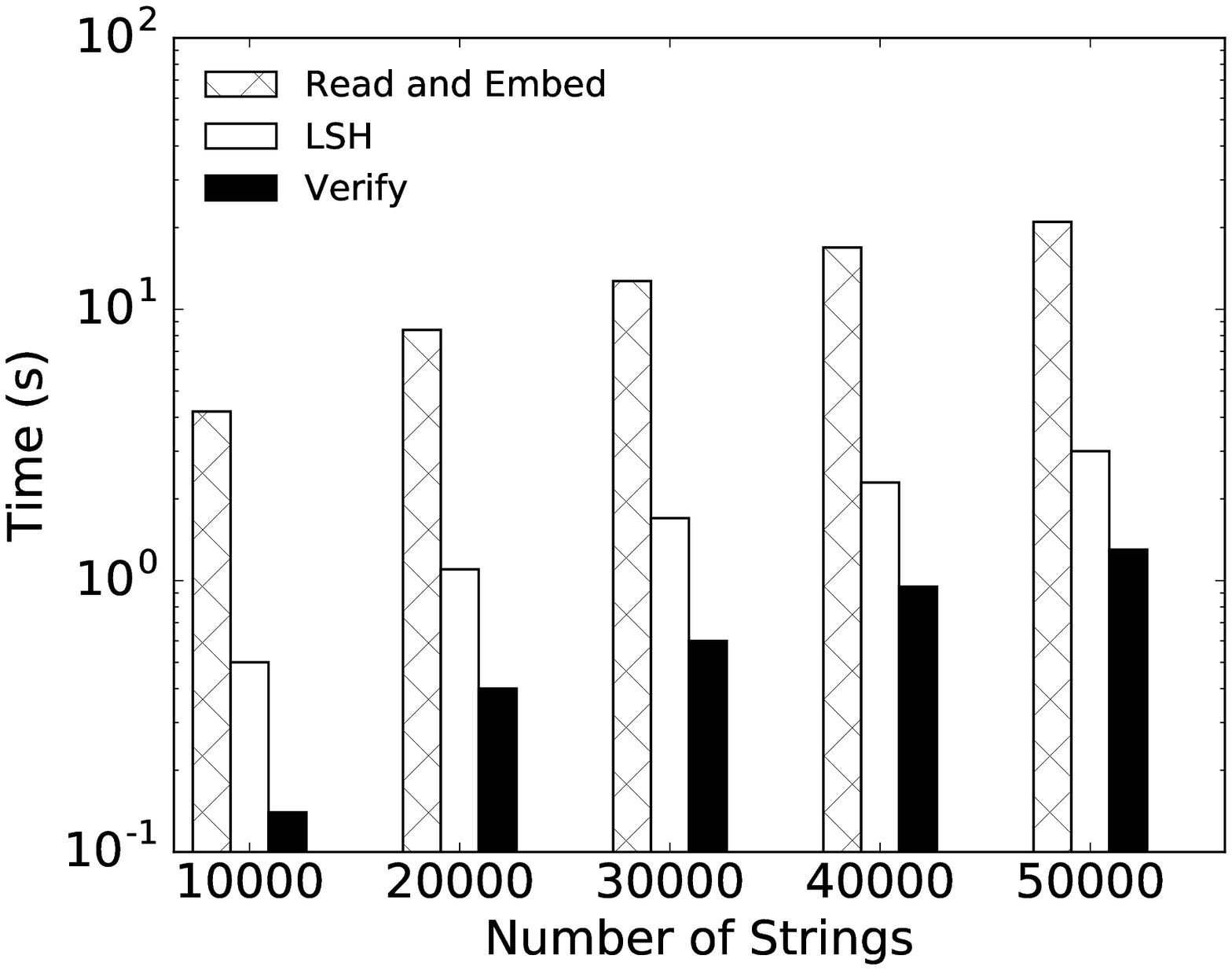}
\centerline{\genoa\ ($K=100$)}
\end{minipage}
\caption{Running time of different parts of \ebdjoin+, varying $n$.}
\label{fig:timen}
\end{figure*}

\begin{figure*}[!ht]
\centering
\begin{minipage}[d]{0.3\linewidth}
\centering
\includegraphics[width=1\textwidth]{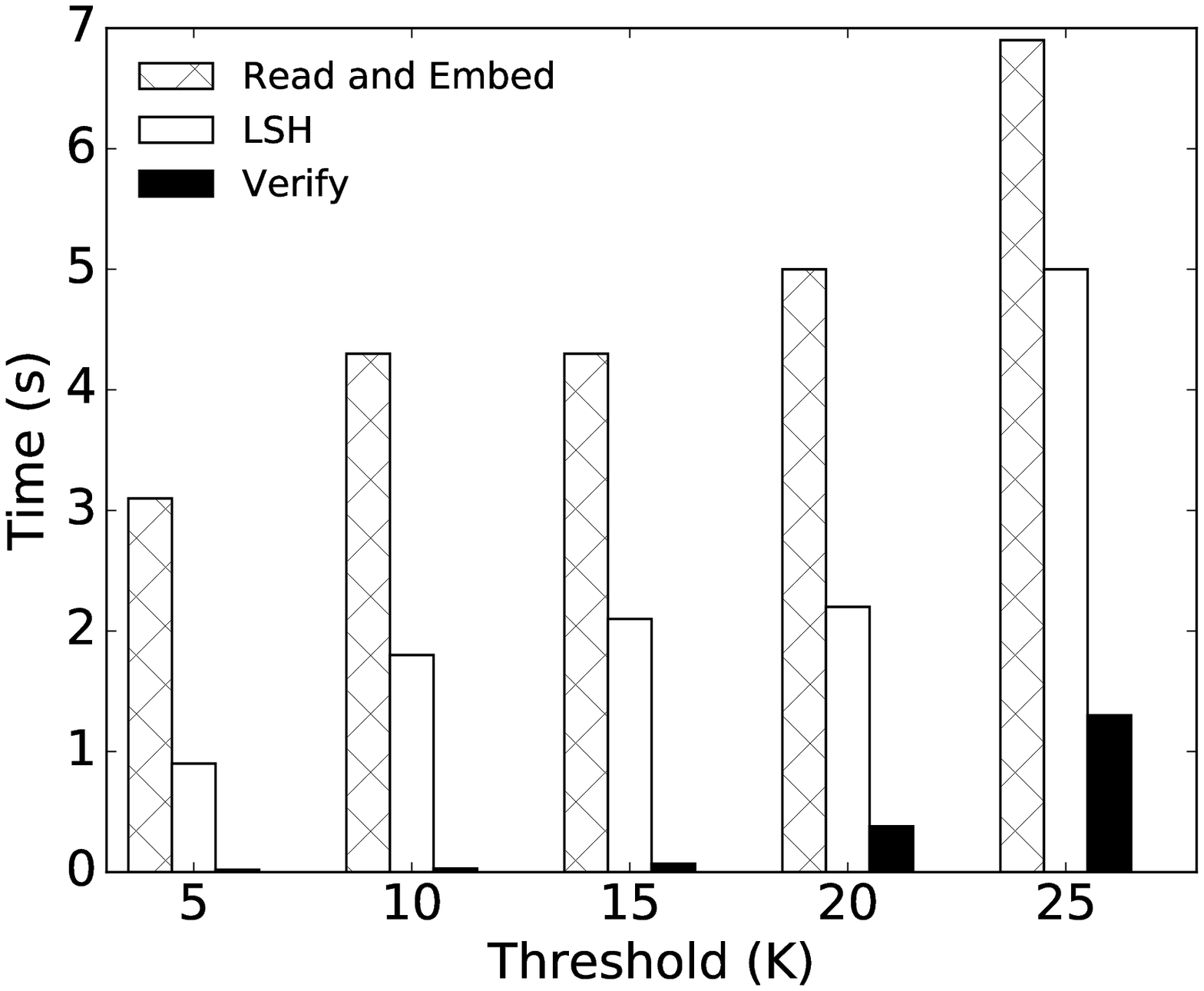}
\centerline{\uniref}
\end{minipage}
\begin{minipage}[d]{0.3\linewidth}
\centering
\includegraphics[width=1\textwidth]{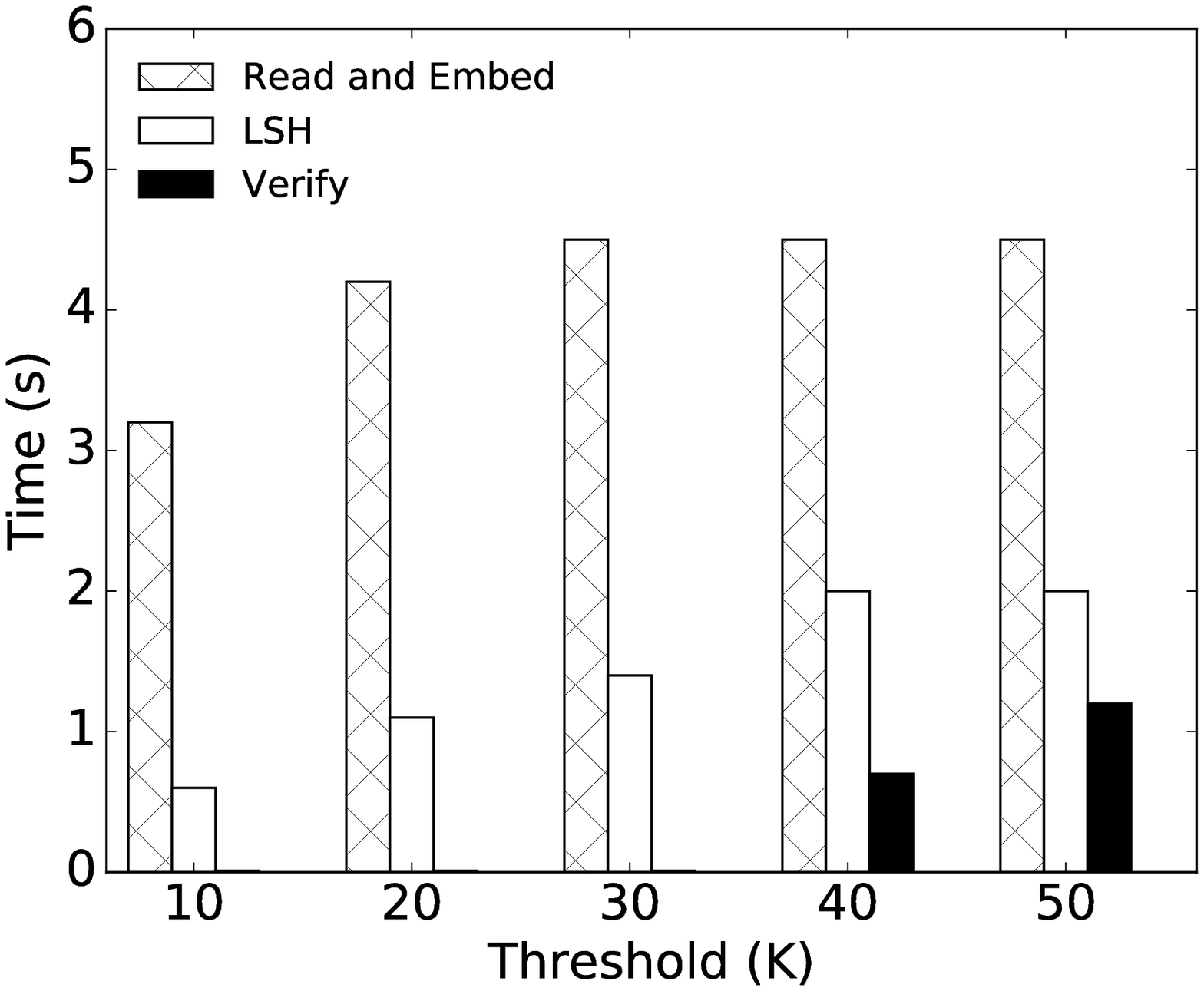}
\centerline{\trec}
\end{minipage}
\begin{minipage}[d]{0.3\linewidth}
\centering
\includegraphics[width=1\textwidth]{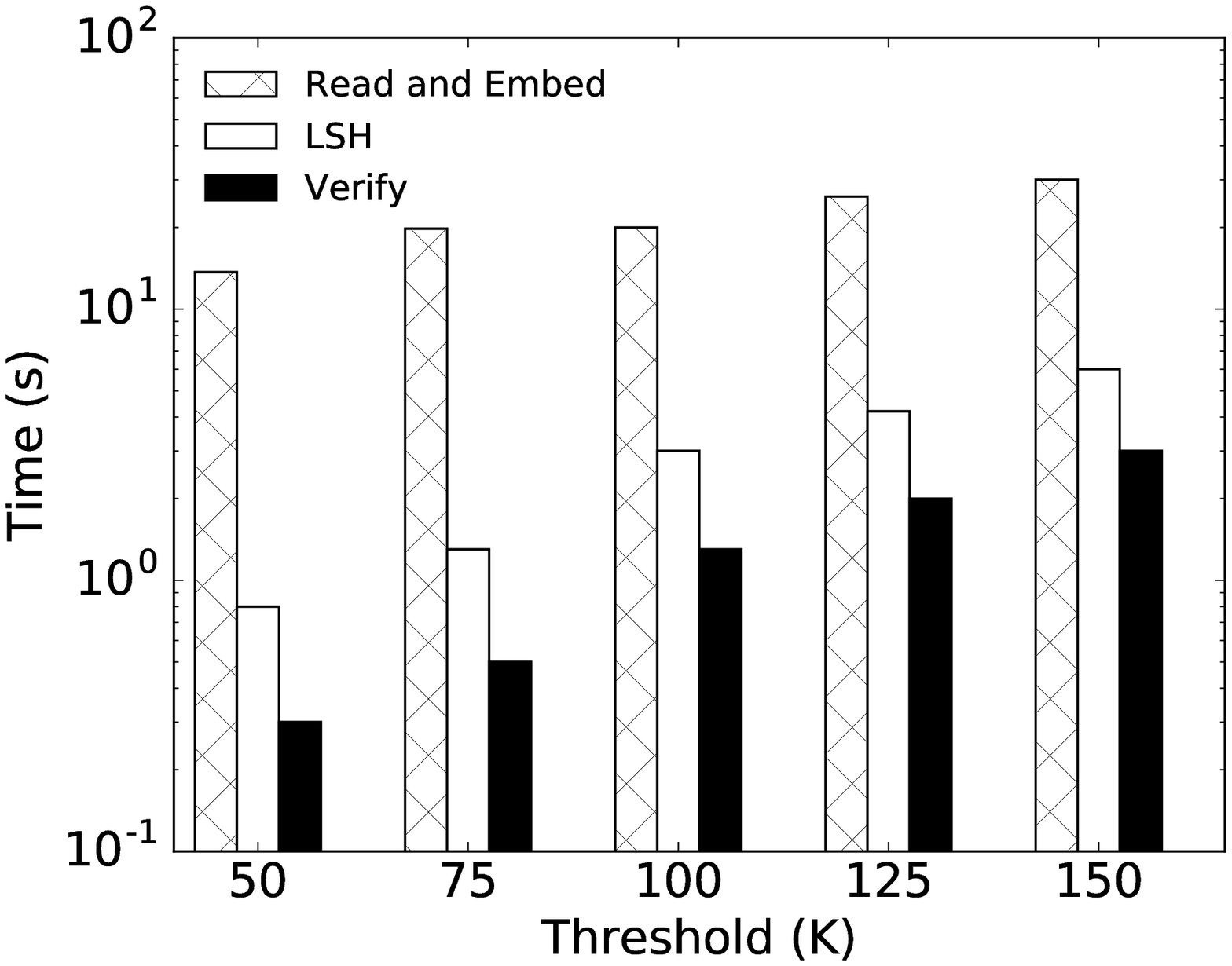}
\centerline{\genoa}
\end{minipage}
\caption{Running time of different parts of \ebdjoin+, varying $K$.}
\label{fig:timek}
\end{figure*}

Figure~\ref{fig:timen} and Figure~\ref{fig:timek} depict the running time of \ebdjoin+\ on (1) reading the input and CGK-embedding, (2) performing LSH, and (3) verification. We vary the number of input strings $n$ and the distance threshold $K$.  We observe that when $n,K$ increases, the time usages of all the three parts increase.  In all cases, the input reading and embedding is the bottleneck.  The first two parts are more sensitive to $n$, which have a higher increasing rate when $n$ increases, and are almost stable when $K$ increases. The verification time increases rapidly when both $n,K$ increase.

\begin{table*}[t]
\centering
\scalebox{0.8}{
  \begin{tabular}{|c|c|c|c|c|c|c|c|c|c|}
    \hline
    \multirow{3}{*}{Time(s)} &
      \multicolumn{3}{c|}{$r=5$} &
      \multicolumn{3}{c|}{$r=7$} &
      \multicolumn{3}{c|}{$r=9$} \\

& $z=8$  &$z=12$ &  $z=16$ & $z=8$  &$z=12$ &  $z=16$ & $z=8$  &$z=12$ &  $z=16$  \\
    \hline
    $m=11$ & 18.3 & 20.3 & 22.8& 25.4& 28.3 & 31.1 & 32.3 & 36.1 & 39.2 \\
    \hline
     $m=13$ & 17.8 & 19.9 & 21.9 & 24.9 & 27.7 & 30.1 & 31.4 & 35.7 & 38.5 \\
    \hline
     $m=15$ & 17.7 & 19.8 & 21.6 & 24.8 & 27.3 & 30.0 & 31.2 & 35.2 & 38.4 \\
    \hline
  \end{tabular}
  }
  \caption{Running time of \ebdjoin+, \genoa\ dataset, $K=100, \Delta = 50$}
\label{tab:genotime-1}
\end{table*}

\begin{table*}[t]
\centering
\scalebox{0.8}{
  \begin{tabular}{|c|c|c|c|c|c|c|c|c|c|}
    \hline
    \multirow{3}{*}{GB} &
      \multicolumn{3}{c|}{$r=5$} &
      \multicolumn{3}{c|}{$r=7$} &
      \multicolumn{3}{c|}{$r=9$} \\

& $z=8$  &$z=12$ &  $z=16$ & $z=8$  &$z=12$ &  $z=16$ & $z=8$  &$z=12$ &  $z=16$  \\
    \hline
    $m=11$ & 2.1 & 2.6 & 3.1 & 2.4 & 3.3 & 3.9 & 2.9 & 4.0 & 4.7\\
    \hline
     $m=13$ & 2.1 & 2.6 & 3.1 & 2.5 & 3.3 & 3.9 & 2.9 & 4.0 & 4.7 \\
    \hline
     $m=15$ & 2.3 & 2.6 & 3.5 & 2.8 & 3.3 & 4.5 & 3.3 & 4.0 & 5.6 \\
    \hline
  \end{tabular}
  }
  \caption{Memory usage of \ebdjoin+, \genoa\ dataset, $K=100, \Delta = 50$. }
\label{tab:genomem}
\end{table*}

\begin{figure*}[!ht]
\centering
\begin{minipage}[d]{0.4\linewidth}
\centering
\includegraphics[width=0.8\textwidth]{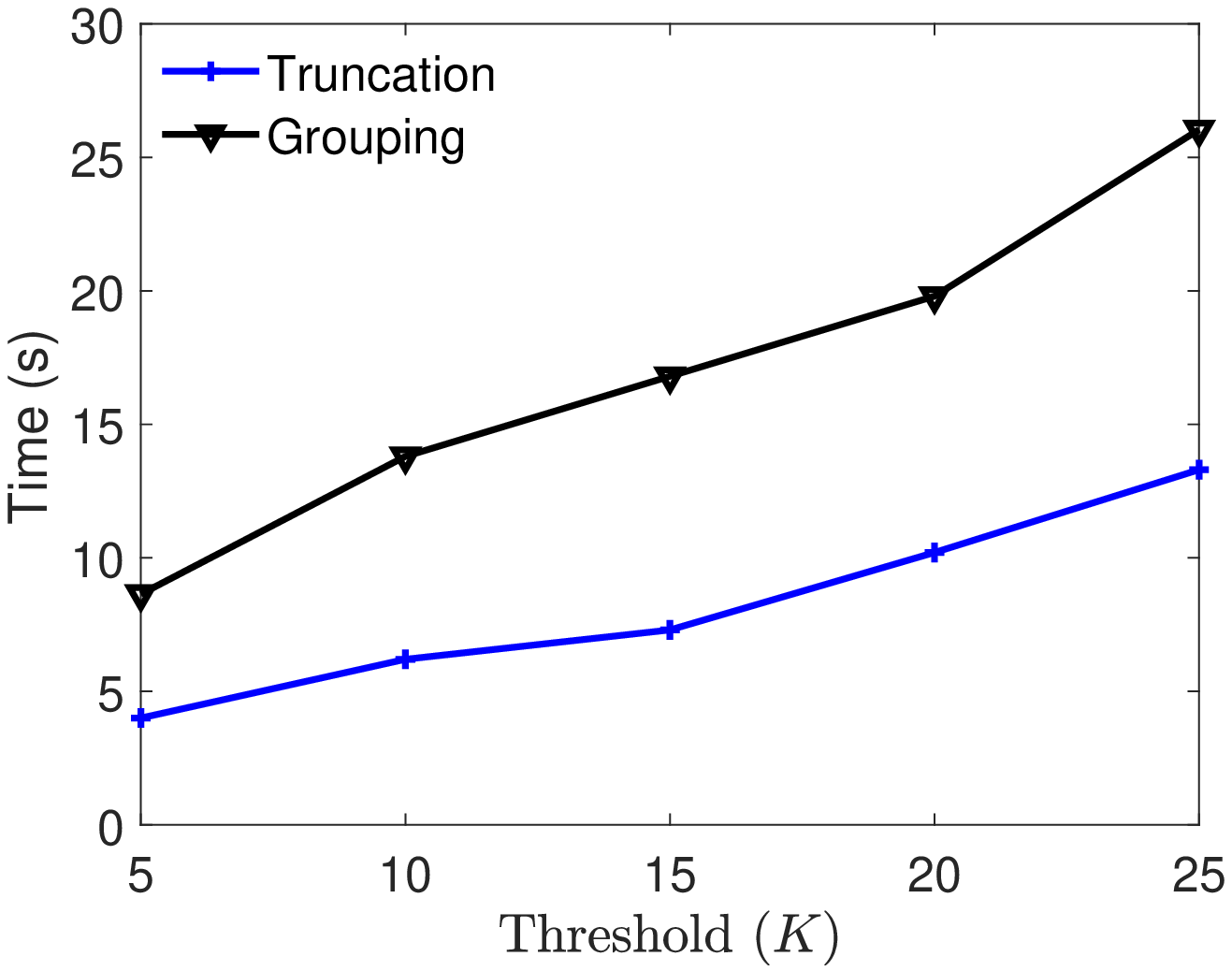}
\centerline{\uniref}
\end{minipage}
\begin{minipage}[d]{0.4\linewidth}
\centering
\includegraphics[width=0.8\textwidth]{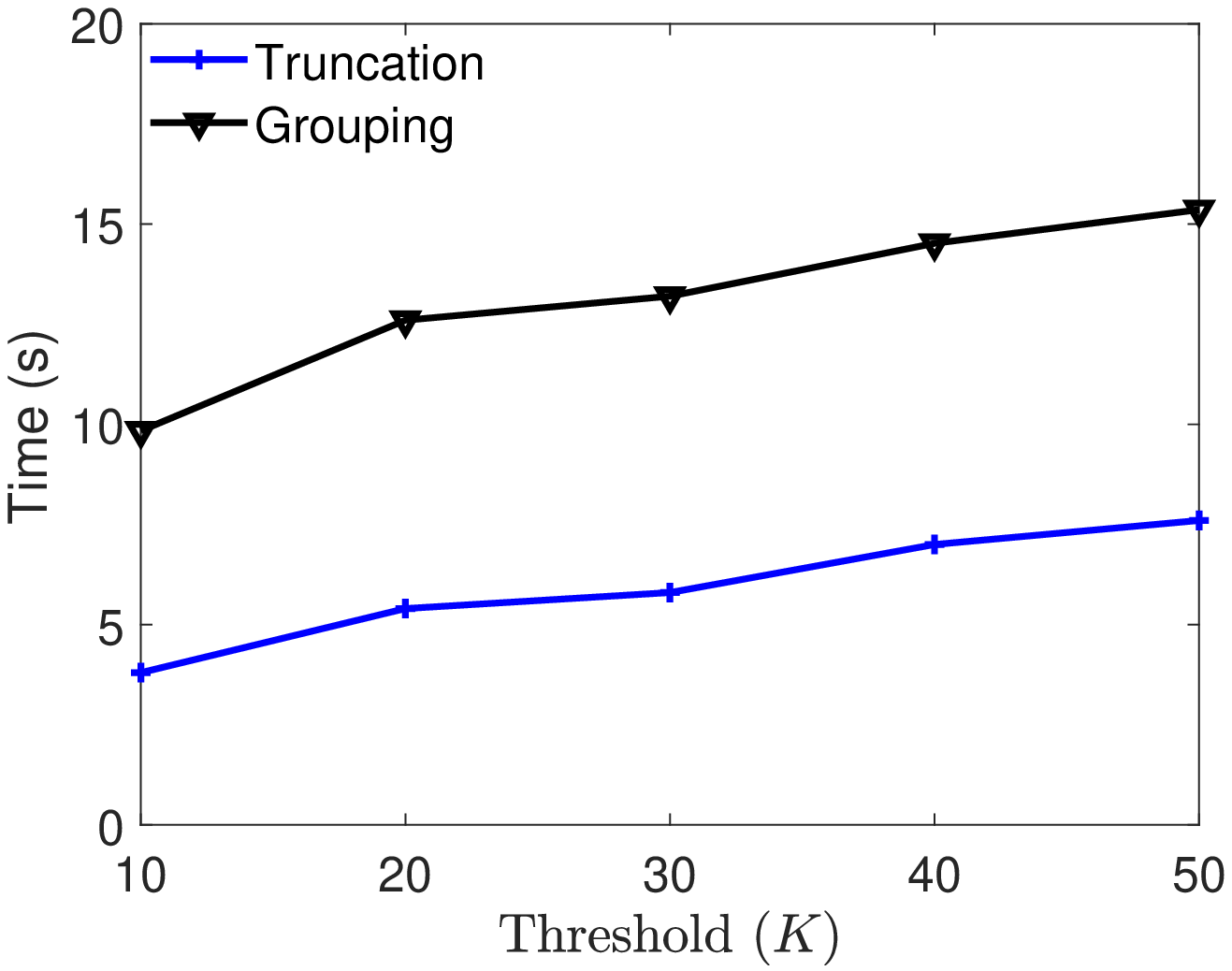}
\centerline{\trec}
\end{minipage}
\caption{Grouping vs Truncation, $r=5$, $z=5$, $m=15$.}
\label{fig:part}
\end{figure*}

 \begin{table*}[!ht]
\centering
\scalebox{0.8}{
  \begin{tabular}{|c|c|c|c|c|c|c|c|c|c|}
    \hline
    \multirow{3}{*}{Time(s)} &
      \multicolumn{3}{c|}{$r=5$} &
      \multicolumn{3}{c|}{$r=7$} &
      \multicolumn{3}{c|}{$r=9$} \\

& $z=8$  &$z=12$ &  $z=16$ & $z=8$  &$z=12$ &  $z=16$ & $z=8$  &$z=12$ &  $z=16$  \\
    \hline
    $\Delta=25$  & 30.2 & 32.5 & 34.8 & 41.6& 45.1 & 48.4 & 53.2 & 58.0 & 62.3 \\
    \hline
     $\Delta=34$  & 26.1 & 28.6 & 31.7 & 36.5 & 40.8 & 45.0 & 46.7 & 51.6 & 57.3 \\
    \hline
     $\Delta=50$  & 17.7 & 19.8 & 21.6 & 24.8 & 27.3 & 30.0 & 31.2 & 35.2 & 38.4 \\
    \hline
  \end{tabular}
  }
  \caption{Running time of \ebdjoin+, \genoa\ dataset, $K=100, m = 13$}
\label{tab:genotimedel}
\end{table*}

Figure~\ref{fig:part} shows the running time of \ebdjoin+\ on datasets with strings of different lengths (\uniref\ and \trec), using the grouping method and the truncation method respectively.  It is clear that truncation is always better than grouping.  
 We thus always use truncation-based \ebdjoin\ and \ebdjoin+ in our (other) experiments.

In Table~\ref{tab:genotimedel} we study how the parameter $\Delta$ influences the running time of \ebdjoin+. We vary $\Delta$ in $\{25, 34, 50\}$ so that the number of substrings for each string are $\{4, 3, 2\}$. We observe that the running time increases when $\Delta$ decreases. This is because when $\Delta$ decreases, there are more substrings to embed and hash for each string, and more candidates to verify.

\paragraph{The Filtering Quality}
Table~\ref{tab:can} shows how different parameters $(r, z, m)$ influence the number of candidates generated by \ebdjoin+.  We use \genoa\ as the test dataset.   We observe that the number of candidates is consistent to the running time, that is, the number increases when $r$ and $z$ increase, and decreases when $m$ increases.  From the table we can see that under different parameters, our numbers of candidates are about $2.80 \sim 5.07$ times of the ground truth $6317$. 

\begin{table*}[!ht]
\centering
\scalebox{0.8}{
  \begin{tabular}{|c|c|c|c|c|c|c|c|c|c|}
    \hline
    \multirow{3}{*}{\# Candidates} &
      \multicolumn{3}{c|}{$r=5$} &
      \multicolumn{3}{c|}{$r=7$} &
      \multicolumn{3}{c|}{$r=9$} \\

& $z=8$  &$z=12$ &  $z=16$ & $z=8$  &$z=12$ &  $z=16$ & $z=8$  &$z=12$ &  $z=16$  \\
    \hline
    $m=11$ & 21498  & 22650  & 24493  & 22652  & 24593  & 28266  & 24002  & 28948   & 32050  \\
    \hline
     $m=13$ & 19245  &20791   & 21256   & 20387  & 21516  & 22163  & 21747  & 22496  & 23390  \\
    \hline
     $m=15$ & 17686  & 19072  & 20269  & 19723   & 21038 & 21746   & 20253  & 21730 & 22377   \\
    \hline
  \end{tabular}
  }
\caption{Number of candidate pairs after filtering of \ebdjoin+, \genoa\ dataset, $K=100, \Delta = 50$; ground truth is $6317$.}
\label{tab:can}
\end{table*}

\subsection{A Comparison with Existing Algorithms}
\label{sec:comparison}

In this section we compare \ebdjoin\ and \ebdjoin+\ with the existing best algorithms introduced in Section~\ref{sec:setup}.  
We note that in some figures some data points for competing algorithms are missing, which is either because these algorithms have implementation limitations (returned wrong answers or triggered memory overflow) or they cannot finish in 24 hours in our computing environment.

\begin{figure*}[!t]
\centering
\begin{minipage}[d]{0.4\linewidth}
\centering
\includegraphics[width=0.9\textwidth]{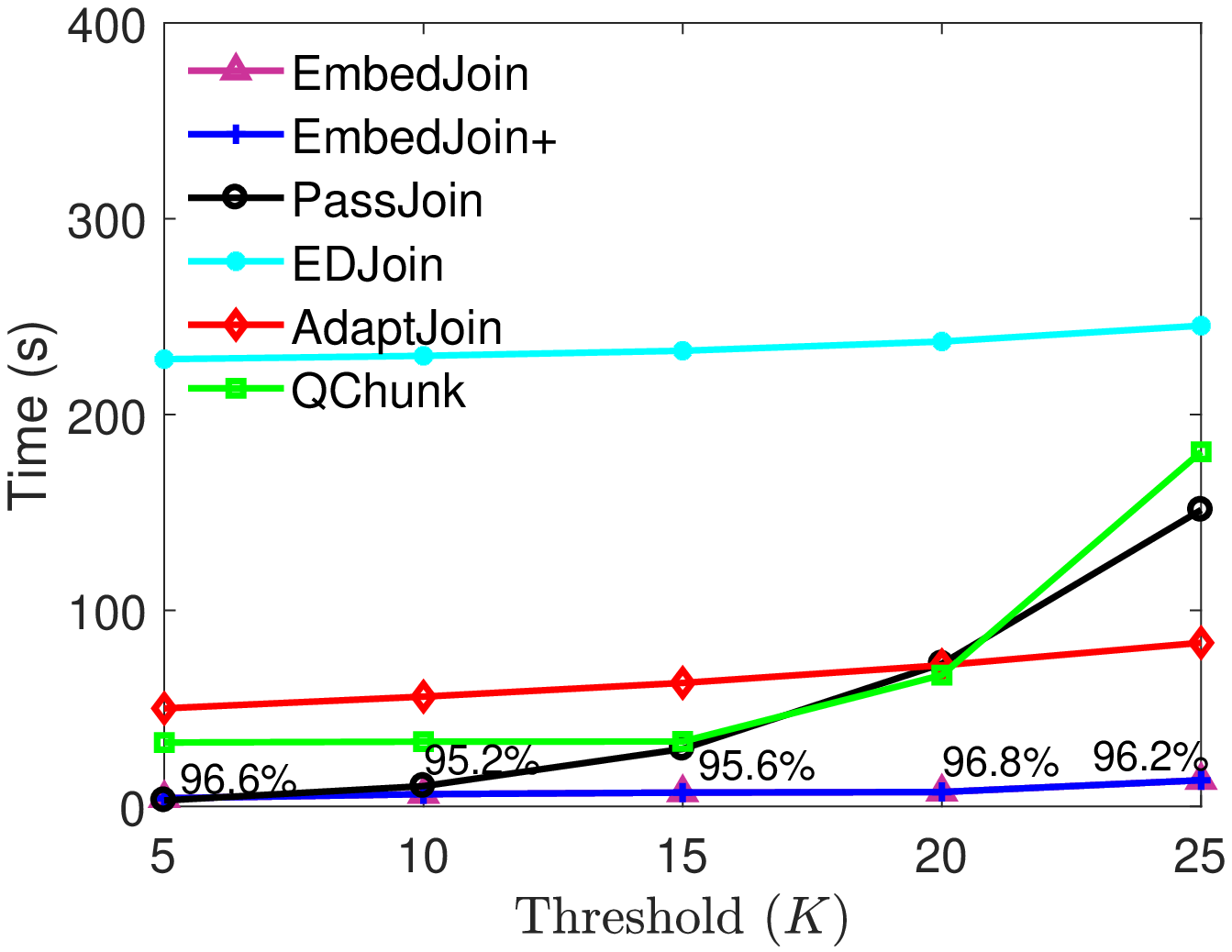}
\centerline{\uniref}
\end{minipage}
\begin{minipage}[d]{0.4\linewidth}
\centering
\includegraphics[width=0.9\textwidth]{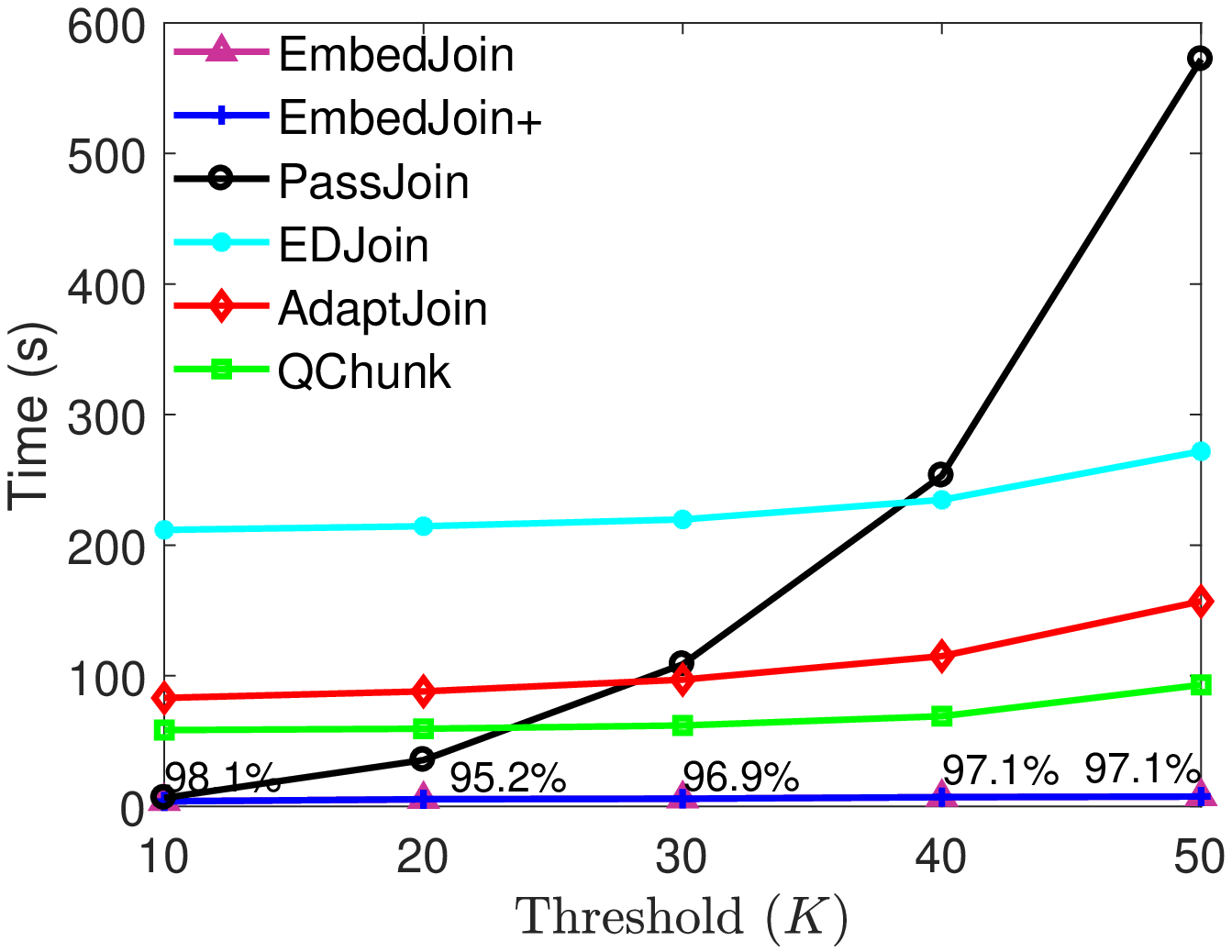}
\centerline{\trec}
\end{minipage}
\begin{minipage}[d]{0.4\linewidth}
\centering
\includegraphics[width=0.9\textwidth]{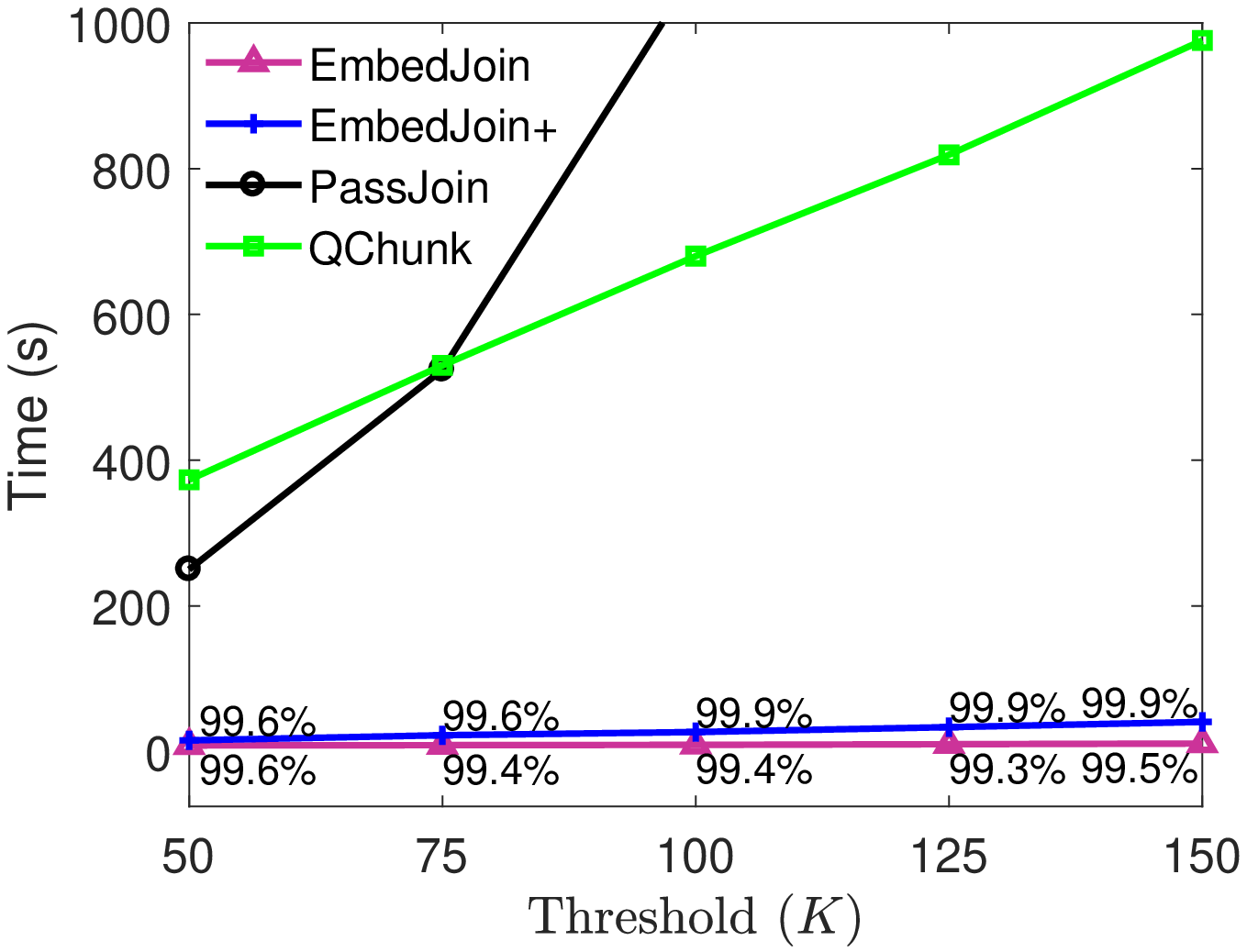}
\centerline{\genoaa}
\end{minipage}
\begin{minipage}[d]{0.4\linewidth}
\centering
\includegraphics[width=0.9\textwidth]{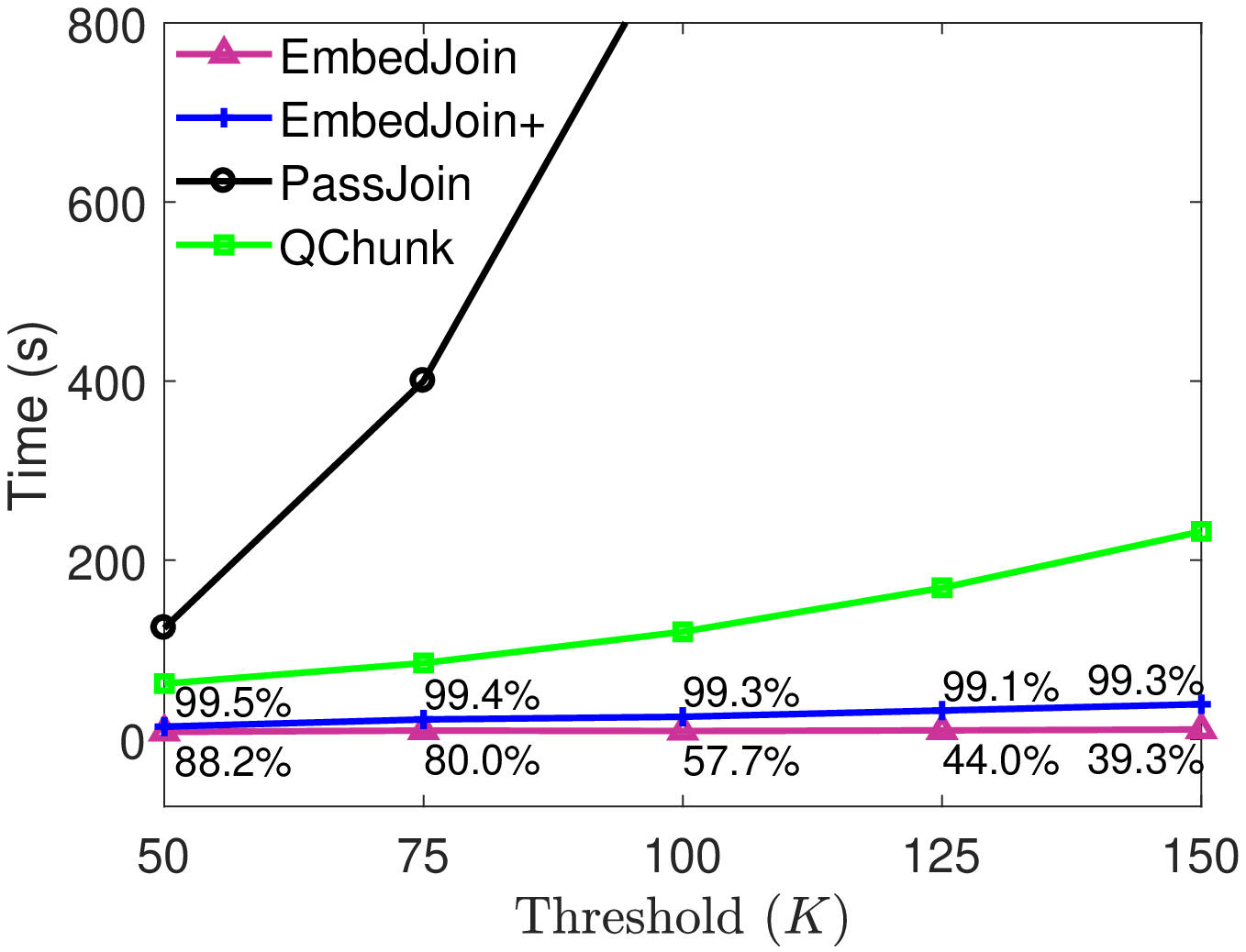}
\centerline{\genoa}
\end{minipage}
\caption{Running time, varying $K$.  Percentages on the curves for \ebdjoin/\ebdjoin+\ are their accuracy
}
\label{fig:ktime}
\end{figure*}

\paragraph{Scalability on the Threshold Distance}
Figure~\ref{fig:ktime} shows the running time of different algorithms when varying the distance threshold $K$ on \uniref, \trec, \genoaa\ and \genoa.   In all experiments we always guarantee that the accuracy of  \ebdjoin\ and \ebdjoin+\ is above $95\%$ on \uniref\ and \trec, and above $99\%$ on \genoaa. We use the same parameters for both algorithms on \genoa\ and \genoaa.  On \uniref\ and \trec\ datasets, where we choose $T = 1$ for \ebdjoin+, \ebdjoin\ and \ebdjoin+\ become the same algorithm, and thus have same accuracy, memory usage and running time.

We observe that \ebdjoin\ and \ebdjoin+\ always have the best time performances: the running time of \ebdjoin\ is better than the best existing algorithm by a factor of $9.2$ on \uniref\  ($K = 20$), 10.2 in \trec\ ($K = 50$), $88.7$ on \genoaa\ ($K = 150$), and $21.4$ on \genoa\ ($K = 150$);  the running time of \ebdjoin+\ is better than the best existing algorithm by a factor of $9.2$ on \uniref\  ($K = 20$), 10.2 in \trec\ ($K = 50$), $23.9$ on \genoaa\ ($K = 150$), and $5.9$ on \genoa\ ($K = 150$).

\begin{figure*}[t]
\centering
\begin{minipage}[d]{0.4\linewidth}
\centering
\includegraphics[width=0.9\textwidth]{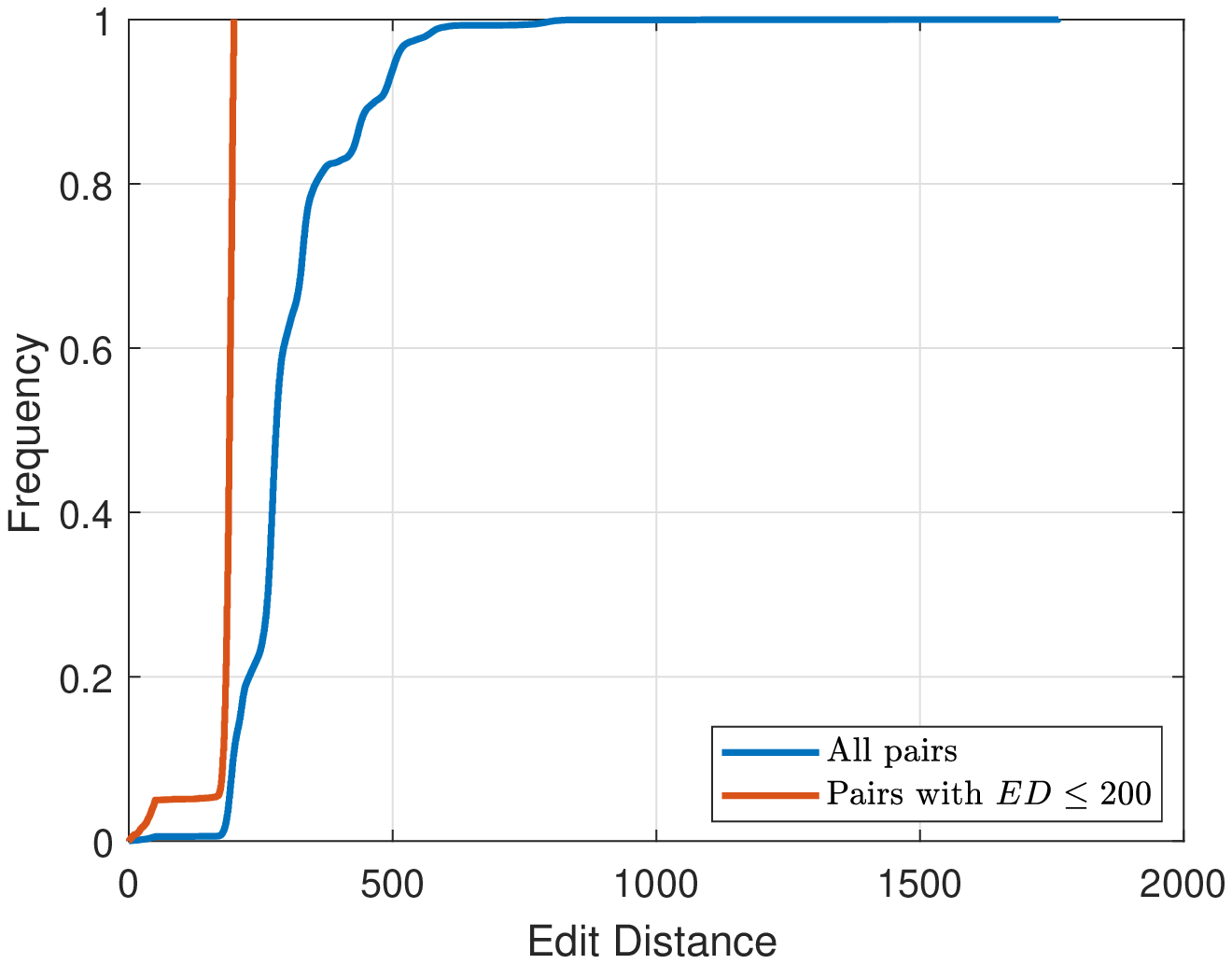}
\centerline{\uniref}
\end{minipage}
\begin{minipage}[d]{0.4\linewidth}
\centering
\includegraphics[width=0.9\textwidth]{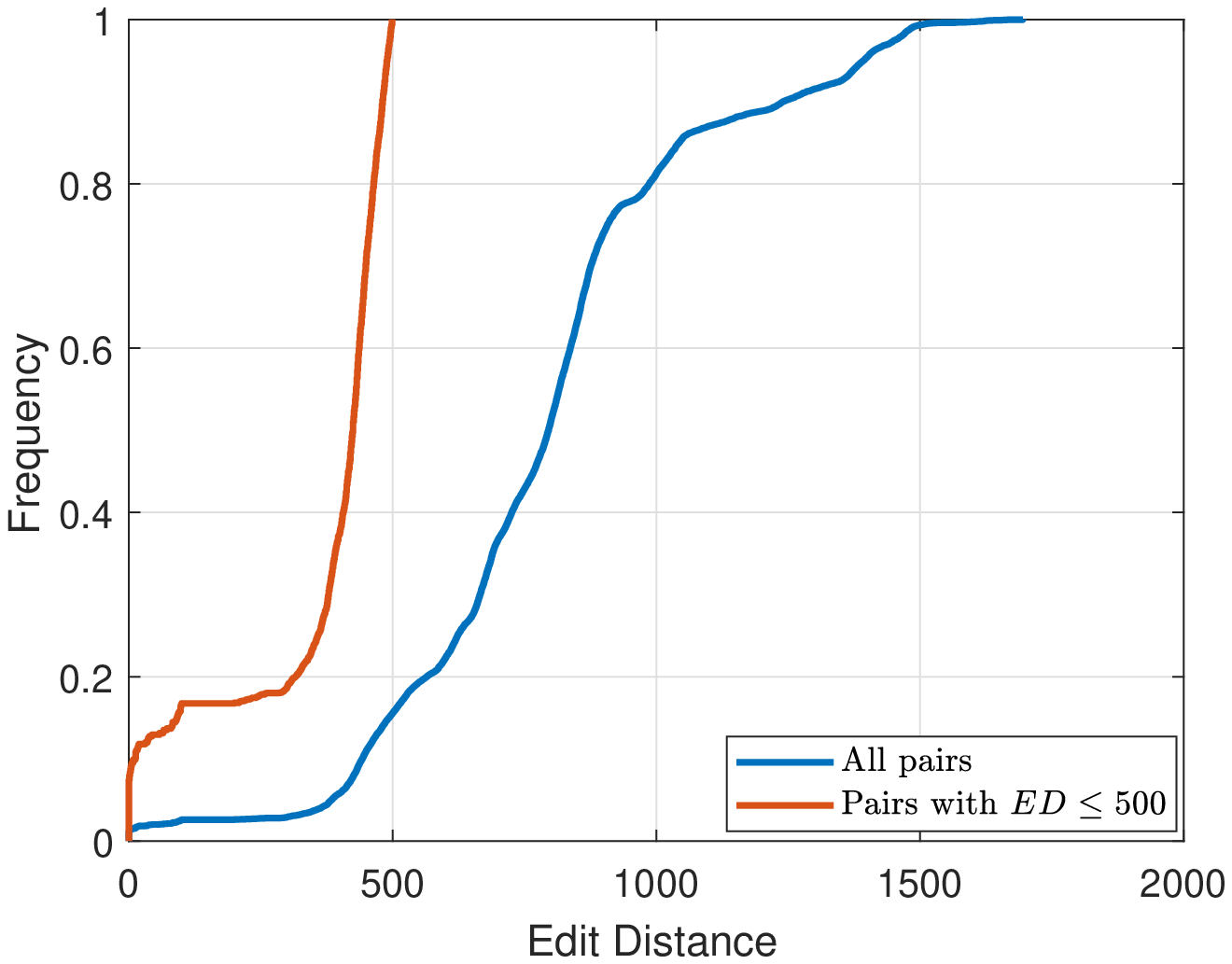}
\centerline{\trec}
\end{minipage}
\begin{minipage}[d]{0.4\linewidth}
\centering
\includegraphics[width=0.9\textwidth]{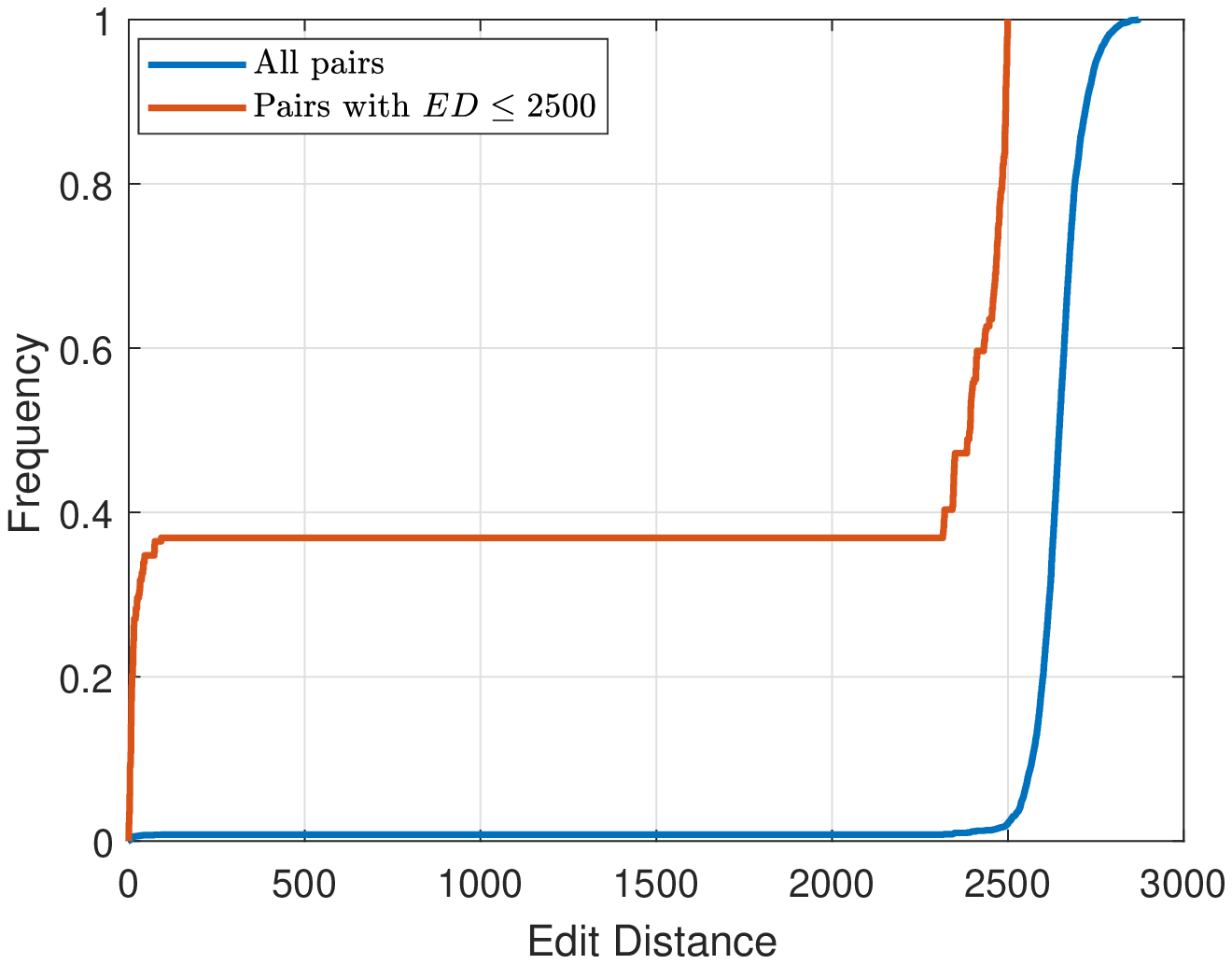}
\centerline{\genoaa}
\end{minipage}
\begin{minipage}[d]{0.4\linewidth}
\centering
\includegraphics[width=0.9\textwidth]{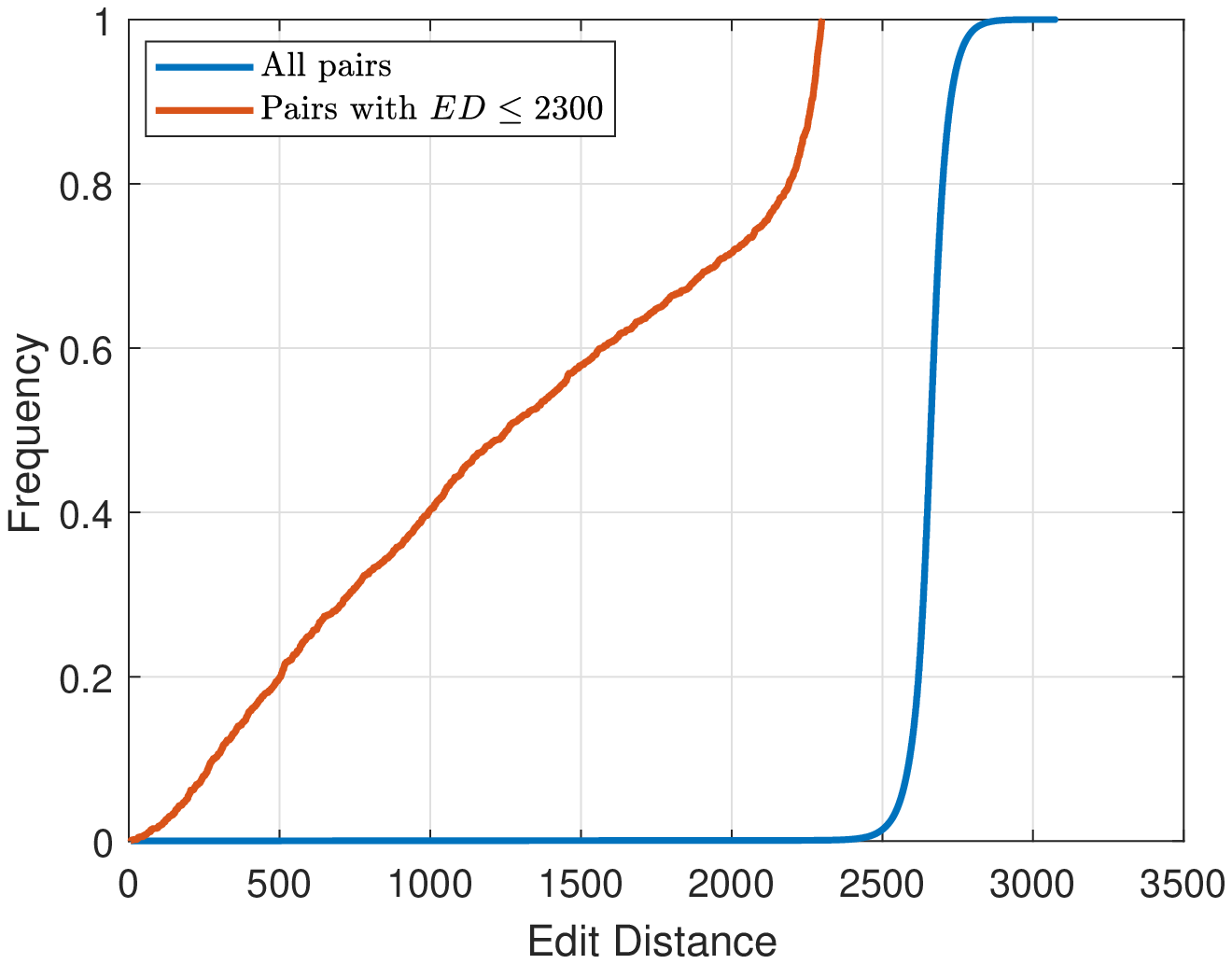}
\centerline{\genoa}
\end{minipage}
\caption{Distance distribution of datasets.}
\label{fig:dist}
\end{figure*}

However, the accuracy of \ebdjoin\ is as low as $39.3\%$ on \genoa\ when $K = 150$. The main reason is that the pairwise edit distance distributes almost uniformly on \genoa\ (and on other random reads genome datasets as well). On the rest of the datasets, there are clear gaps between similar and dissimilar pairs.  See Figure~\ref{fig:dist} for the details.
When the distance gap exists, the distortion generated by the CGK-embedding becomes less critical. Otherwise, in order to maintain a high accuracy, we have to make sure that there are not many false negatives by maintaining a large candidate set, which can be done by adjusting the parameters in LSH.  However, this will make the verification step very expensive, and is thus not a good idea overall.  

The above issue is resolved in \ebdjoin+.  The motivation of proposing \ebdjoin+, as presented in Section~\ref{sec:improved}, is to reduce the shift between a pair of strings so as to reduce the distortion of the CGK-embedding.  Note that when the shift is reduced, the edit distance of the remaining pair of substrings is smaller than original one, which helps to remove false negatives {\em without} changing the LSH module by much (compared with the idea of trying to modify the original \ebdjoin\ mentioned above).  After such a procedure the number of false positives in the candidate set will still increase, but only at a modest amount.

The \pass\ algorithm does not scale well on $K$: when $K$ increases, the running time jumps sharply.  This may due to the fact that the time complexity in the filtering step of \pass\ is $O(n K^3)$ -- a cubic dependence on $K$.  The other three algorithms, \edjoin, \adpjoin\ and \qchunk, are all based on $q$-gram or its variants; they generally have similar running time curves, which rise much slower compared with \pass\ when $K$ increases.  One exception is that on the \uniref\ dataset the running time of \qchunk\ increases sharply when $K$ passes $20$, which may due to the sudden increase of the number of candidate pairs that \qchunk\ produces.  On \genoaa\ and \genoa, the running time of \edjoin\ is too large ($>10000$s when $K = 50$) and thus does not fit the figure, and \adpjoin\ reports erroneous results.


\begin{figure*}[t]
\centering
\begin{minipage}[d]{0.4\linewidth}
\centering
\includegraphics[width=0.9\textwidth]{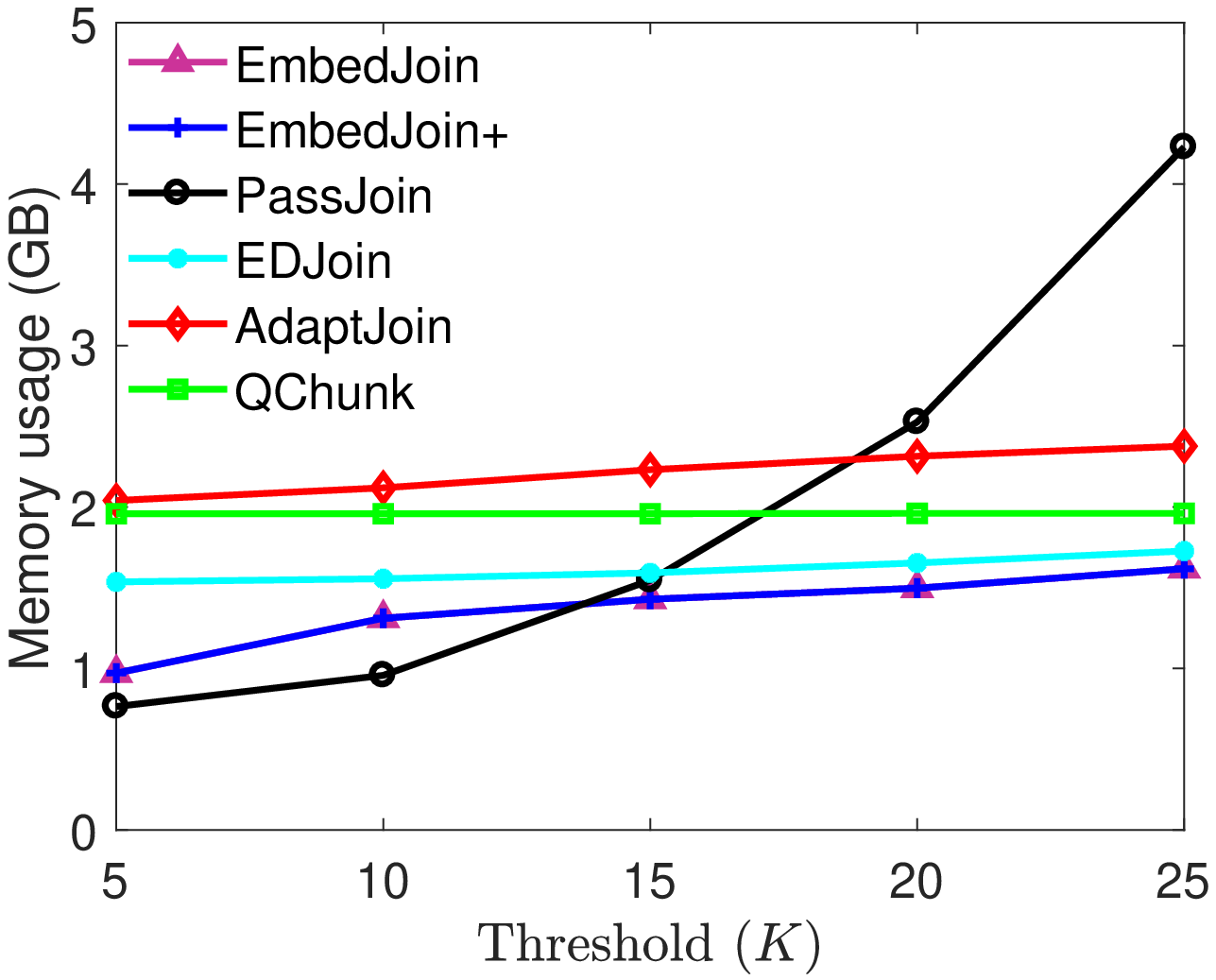}
\centerline{\uniref}
\end{minipage}
\begin{minipage}[d]{0.4\linewidth}
\centering
\includegraphics[width=0.9\textwidth]{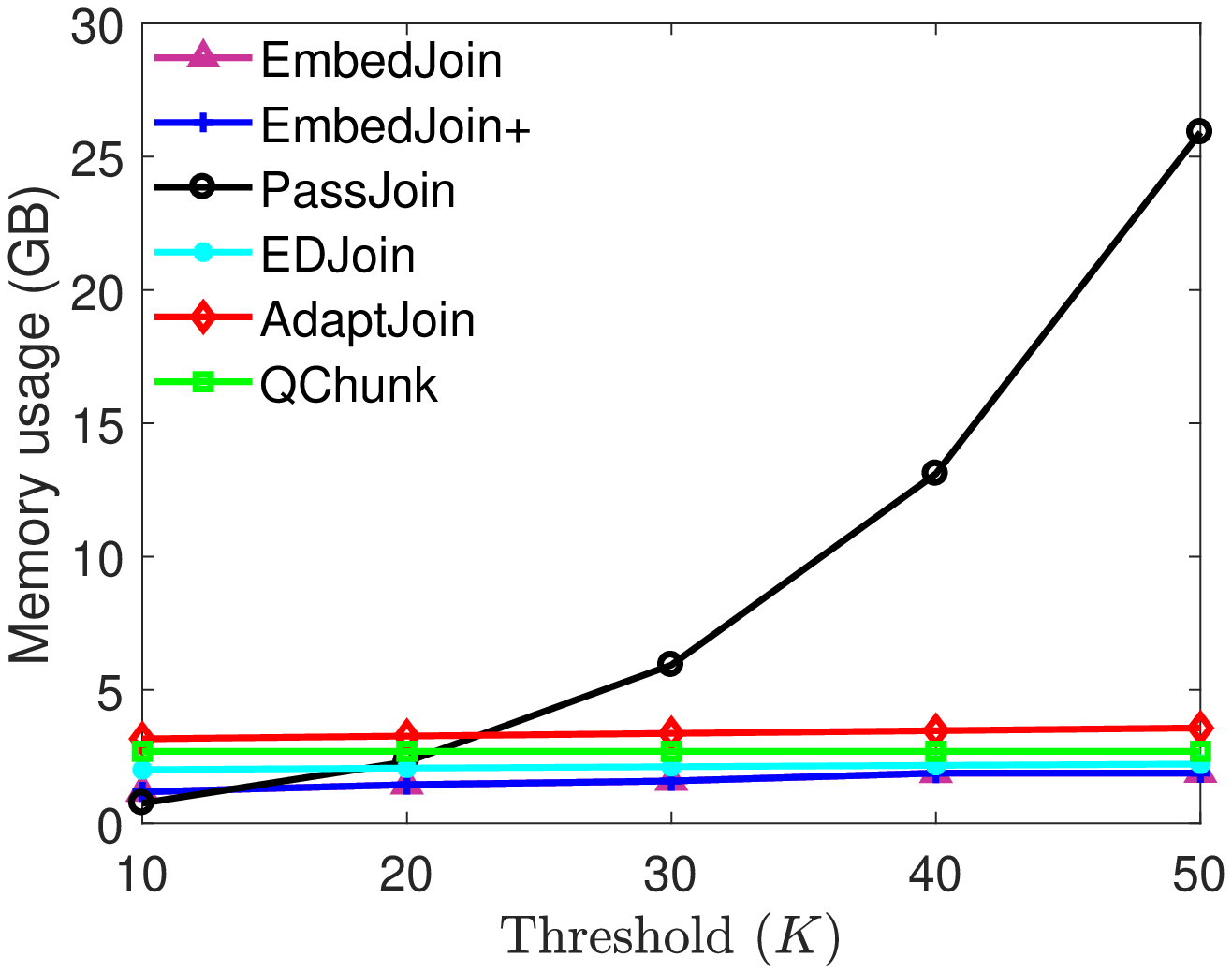}
\centerline{\trec}
\end{minipage}
\begin{minipage}[d]{0.4\linewidth}
\centering
\includegraphics[width=0.9\textwidth]{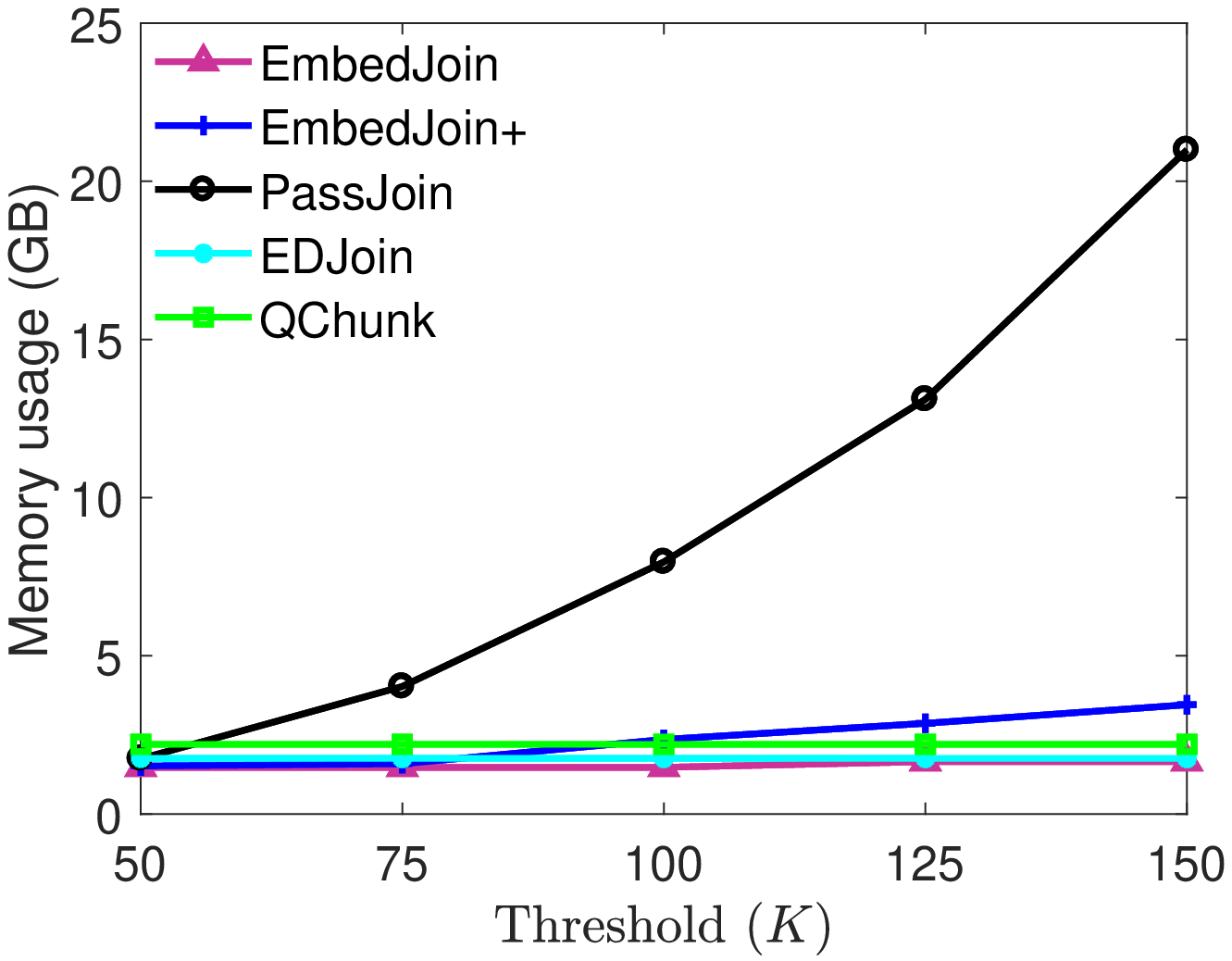}
\centerline{\genoaa}
\end{minipage}
\begin{minipage}[d]{0.4\linewidth}
\centering
\includegraphics[width=0.9\textwidth]{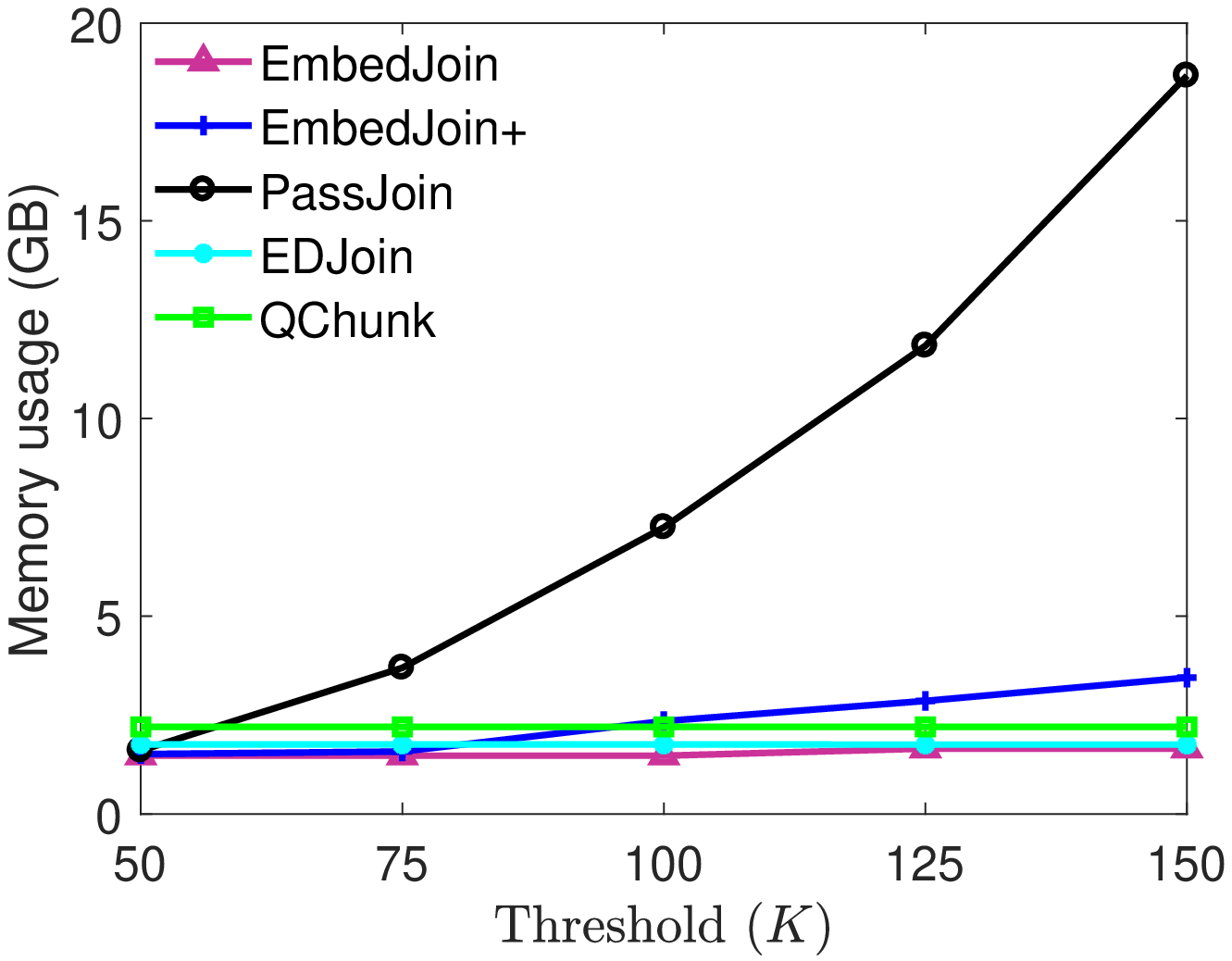}
\centerline{\genoa}
\end{minipage}
\caption{Memory usage, varying $K$. 
}
\label{fig:kmem}
\end{figure*}

Figure~\ref{fig:kmem} shows the memory usages of different algorithms in the same settings as Figure~\ref{fig:ktime}.  The memory used by \ebdjoin\ is the smallest among all in most cases, and \ebdjoin+\ uses a little bit more memory when $K$ is relatively large. Note that the memory usage of \ebdjoin+\ has a linear dependency on $K$, which is because the number of substrings for each string is $\lceil K/\Delta \rceil$ and we need to store signatures for each of them.  The memory usage of \pass\ is also small at the beginning, but deteriorates fast when $K$ increases.  The three $q$-gram based algorithms have similar trends in memory usage.

\begin{figure*}[t]
\centering
\begin{minipage}[d]{0.4\linewidth}
\centering
\includegraphics[width=0.9\textwidth]{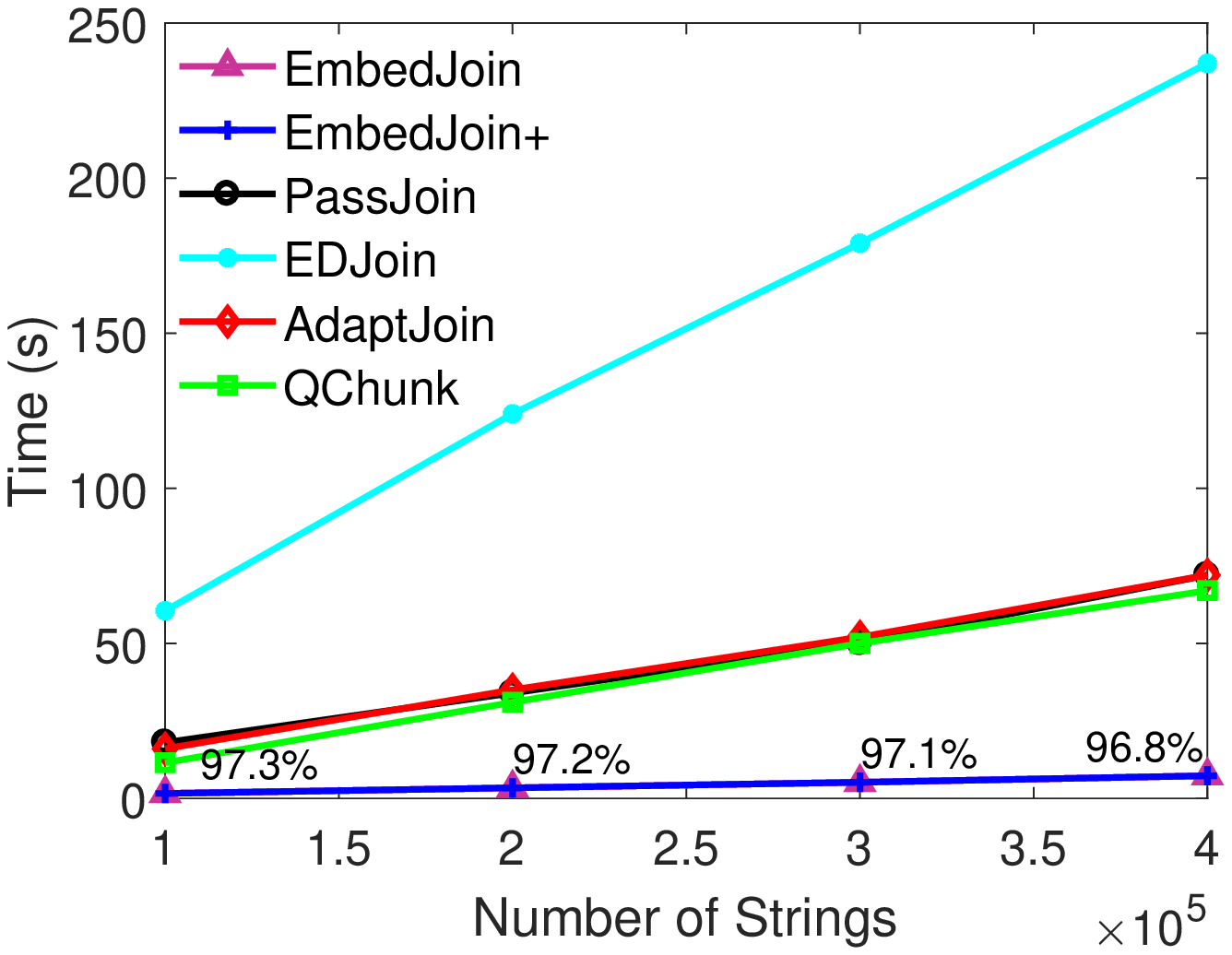}
\centerline{\uniref\ ($K=20$)}
\end{minipage}
\begin{minipage}[d]{0.4\linewidth}
\centering
\includegraphics[width=0.9\textwidth]{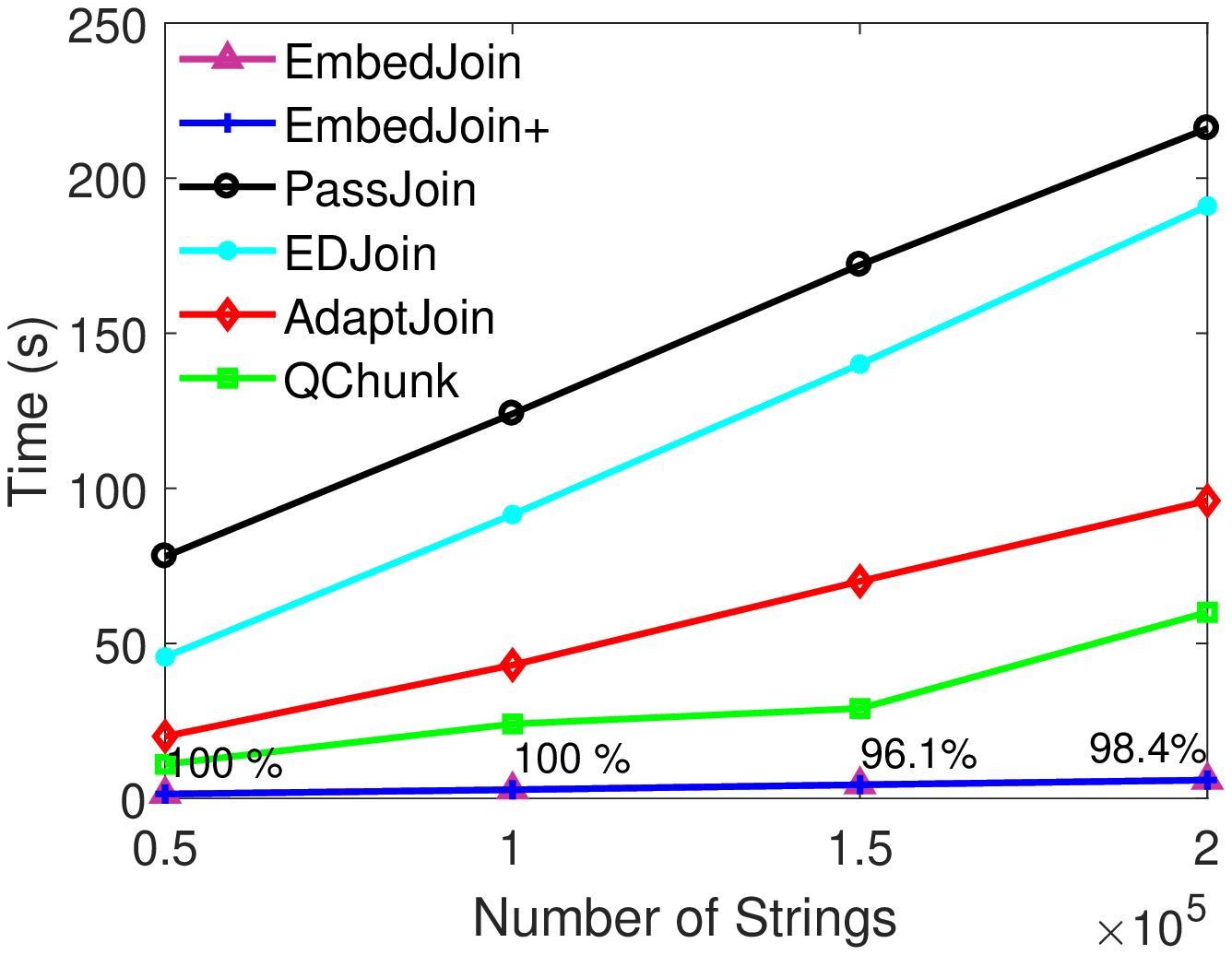}
\centerline{\trec\ ($K=40$)}
\end{minipage}
\begin{minipage}[d]{0.4\linewidth}
\centering
\includegraphics[width=0.9\textwidth]{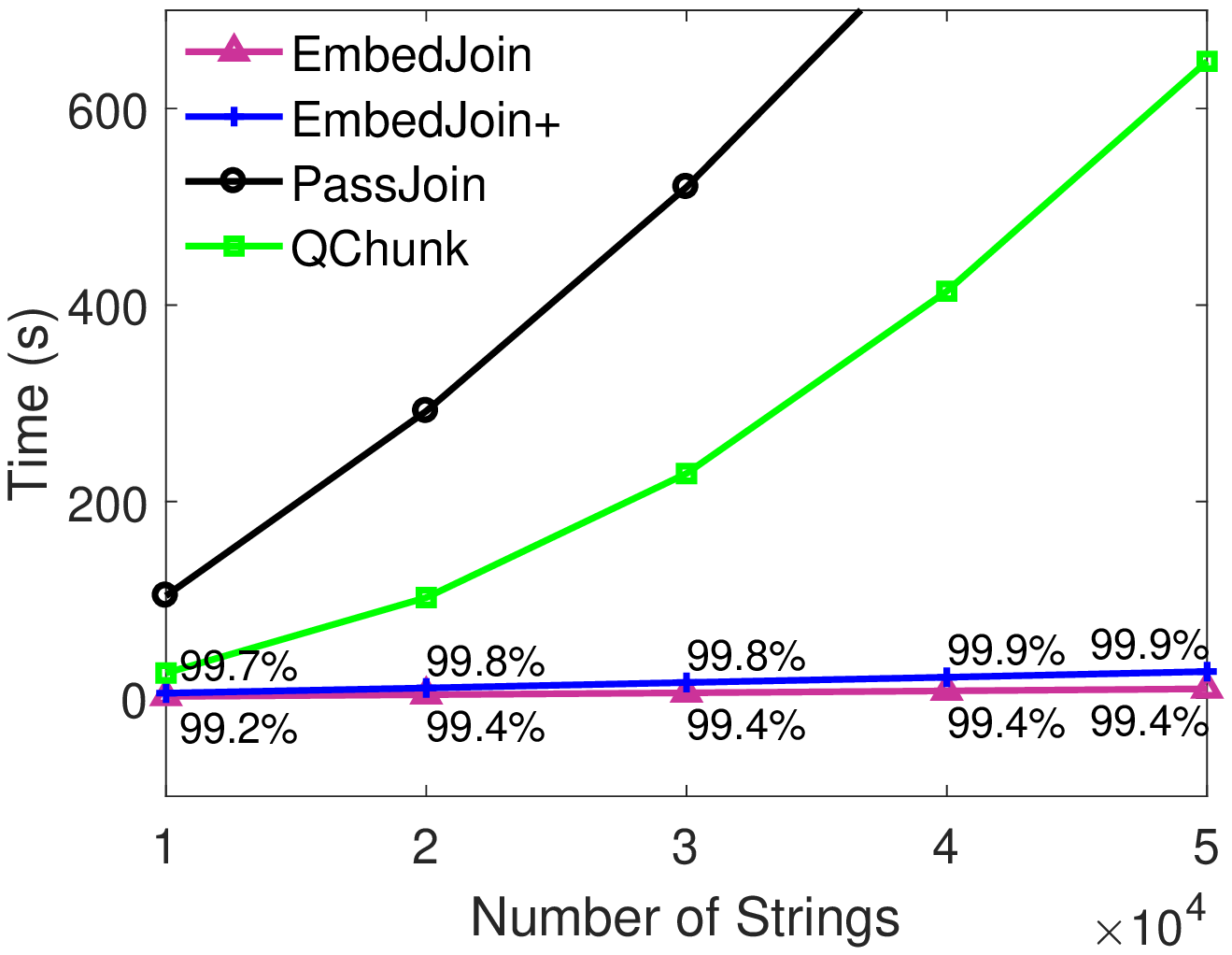}
\centerline{\genoaa\ ($K=100$)}
\end{minipage}
\begin{minipage}[d]{0.4\linewidth}
\centering
\includegraphics[width=0.9\textwidth]{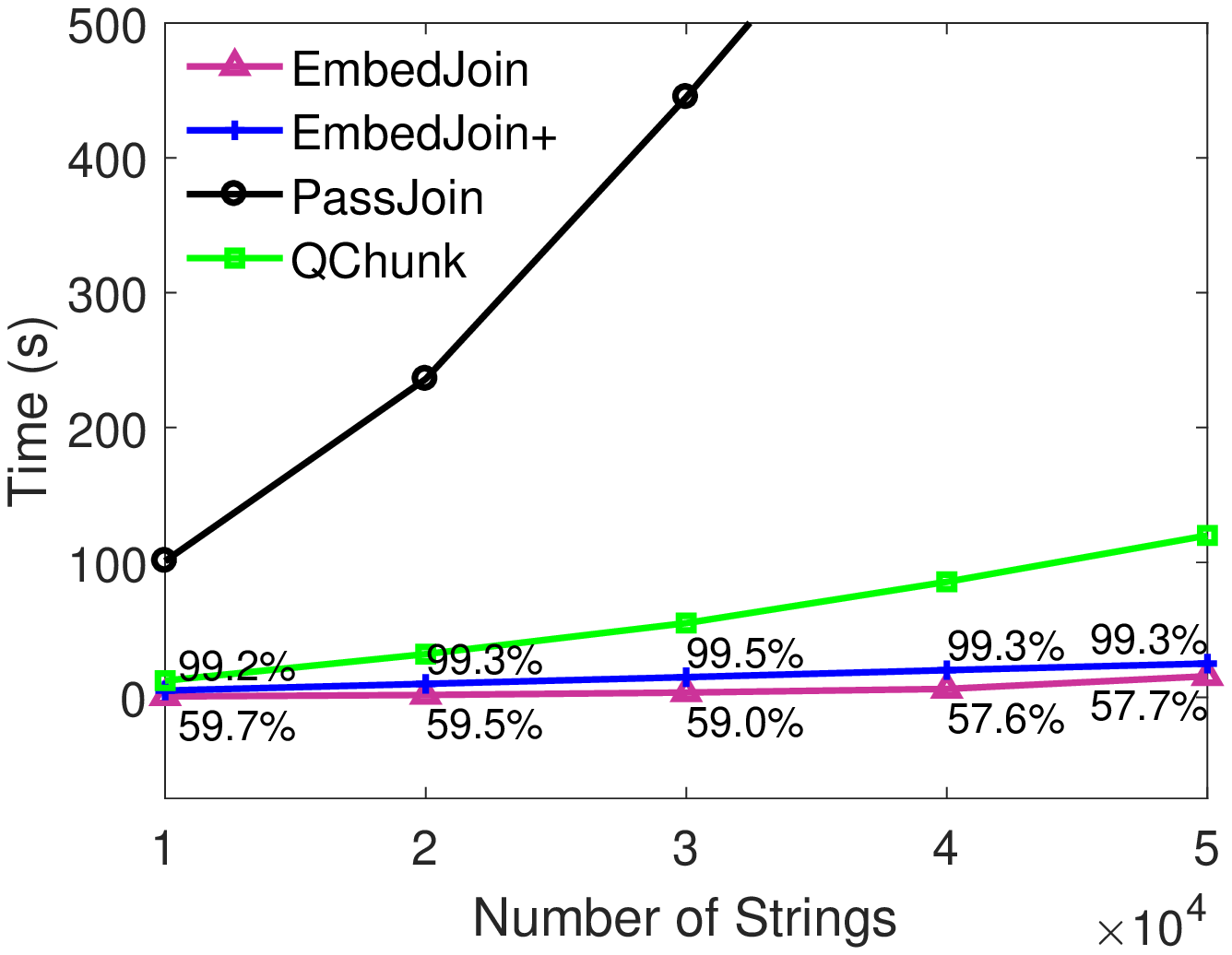}
\centerline{\genoa\ ($K=100$)}
\end{minipage}
\caption{Running time, varying $n$.  Percentages on the curves for \ebdjoin/\ebdjoin+\  are their accuracy}
\label{fig:ntime}
\end{figure*}

\paragraph{Scalability on the Input Size}
Figure~\ref{fig:ntime} shows the running time of different algorithms on the \uniref, \trec, \genoaa\ and \genoa\ datasets when varying input size $n$. The trends of the running time of all algorithms are similar; they increase with respect to $n$.   It is clear that \ebdjoin\ and \ebdjoin+\ perform much better than all the other algorithms: \ebdjoin\ performs better than the best existing algorithm by a factor of $9.2$ on \uniref\ ($N = 4 \times 10^5$), $11.5$ on \trec\ ($N = 2 \times 10^5$), $69.7$ on \genoaa\ ($N = 5 \times 10^4$), and $7.7$ on \genoa\ ($N = 5 \times 10^4$);  the running time of \ebdjoin+\ is better than the best existing algorithm by a factor of $9.2$ on \uniref\ ($N = 4 \times 10^5$), $11.5$ on \trec\ ($N = 2 \times 10^5$), $22.1$ on \genoaa\ ($N = 5 \times 10^4$), and $4.8$ on \genoa\ ($N = 5 \times 10^4$) . 

\begin{figure*}[t]
\centering
\begin{minipage}[d]{0.4\linewidth}
\centering
\includegraphics[width=0.9\textwidth]{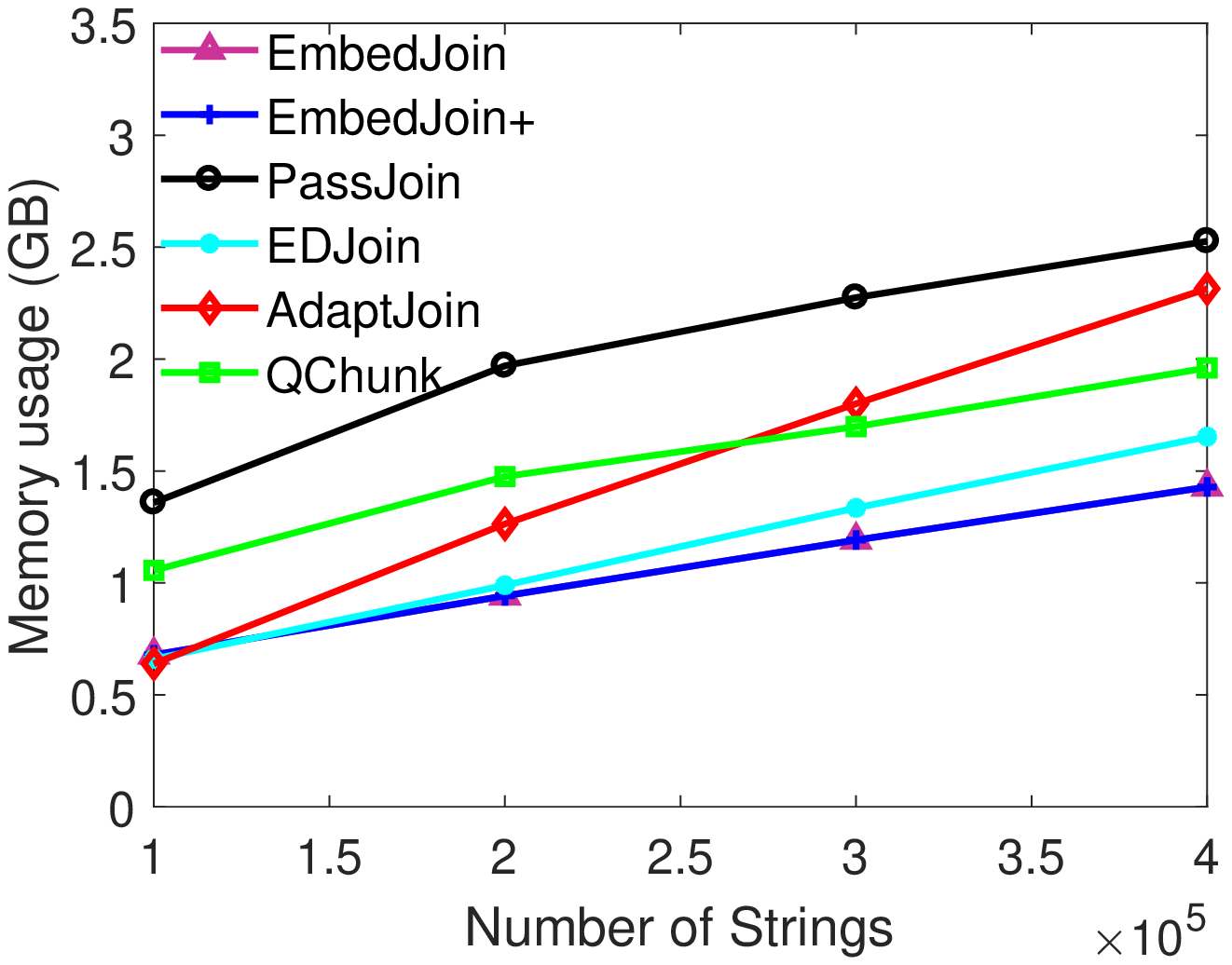}
\centerline{\uniref\ ($K=20$)}
\end{minipage}
\begin{minipage}[d]{0.4\linewidth}
\centering
\includegraphics[width=0.9\textwidth]{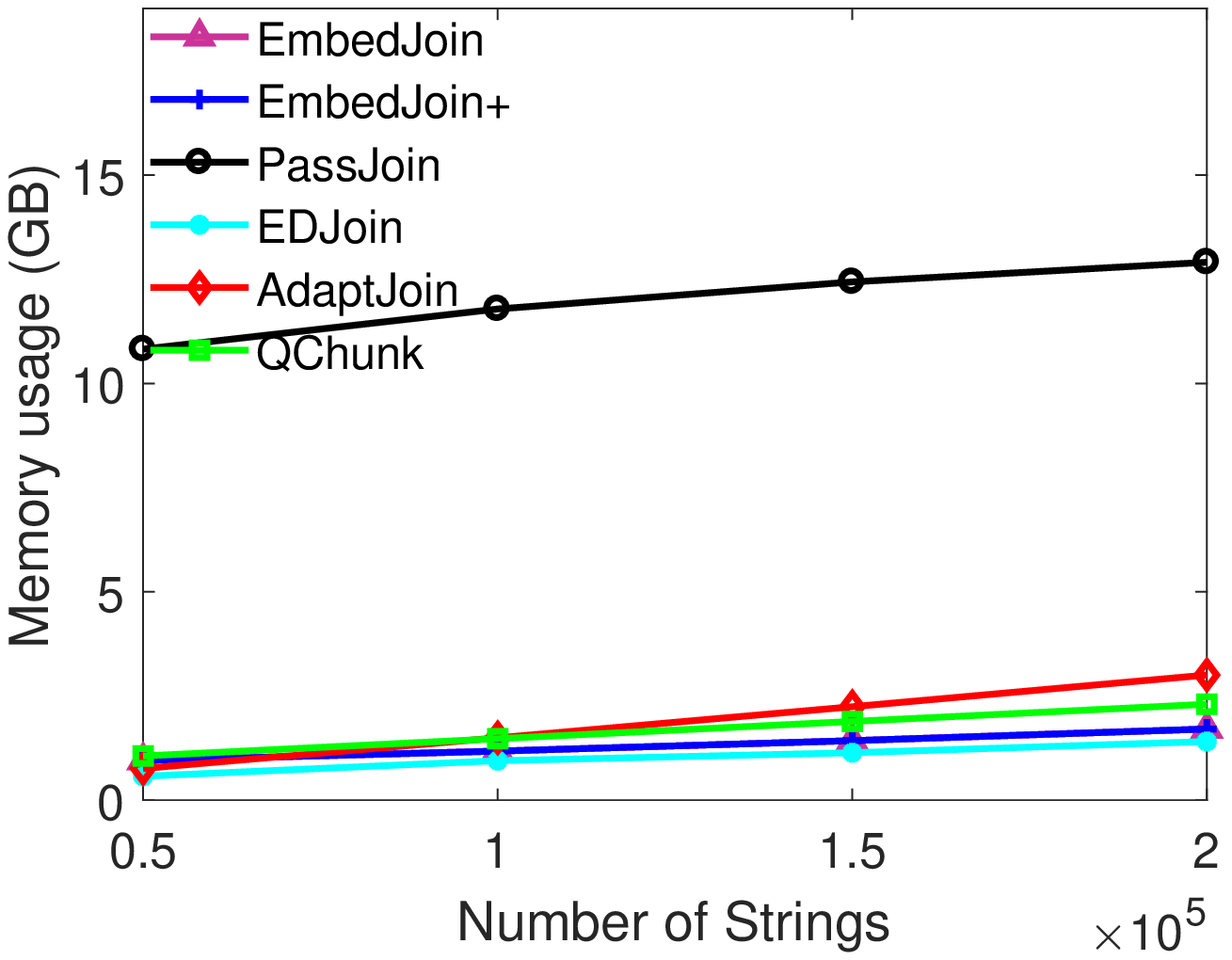}
\centerline{\trec\ ($K=40$)}
\end{minipage}
\begin{minipage}[d]{0.4\linewidth}
\centering
\includegraphics[width=0.9\textwidth]{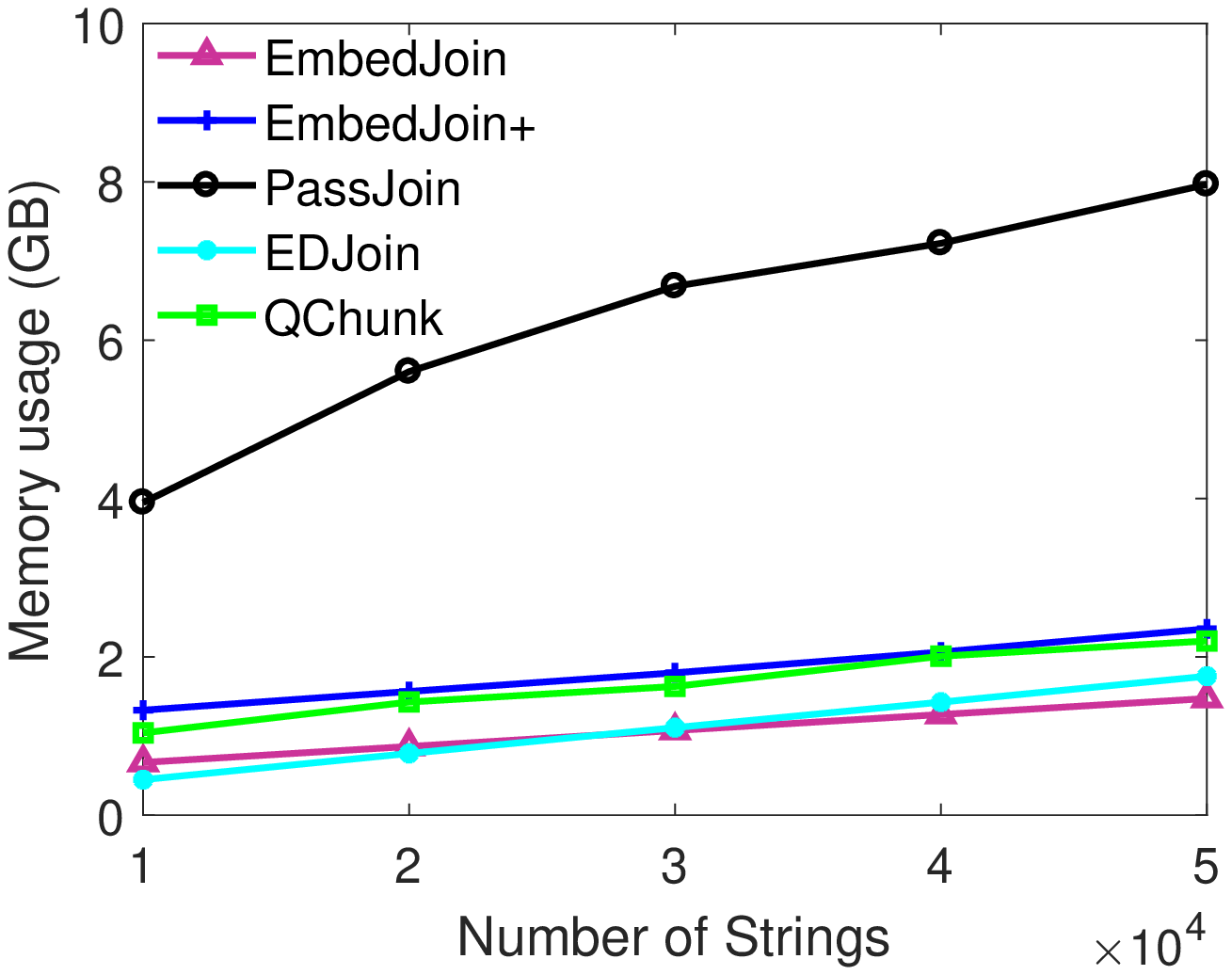}
\centerline{\genoaa\ ($K=100$)}
\end{minipage}
\begin{minipage}[d]{0.4\linewidth}
\centering
\includegraphics[width=0.9\textwidth]{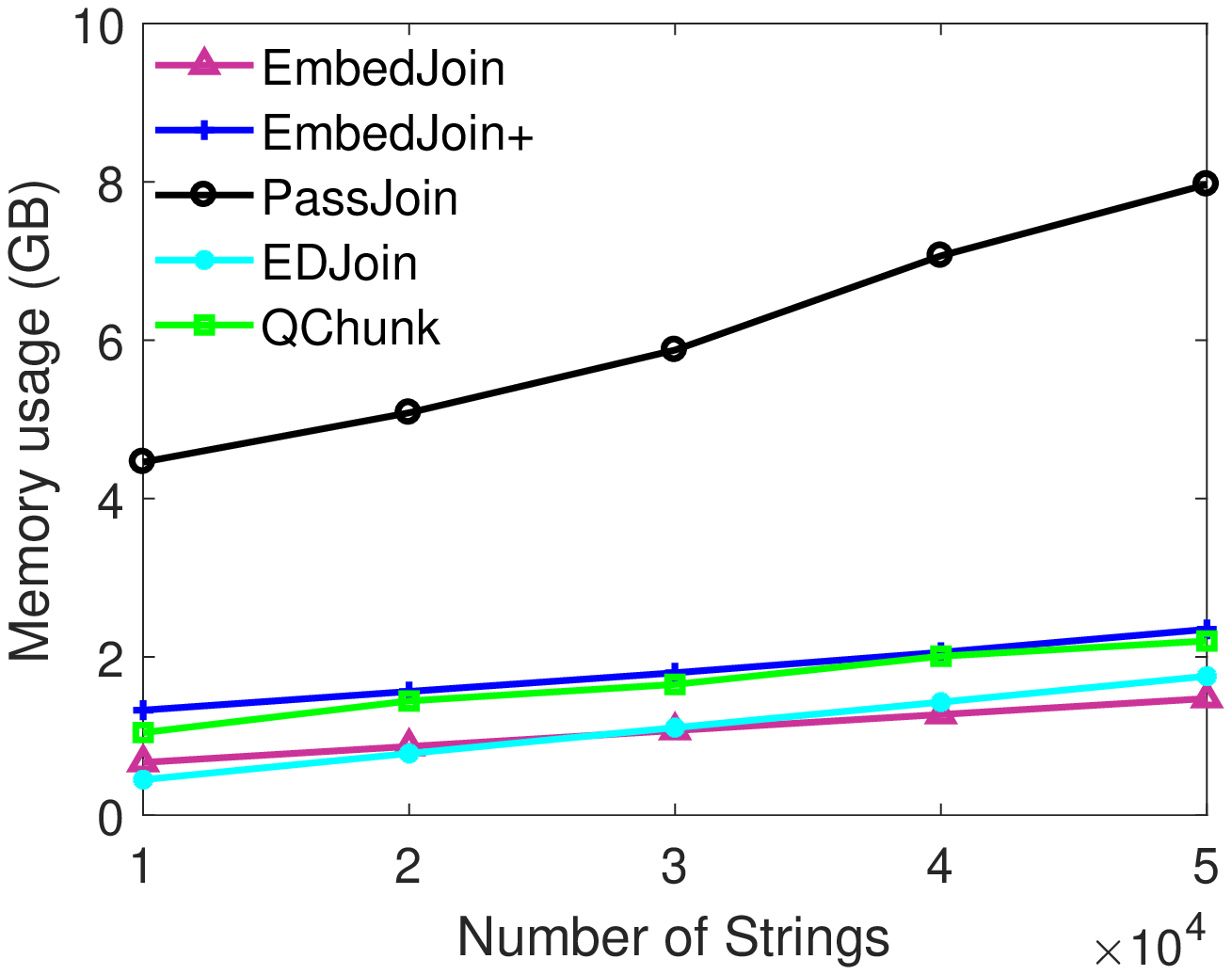}
\centerline{\genoa\ ($K=100$)}
\end{minipage}
\caption{Memory usage, varying $n$.}
\label{fig:nmem}
\end{figure*}

Figure~\ref{fig:nmem} shows the memory usages of different algorithms in the same settings as Figure~\ref{fig:ntime}. The trends of the memory usages of all algorithms are similar; they increase almost linearly with respect to $n$.

\paragraph{The Ultimate Scalability of EmbedJoin+}   
Finally, we present a set of experiments that distinguish \ebdjoin+\ from all the competing algorithms. We test all the algorithms on longer strings (length ranges from 5,000 to 20,000) with larger distance thresholds ($1\% \sim 20\%$ of the corresponding string length).  The numbers of strings in the datasets range from 20,000 to 320,000.  For \ebdjoin\ we fix $r = z = 7$, and set $m= 15 - \lfloor \log_2 x \rfloor$ where $x \%$ is the threshold.  For \ebdjoin+\ we fix $r =  7,z = 16, \Delta = 50$, and set $m= 15 - \lfloor \log_2 x \rfloor$ where $x \%$ is the threshold.  Result points are only depicted for those that can finish in 24 hours, {\em and} return correct answers.  

\begin{figure*}[t]
\centering
\begin{minipage}[d]{0.32\linewidth}
\centering
\includegraphics[width=1\textwidth]{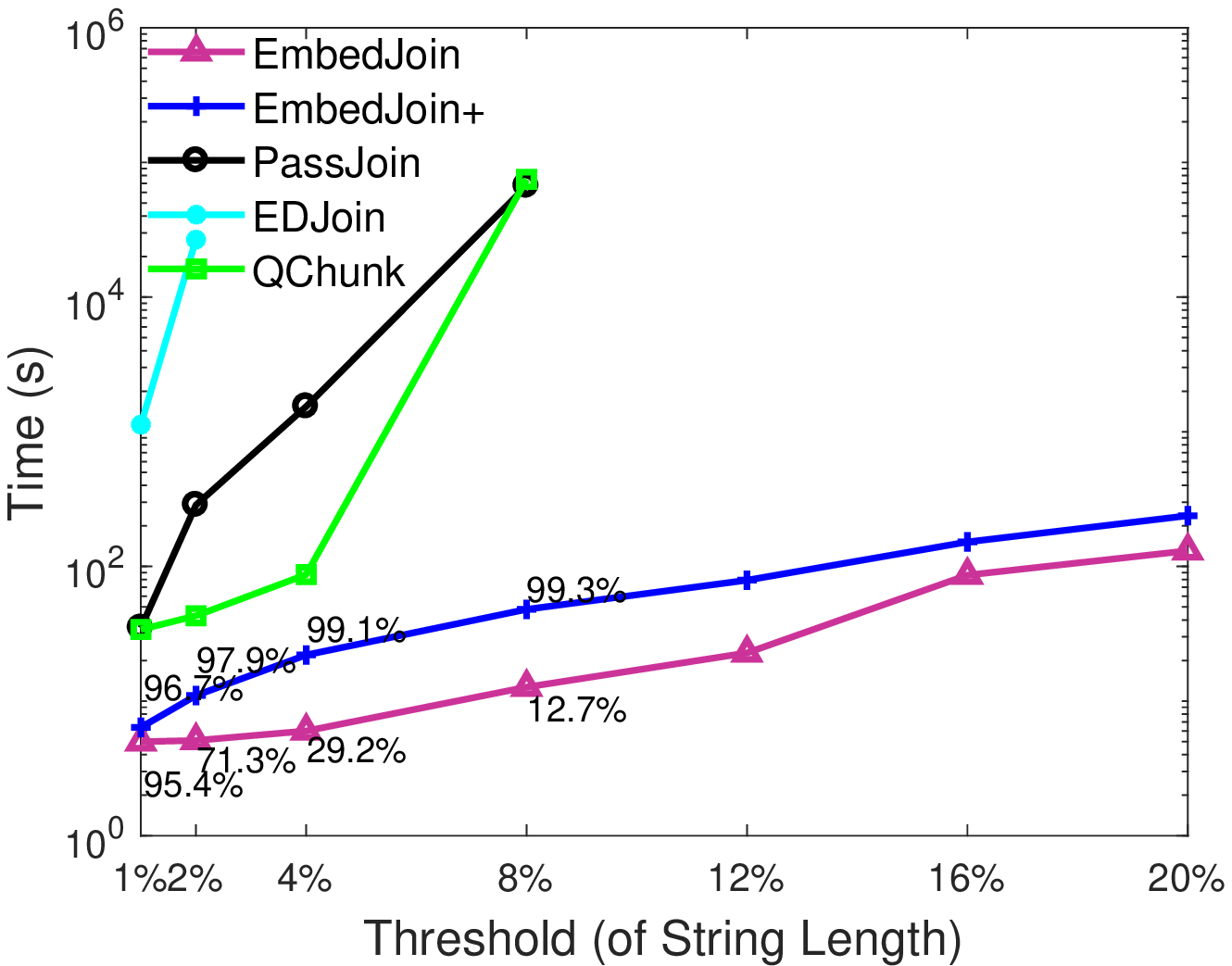}
\centerline{\genob}
\end{minipage}
\begin{minipage}[d]{0.32\linewidth}
\centering
\includegraphics[width=1\textwidth]{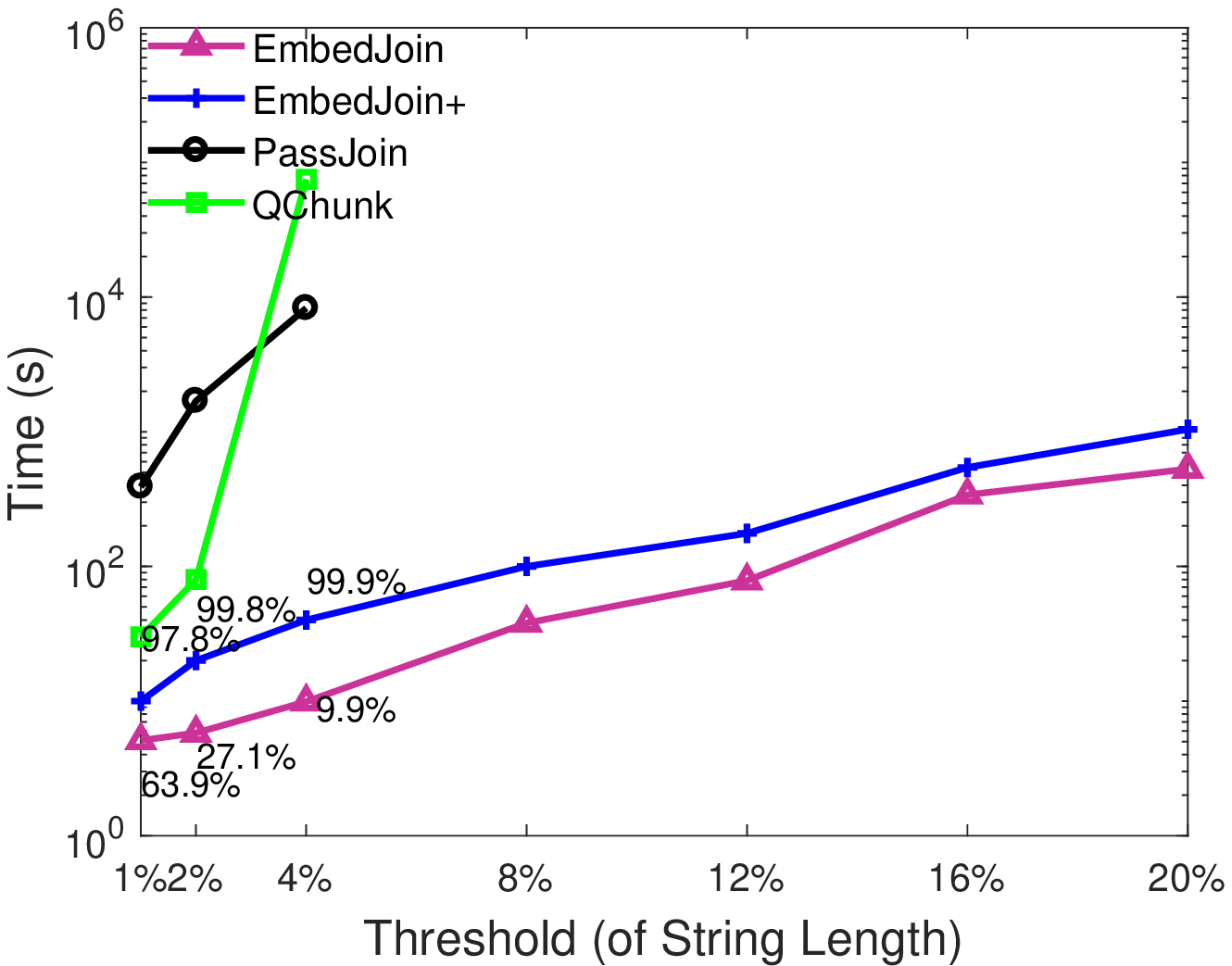}
\centerline{\genoc}
\end{minipage}
\begin{minipage}[d]{0.32\linewidth}
\centering
\includegraphics[width=1\textwidth]{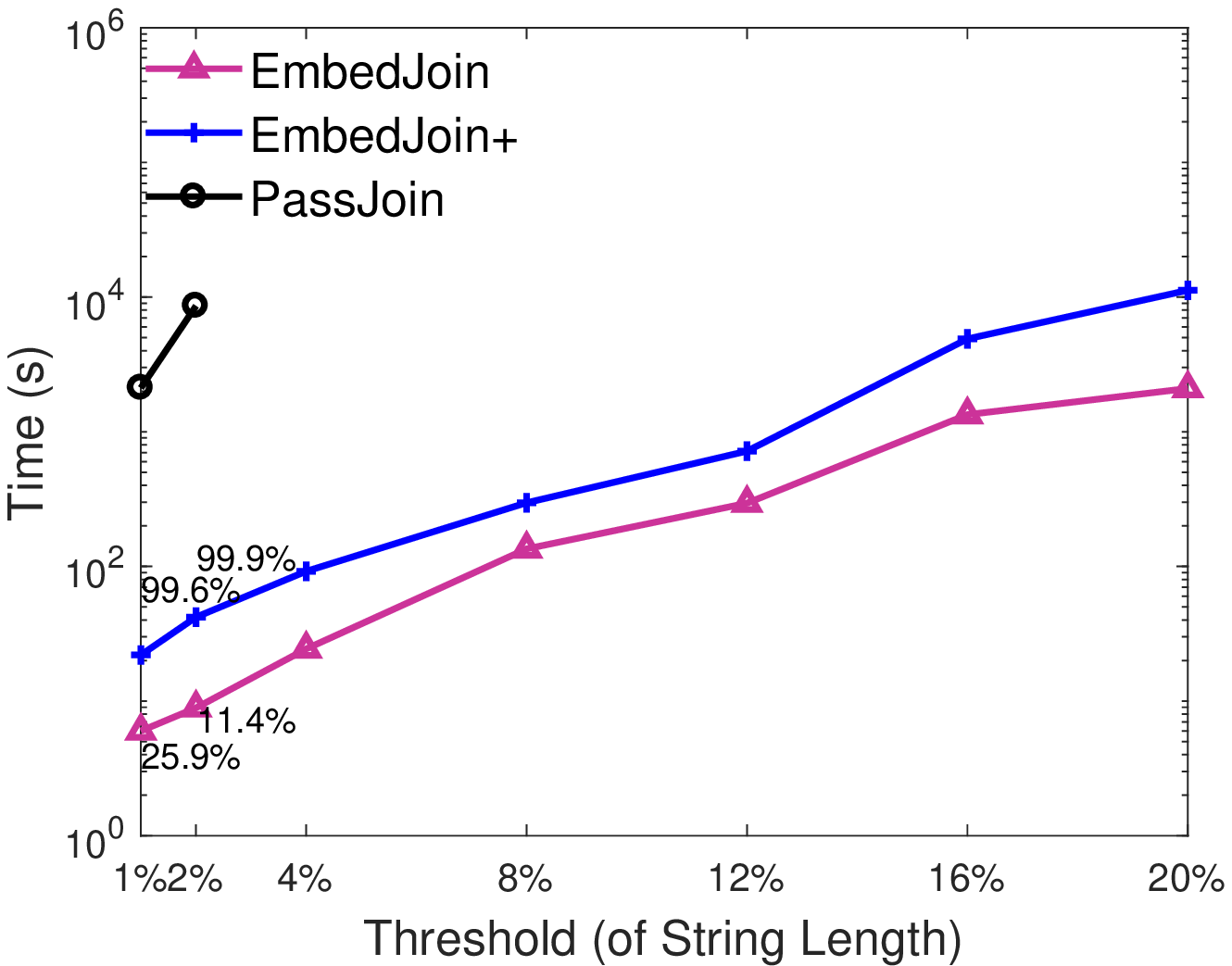}
\centerline{\genod}
\end{minipage}
\caption{Scalability on string length. Percentages on the curves for \ebdjoin/\ebdjoin+\  are their accuracy}
\label{fig:scalek}
\end{figure*}

When varying the string length $N$ (see Figure~\ref{fig:scalek}), there are three other algorithms that can produce data points in the \genob\ dataset: \edjoin\ can report answer up to the $2\%$ distance threshold, and \pass\ and \qchunk\ can go up to $8\%$.  We observe a sharp time jump of \qchunk\  from $4\%$ to $8\%$ -- at the $8\%$ distance threshold \qchunk\ barely finished within 24 hours.  
On \genod, unfortunately, the program for \qchunk\ that we have used cannot produce any data point due to memory overflow.  \pass\ only succeeds at the $2\%$ distance threshold.  

\begin{figure*}[t]
\centering
\begin{minipage}[d]{0.32\linewidth}
\centering
\includegraphics[width=1\textwidth]{dnasmall.eps}
\centerline{\genob}
\end{minipage}
\begin{minipage}[d]{0.32\linewidth}
\centering
\includegraphics[width=1\textwidth]{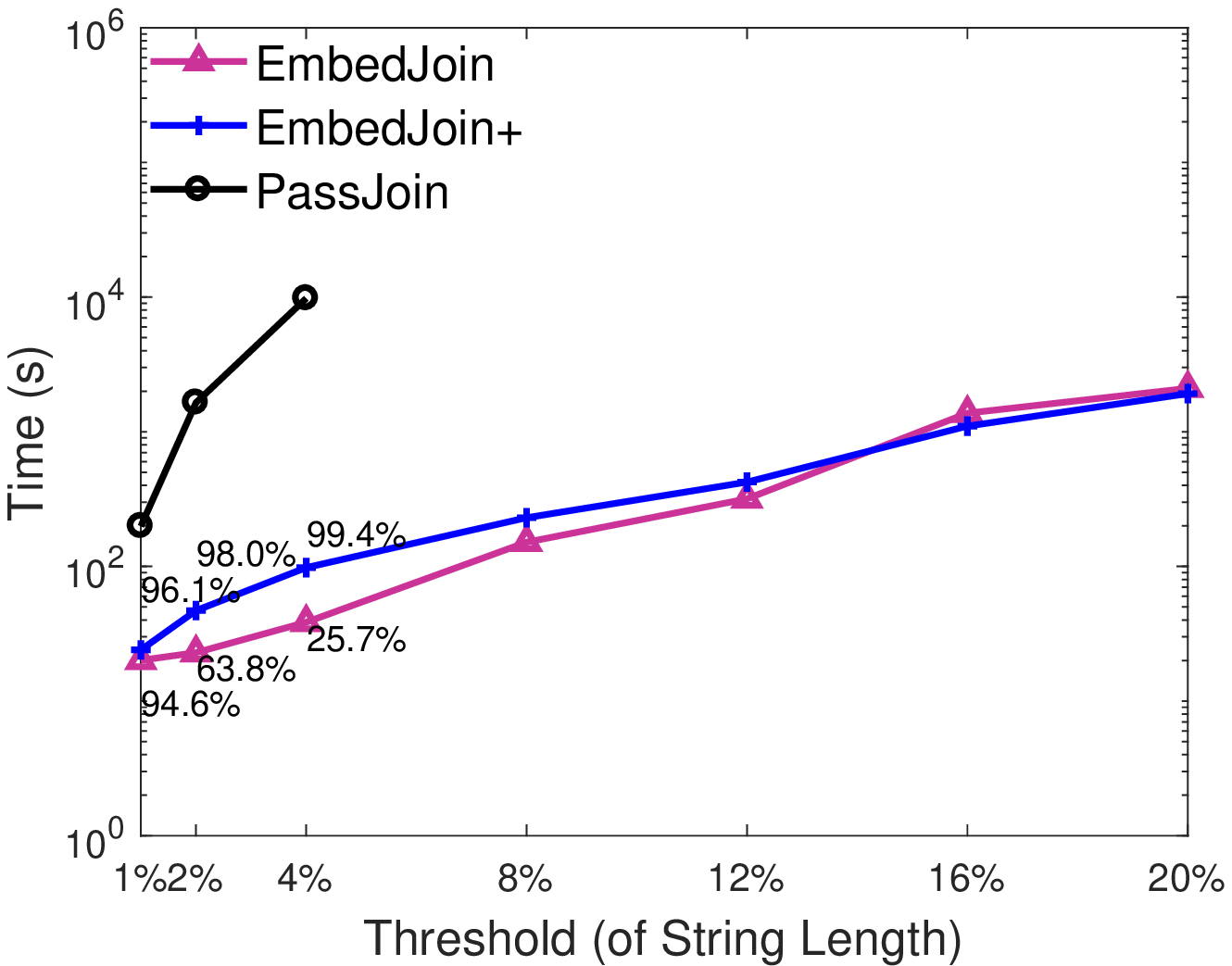}
\centerline{\genoe}
\end{minipage}
\begin{minipage}[d]{0.32\linewidth}
\centering
\includegraphics[width=1\textwidth]{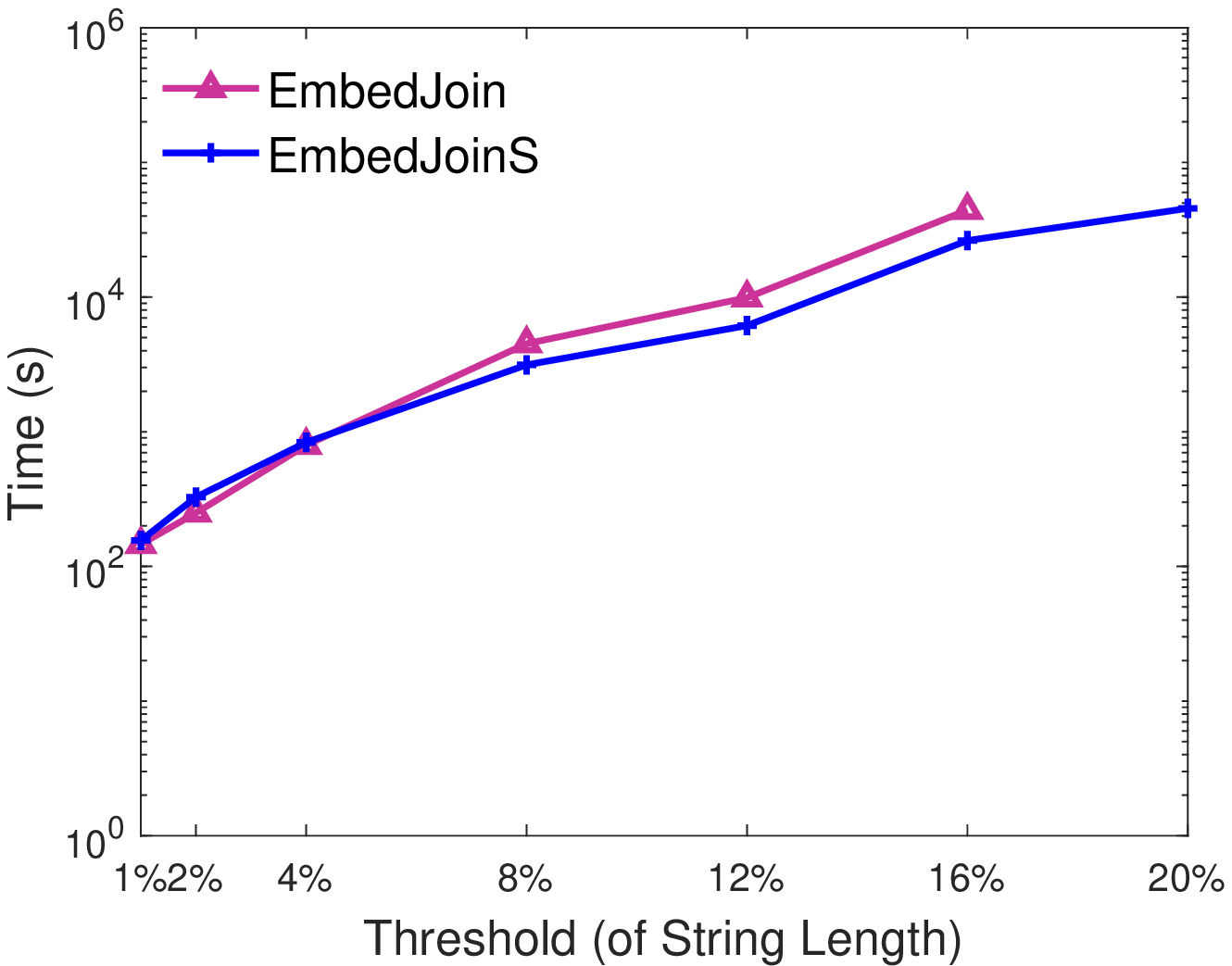}
\centerline{\genof}
\end{minipage}
\caption{Scalability on number of strings. Percentages on the curves for \ebdjoin/\ebdjoin+\  are their accuracy. }
\label{fig:scalen}
\end{figure*}

When varying the number of input strings $n$ (see Figure~\ref{fig:scalen}; the first subfigure of Figure~\ref{fig:scalen} is simply a  repeat of the first subfigure of Figure~\ref{fig:scalek}), all the other computing algorithms cannot produce anything on \genof.  \pass\ manages to produce results on \genoe\ up to $4\%$ distance threshold.  On the other hand, \ebdjoin\ and \ebdjoin+\ scales smoothly on all the datasets. 

The accuracy of \ebdjoin\ decreases sharply with $K$, while \ebdjoin+\ always maintains a good accuracy. The accuracy of \ebdjoin+\ even increases with $K$.  This is because we use a fixed $\Delta$ value for different thresholds $K$, and as a result the number of substrings for each string increases with $K$, which means that the chance for a pair of strings to be chosen as a candidate increases, and consequently the number of false negatives decreases.    We observe that on $\genoe$ and $\genof$ datasets, \ebdjoin+\ has a better time performance than \ebdjoin\ when $K$ is large, even that it needs to spend more time on embedding and hashing. This is because \ebdjoin+\ requires similar pairs to have a pair of substrings with at least $T$ hash signature matches, which decreases number of false positives and consequently saves the verification time. 

To summarize, it is clear that on large datasets with long string,  \ebdjoin+\ performs much better than all the competing algorithms, and scales well up to distance threshold $20\%$.  Unfortunately, we do not know the exact accuracy of \ebdjoin+\ in many points where other exact computation algorithms cannot finish, but from the trends that we have observed on shorter strings and smaller distance thresholds, we would expect that its accuracy will be consistently high.

\section{Conclusion}
\label{sec:conclude}

We propose an algorithm named \ebdjoin+ for computing edit similarity join, one of the most important operations in database systems.  Different from all previous approaches, we first embed the input strings from the edit space to the Hamming space, and then try to perform a filtering (for reducing candidate pairs) in the Hamming space where efficient tools like locality sensitive hashing are available.  Our experiments have shown that \ebdjoin+ significantly outperforms, at a very small cost of accuracy, all existing algorithms on long strings and large thresholds.

\section{ACKNOWLEDGMENT}
The authors would like to thank Djamal Belazzougui and Michal Kouck\'y for many helpful discussions, and Haixu Tang and Diyue Bu for their help on preparing the human genome datasets.  The authors would also like to thank Michal Kouck\'y for introducing us the efficient implementation of the algorithm for computing exact edit distance by Ukkonen~\cite{Ukkonen85}.

\bibliographystyle{acm}
\bibliography{paper}


\end{document}